\documentclass[aps,pre,superscriptaddress,preprint,11pt]{revtex4-1}

\usepackage{amsmath}
\usepackage{color}
\usepackage[final]{graphicx}
\usepackage[english]{babel}
\usepackage{bm,bbold}

\def\al{\alpha}
\def\la{\label}
\def\o{\overline}

\def\r{\rho}
\def\hr{\hat\rho}
\def\del{\delta}

\def\h{\hat}
\def\b{\bf}

\def\l{\lambda}
\def\Gam{\Gamma}

\def\Lam{\Lambda}
\def\S{\Sigma}
\DeclareMathOperator{\Det}{\textsc{Det}}

\begin{document}
\title[Colloidal fluid in a Gaussian random potential]{Dynamics of a
  noninteracting colloidal fluid in a quenched Gaussian random
  potential: A time-reversal-symmetry-preserving field-theoretic
  approach}

\author{Bongsoo Kim}
\affiliation{Department of Physics, Changwon National University,
  Changwon 51140, Korea}
\affiliation{Institute for Soft and Bio Matter Science, Changwon
  National University, Changwon 51140, Korea}

\author{Matthias Fuchs} \affiliation{Fachbereich Physik,
  Universit{\"a}t Konstanz, 78457 Konstanz, Germany}

\author{Vincent Krakoviack} \email{vincent.krakoviack@ens-lyon.fr}
\affiliation{Universit{\'e} de Lyon, ENS de Lyon, Universit{\'e}
  Claude Bernard Lyon 1, CNRS, Laboratoire de Chimie and Centre Blaise
  Pascal, F-69342 Lyon, France}

\date{\today}

\begin{abstract}
We develop a field-theoretic perturbation method preserving the fluctuation-dissipation relation (FDR) for the dynamics of the density fluctuations of a noninteracting colloidal gas plunged in a quenched Gaussian random field.  It is based on an expansion about the Brownian noninteracting gas and can be considered and justified as a low-disorder or high-temperature expansion.  The first-order bare theory yields the same memory integral as the mode-coupling theory (MCT) developed for (ideal) fluids in random environments, apart from the bare nature of the correlation functions involved.  It predicts an ergodic dynamical behavior for the relaxation of the density fluctuations, in which the memory kernels and correlation functions develop long-time algebraic tails.  A FDR-consistent renormalized theory is also constructed from the bare theory.  It is shown to display a dynamic ergodic-nonergodic transition similar to the one predicted by the MCT at the level of the density fluctuations, but, at variance with the MCT, the transition does not fully carry over to the self-diffusion, which always reaches normal diffusive behavior at long time, in agreement with known rigorous results.
\end{abstract}

\maketitle

\section{Introduction}
\setcounter{equation}{0} 

In a number of circumstances, simple fluids may generically develop
slow and complex dynamics.  For instance, glassy dynamics unfolds in
the low-temperature and high-density regimes corresponding to
supercooled or overcompressed liquid states.  It is characterized by a
considerable slowing-down of the structural relaxation, eventually
leading to the fluid falling out of equilibrium at the glass
transition \cite{Cav2009PR, BerBir11RMP, WolLub12book}.  Another
example is provided by fluids in quenched-random environments, with
either geometric or energetic disorder.  Their single-particle
dynamics is often characterized by diffusion anomalies, possibly
leading to diffusion-localization transitions or other types of
nonergodic behaviors \cite{HavBen87AP, BouGeo90PR, HofFra13RPP}.

A versatile framework to investigate such problems on unified grounds
from first principles is provided by the mode-coupling theory (MCT),
more specifically a self-consistent current-relaxation theory, as
termed by G{\"o}tze \cite{LesHouches, GotzeBook}.  In its first few
years, this very scheme could indeed be successively applied to liquid
helium at zero temperature \cite{GotLuc76PRB, GotLucSzp79PRB}, to
noninteracting electrons in a random impurity potential
\cite{Got78SSC, Got79JPC, Got81PMB}, to the random Lorentz gas
\cite{GotLeuYip81aPRA, GotLeuYip81bPRA, Leu83aPRA}, and to simple
glassforming liquids \cite{BenGotSjo84JPC, Leu84PRA}.

It is in the field of glassy dynamics that the MCT has had the
strongest influence.  It was indeed quickly realized that the theory
seems to satisfactorily capture many nontrivial aspects of the
dynamics of simple glassforming liquids, at least on a qualitative or
semi-quantitative level \cite{GotSjo92RPP,Got99JPCM}.  This triggered
and shaped an intensive experimental and computational effort and
stimulated numerous further theoretical developments.  A difficulty
has however been nagging all along, for the main results of the MCT
essentially follow from the analysis of a predicted sharp transition
between a fluid-like ergodic state and a glass-like nonergodic one.
In fact, such a kinetic transition is absent in the actual dynamics of
glassforming liquids, and it must be interpreted as giving rise to a
dynamical crossover in the moderately supercooled or overcompressed
regimes in order to make contact between observations and theory.

From this unsettling situation and the need to clear it up emerged an
interest for theoretical approaches in which the MCT, or a MCT-like
theory, would be the outcome of a well-defined and controlled
approximation scheme, amenable to systematic corrections and
improvements.  Indeed, the original derivation of the MCT within the
Mori-Zwanzig projection-operator formalism does not really lend itself
to such a program, although proposed extensions exist
\cite{GotSjo87ZPB, Sza03PRL, MayMiyRei06PRL, JanMayRei14PRE,
  JanRei15PRL, JanMayRei16JSMTE}.  By contrast, field-theoretic
approaches appear as methods of choice for such a purpose, and a
number of them have accordingly been developed
\cite{DasMazRamTon85PRL, DasMaz86PRA, SchDufDe93PRL, KawMiy97ZPB,
  MiyRei05JPA, AndBirLef06JSMTE, KimKaw07JPA, KimKaw08JSMTE,
  BasRam07JSMTE, JacWij11PRL, KimKawJacWij14PRE}.  In particular, the
most recent studies have paid special attention to the symmetries of
the dynamical action, from which crucial equilibrium results readily
stem, such as the fluctuation-dissipation relation (FDR)
\cite{AndBirLef06JSMTE, KimKaw07JPA, KimKaw08JSMTE, BasRam07JSMTE,
  JacWij11PRL, KimKawJacWij14PRE, VelChaCugKre08JPA}.  It is actually
one of the great strengths of field theories to offer command on these
aspects.

In the present work, we follow the lead of the latter studies, but,
instead of glassforming liquids, we focus on noninteracting fluids in
quenched-random environments.  This indeed appears as an interesting
new window on the use of field theory and its relation with MCT,
complementary to what has already been done.  Note that, within MCT,
the presence of interactions does not actually lead to any particular
technical difficulty \cite{Kra05PRL, Kra07PRE, Kra09PRE, KonKra17SM}.
However, for more general considerations, it clearly seems advisable
to first isolate the effects of disorder from those of interactions,
hence the present restriction to noninteracting systems.  In this
respect, it should be borne in mind that the dynamics of a pure
noninteracting gas, while essentially trivial in a particle-based
formalism, is not so simple from a field-theoretic perspective
\cite{VelChaCugKre08JPA}.

Before being more specific about our approach, it is worth mentioning
that fluids in quenched-random environments have recently received
renewed attention, thanks to ingenious experimental developments
leading to novel realizations of such systems.  Important examples,
further investigated by computer simulations, include colloids and
aerosols in optical speckle patterns \cite{HanDalSchJenEge12SM,
  HanEge12JPCM, EveZunHanBewLadHeuEge13PRE, HanSchEge13PRE,
  BewEge16PRA, BewLadPlaZunHeuEge16PCCP, BewSenCapPlaSenEge16JCP,
  ShvRodIzdDesKroKiv10OE, ShvRodIzdLeyDesKroKiv10JO, VolVolGig14SR,
  VolKurCalVolGig14OE, YokAiz17OLT}, binary mixtures of
superparamagnetic particles squeezed between glass slides
\cite{SkiSchAarHorDul13PRL, SchSpaHofFraHor15SM,
  SchSkiThoAarHorDul17PRE, SchHor18PRL}, and colloids diffusing over
rough randomly packed colloidal monolayers \cite{SuMaLaiTon17SM}.
Therefore, beyond purely technical considerations, it also seems
timely to try and achieve further theoretical progress in this field.

In practice, we here study the equilibrium dynamics of the density
fluctuations of a gas of noninteracting Brownian particles plunged in
a random external potential-energy landscape with Gaussian statistics.
This specific nature of the disorder indeed appears as particularly
well suited for our initial field-theoretic developments, being itself
formulated as a very simple and nonsingular field theory.

The time evolution of the density fluctuations is governed by the
so-called Dean-Kawasaki (DK) equation (generalized to include the
random potential), a nonlinear Langevin equation for the density field
with a multiplicative thermal noise \cite{Dea96JPA, Kaw94PA}.  Using
the functional formalism of Martin-Siggia-Rose-Janssen-de Dominicis
(MSRJD) \cite{MarSigRos73PRA, Jan79LNP, Dom76JPColl}, this equation
can be turned into a dynamical action functional.  As alluded above,
it was recently recognized that such an action possesses properties of
time-reversal (TR) invariance under specific sets of field
transformations, intimately connected to the FDR
\cite{AndBirLef06JSMTE, AroBirCug10JSMTE}.  These TR symmetries can
play the role of guiding principles as to how to develop perturbation
theories consistent with the FDR at each order of expansion.  Indeed,
a difficulty that defeats too naive approaches is that the Gaussian
and non-Gaussian components of the action are not separately invariant
under these field transformations \cite{MiyRei05JPA,AndBirLef06JSMTE}.
One such a FDR-preserving theory for the full DK equation (with
interactions and without random potential) has recently been developed
via the linearization of one of the TR transformations, called the
$\mathcal{U}$-transformation, at the expense of introducing a new set
of conjugated fields.  Further details can be found in
Ref.~\cite{KimKawJacWij14PRE}.

Another TR transformation, known as the $\mathcal{T}$-transformation,
suggests an expansion about the pure noninteracting system as a
possible approach.  The action is thus decomposed into its free and
disorder-induced components, then the latter is treated perturbatively
around the former which is non-Gaussian.  This procedure can actually
be considered and motivated as a weak-disorder or high-temperature
expansion.  The rationale behind this scheme is that the corresponding
nonlinear TR field transformation leaves separately invariant the two
decomposed parts of the action.  Consistency with the FDR however
requires the free part of the action to be treated \emph{exactly}.
Notwithstanding its non-Gaussianity due to the multiplicative nature
of the thermal noise, this is made possible thanks to the special form
of its cubic nonlinearity (quadratic in the noise-response field) and
to causality.  This aspect is a novel feature of this perturbation
method.  It is also advantageous that one is freed from introducing
extra fields into the problem.

The present method \emph{per se} is a bare perturbation theory, that
is, the perturbative corrections are naturally expressed in terms of
the bare correlation and response functions.  It is not a loop
expansion, and it would be a challenge to develop the
two-particle-irreducible effective action method for the strongly
non-Gaussian noninteracting gas.  Note that the approach could also be
applied to the full DK equation, with a perturbative treatment of the
interactions.  This will be examined separately in the future.

We now summarize the main results of our work.  The first-order bare
theory (FOBT) gives a dynamical equation for the density correlation
function that can be put in the same form as that of the
self-consistent MCT developed by one of us \cite{Kra05PRL, Kra07PRE,
  Kra09PRE, KonKra17SM}, albeit with a memory term written in terms of
the bare correlation function [see Eqs.~\eqref{eqBMCT2} and
  \eqref{eqFOBTM} below].  From this equation, one can compute the
mean-squared displacement (MSD) and characterize the long-time tails
that develop due to the quenched randomness.  The corresponding
dynamics is found to always remain ergodic, until the theory breaks
down at too strong disorder.

A first-order renormalized theory (FORT) has also been developed out
of the bare perturbation theory.  It is self-consistently derived from
a second-order bare calculation, with empirical adjustments
constrained by the requirements of consistency with the FDR and with
the FOBT, and eventually singled out through numerical considerations.
This theory is distinct from the MCT, but shows some structural
similarity with it.  In particular, a self-closed dynamical equation
for the density correlation function is again obtained [see
  Eqs.~\eqref{eqFORT17}--\eqref{eqFORT19} below].  However, its
predictions noticeably improve upon those of the MCT.  Indeed, an
ergodicity-breaking transition is still predicted for the density
fluctuations, but, at variance with the MCT, it does only partially
carry over to the MSD, which always reaches a normal diffusive
behavior at long time, in agreement with the known rigorous results
\cite{MasFerGolWic89JSP}.  Note that, if the Brownian dynamics is
replaced with a Newtonian energy-conserving one, then a
diffusion-localization transition does occur \cite{Zim68JPC,
  ZalSch71PRB, Isi92RMP}, as found in the MCT.  Therefore, the
confrontation of the MCT and of the present theory might well
represent a first step towards an understanding of the phenomenon of
avoided or rounded kinetic transitions.

The paper is organized as follows.  In Secs.~\ref{sec:DK} and
\ref{sec:GRF}, we present the time evolution equation for the density
fluctuations of Brownian particles in a frozen Gaussian random
potential, and the corresponding dynamical action.  The time-reversal
symmetries of the action and the resulting FDRs are contained in
Sec.~\ref{sec:TRS}.  Section \ref{sec:expansion} describes the
FDR-preserving perturbation expansion method about the non-Gaussian
pure noninteracting state.  We write down in Sec.~\ref{sec:nonpert}
the nonperturbative form of the dynamical equations for the
correlation and response functions.  Section \ref{sec:zeroth} recalls
the solution for the pure noninteracting reference state.  Sections
\ref{sec:firstorder} to \ref{sec:FORT} present the main results of the
paper, namely, the first-order perturbation corrections to the simple
free diffusion.  Summary and outlook are given in the last section.

\section{Time-evolution equation for the density fluctuations of
  colloidal particles moving in a (random) external potential}
\setcounter{equation}{0}
\label{sec:DK}

In the present work, we investigate a situation where $N$ colloidal
particles in a volume $V$, hence the average fluid density $\r_0 =
N/V$, move in a (random) external potential.  The particle positions
are denoted by $\{{\bf r}_i\}$, $i=1,2,\dots,N$.  As a first step, the
derivation of the time-evolution equation for the density fluctuations
of these particles is required. This task can be carried out in a
rather general way, following an approach due to Dean \cite{Dea96JPA}.
We consider the case of interacting particles, as this does not
introduce any particular difficulty at this stage.

The motion of the individual particles is assumed to be described by
the overdamped Langevin equation,
\begin{equation} \la{eq1.1}
  \dot {\bf r}_i(t) = \frac{D_0}{T} {\bf F}_i(t) + {\bf f}_i(t),
\end{equation}
where $D_0$ is the bare diffusion coefficient, $T$ is the temperature
of the system (the Boltzmann constant $k_B$ is set to unity
throughout), and ${\bf f}_i(t)$ is a Gaussian thermal noise with zero
mean and variance
\begin{equation} \la{eq1.2}
  \langle f_i^\alpha(t) \, f_j^\beta(t') \rangle = 2 D_0 \del_{ij}
 \delta_{\alpha\beta} \del(t-t'),
\end{equation}
$\alpha$ and $\beta$ denoting vector components in Cartesian
coordinates.  The force ${\bf F}_i(t) $ acting on the $i$th particle
is given by
\begin{equation} \la{eq1.3}
  {\bf F}_i(t) = {\bf F}^\text{int}_i(t) + {\bf F}^\text{ext}_i(t),
\end{equation}
where 
\begin{equation} \la{eq1.4}
  {\bf F}^\text{int}_i(t) = -\frac{\partial}{\partial {\bf r}_i(t)}
  \sum_{j=1}^{N} u(|{\bf r}_i(t)-{\bf r}_j(t)|)
\end{equation}
is due to the interactions between the fluid particles with pair
potential $u(r)$ [for simplicity, $\nabla u(0) = \boldsymbol{0}$ is
  assumed], and
\begin{equation} \la{eq1.5}
  {\bf F}^\text{ext}_i (t) = -\frac{\partial v({\bf r}_i(t))}{\partial
    {\bf r}_i(t)}
\end{equation}
derives from the external potential with one-body potential energy
$v({\bf r})$.

The microscopic fluid density is defined as
\begin{equation} \la{eq1.6}
  \r ({\bf r},t) \equiv \sum_{i=1}^{N} \del ({\bf r}-{\bf r}_i (t)) =
  \sum_{i=1}^{N} \r_i({\bf r},t), 
\end{equation}
where we introduced the single-particle densities, $\r_i({\bf r},t)
\equiv \del ({\bf r}-{\bf r}_i (t))$, $i=1,2,\dots,N$.  Its
fluctuations about the average fluid density are denoted by
\begin{equation}
  \del \r ({\bf r},t) \equiv \r({\bf r},t)-\r_0.
\end{equation}

In order to derive the dynamical equation for $\r({\bf r},t) $, we
follow the It\^o prescription.  Consider the following set of
stochastic equations for the variables $x_{a}(t)$ (using the summation
convention),
\begin{equation} \la{eq1.7}
  \frac{d x_{a}(t)}{dt} = h_{a} + g_{ab} \xi_{b}(t), 
\end{equation}
where the correlation of the Gaussian white noise $\xi_{b}(t)$ is
defined as
\begin{equation} \la{eq1.8}
  \langle \xi_{b}(t) \xi_{b'}(t') \rangle =\del_{b b'}\del(t-t').
\end{equation}
The It\^o chain rule then gives the stochastic equation for a variable
$y[\{x\}]$ in the form
\begin{equation} \la{eq1.9}
  \frac{d y(t)}{dt} = \frac{d x_{a}(t)}{dt} \frac{\partial y}{\partial
    x_{a}} + \frac{1}{2} \frac{\partial^2 y}{\partial x_{a} \partial
    x_{b}} g_{ac} g_{cb}.
\end{equation}
Using this rule, we get the dynamical equation
\begin{equation} \begin{split} \la{eq1.10}
  \partial_t \r ({\bf r},t) & = D_0 \nabla^2 \r ({\bf r},t) -
  \sum_{i=1}^{N} \dot {\bf r}_i(t) \cdot \nabla \r_i ({\bf r},t) \\
  & = D_0 \nabla^2 \r ({\bf r},t) - \sum_{i=1}^{N} \nabla \r_i ({\bf
    r},t) \cdot \left[ \frac{D_0}{T}{\bf F}_i(t) + {\bf f}_i(t)
    \right].
\end{split} \end{equation}
One can express the force contributions in Eq.~\eqref{eq1.10} in terms
of the fluid density, as
\begin{equation} \la{eq1.11}
  \begin{split}
    -\sum_{i=1}^{N} \nabla \r_i({\bf r},t) \cdot {\bf
      F}^\text{int}_i(t)
  & = \nabla \cdot \left[ \sum_{i=1}^{N} \del ({\bf r}-{\bf r}_i(t))
      \frac{\partial}{\partial {\bf r}_i(t)} \int d{\bf r}' \, u(|{\bf
        r}_i(t)-{\bf r}'|) \sum_{j=1}^{N} \del ({\bf r}'-{\bf r}_j(t))
      \right] \\
 & = \nabla \cdot \left[ \r({\bf r},t) \nabla \int d{\bf r}' u(|{\bf
        r}-{\bf r}'|) \r({\bf r}',t) \right]
 \end{split}
\end{equation}
and
\begin{equation} \la{eq1.12}
  -\sum_{i=1}^{N} \nabla \r_i({\bf r},t) \cdot {\bf F}^\text{ext}_i(t)
  = \nabla \cdot \left[ \sum_{i=1}^{N} \del ({\bf r}-{\bf r}_i(t))
    \frac{\partial v({\bf r}_i(t))}{\partial {\bf r}_i(t)} \right]
  = \nabla \cdot \left[ \r({\bf r},t) \nabla v({\bf r}) \right].
\end{equation}
Also, the thermal noise defined as $\boldsymbol{\eta}({\bf r},t)
\equiv - \sum_{i=1}^{N} \r_i({\bf r},t) {\bf f}_i(t)$ keeps a Gaussian
character with zero mean and correlations given by
\begin{equation} \la{eq1.13}
 \begin{split}
  \langle \eta^\alpha({\bf r},t) \eta^\beta ({\bf r}',t') \rangle & =
  \sum_{i=1}^{N} \sum_{j=1}^{N} \del({\bf r}-{\bf r}_i(t)) \del ({\bf
    r}'-{\bf r}_j(t')) \langle f_i^\alpha(t) f_j^\beta(t') \rangle \\
  & = 2 D_0 \r({\bf r},t) \delta_{\alpha\beta} \del({\bf r}-{\bf r}')
  \del(t-t').
 \end{split}
\end{equation}
Substituting Eqs.~\eqref{eq1.11}-\eqref{eq1.13} into
Eq.~\eqref{eq1.10}, one obtains the desired dynamical equation,
\begin{multline} \la{eq1.14}
  \partial_t \r ({\bf r},t) = D_0 \nabla^2 \r ({\bf r},t) +
  \frac{D_0}{T} \nabla \cdot \left[ \r({\bf r},t) \nabla \int d{\bf
      r}' u(|{\bf r}-{\bf r}'|) \r ({\bf r}',t) \right] \\ +
  \frac{D_0}{T} \nabla \cdot \left[ \r({\bf r},t) \nabla v({\bf r})
    \right] + \nabla \cdot \left[ \sqrt{ \r ({\bf r},t)} {\boldsymbol
      \xi}({\bf r}, t) \right],
\end{multline}
where ${\boldsymbol \xi}({\bf r}, t)$ is a Gaussian thermal noise with
zero mean and variance
\begin{equation}
  \langle \xi^\alpha({\bf r}, t) \, \xi^\beta({\bf r}', t') \rangle =
  2 D_0 \delta_{\alpha\beta} \del({\bf r}-{\bf r}') \del(t-t').
  \la{eq1.15}
\end{equation}

Equation~\eqref{eq1.14} can be expressed in terms of a free-energy
density functional ${\cal F}[\r;v]$ as
\begin{equation} \la{eq1.16}
  \partial_t \r ({\bf r},t) = \frac{D_0}{T} \nabla \cdot \left[
    \r({\bf r},t) \nabla \left. \frac{\del {\cal F}[\r;v]}{\del
      \r({\bf r})} \right|_{\rho({\bf r},t)} \right] + \nabla \cdot
  \left[ \sqrt{ \r ({\bf r},t)} {\boldsymbol \xi}({\bf r}, t) \right],
\end{equation}
where
\begin{subequations} \la{eq1.17}
\begin{align}
  {\cal F}[\r;v] & = {\cal F}_\text{id}[\r] + {\cal F}_\text{int}[\r]
  + {\cal F}_\text{ext}[\r;v], \\
  {\cal F}_\text{id}[\r] & = T \int d {\bf r} \, \r({\bf r}) \left[\ln
    \left(\r ({\bf r})/\r_0 \right) -1 \right], \\
  {\cal F}_\text{int}[\r] & = \frac{1}{2} \iint d{\bf r} d {\bf r}'
  u(|{\bf r}-{\bf r}'|) \del\r({\bf r}) \del\r({\bf r}'), \\
  {\cal F}_\text{ext}[\r;v] & = \int d{\bf r} \, v({\bf r}) \del\r({\bf
    r}).
\end{align}
\end{subequations}
The Fokker-Planck equation for
Eq.~\eqref{eq1.16} reads
\begin{equation} \la{eq1.18}
  \frac{\partial }{\partial t} P[\r,t;v] = -D_0 \int d{\bf r}
  \frac{\del}{\del \r({\bf r})} \nabla \cdot \r ({\bf r}) \nabla
  \left[ \frac{\del}{\del \r({\bf r})} +\frac{1}{T}\frac{\del {\cal
        F}[\r;v]}{\del \r({\bf r})} \right] P[\r,t;v].
\end{equation}
Evidently, the equilibrium Boltzmann distribution $P_\text{eq}[\r;v]
\propto \exp \left( -{\cal F}[\r;v]/T \right)$ is a stationary
solution of this equation.

Finally, within the functional formalism of MSRJD
\cite{MarSigRos73PRA, Jan79LNP, Dom76JPColl}, the time evolution
described by Eqs.~\eqref{eq1.14}-\eqref{eq1.16} can be recast into a
dynamical generating functional
\begin{equation} \la{eq1.19}
  Z[l,\hat{l};v] = \int D\r \int D\hr \, J(\r) e^{S[\r, \hr;v]}
  e^{\int_{{\bf r},t} [\r({\bf r},t) l({\bf r},t) + \hr({\bf r},t)
      \hat{l}({\bf r},t)]},
\end{equation}
where the action $S[\r, \hr;v]$ takes the form
\begin{equation} \la{eq1.20}
  S[\r,\hr;v] = \int_{{\bf r},t} \left\{ i\hr({\bf r},t) \left(
  \partial_t \r({\bf r},t) - \frac{D_0}{T} \nabla \cdot \left[ \r({\bf
      r},t) \nabla \left. \frac{\del {\cal F}[\r;v]}{\del \r({\bf r})}
    \right|_{\rho({\bf r},t)} \right] \right) -D_0 \r({\bf r},t)
          [\nabla \hr({\bf r},t)]^2 \right\},
\end{equation}
with $\int_{{\bf r},t} \equiv \int d{\bf r} \int dt$.  Here, the
thermal average has already been performed, and the term proportional
to $\r(\nabla\hr)^2$ comes from the average over the multiplicative
thermal noise.  The Jacobian $J(\r)$ guarantees that the normalization
condition $Z[l=0,\hat{l}=0;v]=1$, of critical importance in
applications of the formalism to quenched-disordered systems, indeed
holds.  In the It\^o discretization scheme, $J(\r)$ becomes a constant
and can be absorbed into the functional measure. From the knowledge of
$Z[l,\hat{l};v]$, the time-dependent correlation functions of the
fields $\r$ and $\hr$ can be straightforwardly obtained as functional
derivatives with respect to $l$ and $\hat{l}$ at $l=0$, $\hat{l}=0$.
More generally, dynamical quantities averaged over the thermal noise
can be evaluated with respect to the action $S[\r,\hr;v]$ as
\begin{equation} \la{eq1.21}
  \langle A[\r,\hr] \rangle = \int_{\r,\hr} A[\r,\hr] e^{S[\r,\hr;v]},
\end{equation}
where $\int_{\r, \hr} \equiv \int D\r \int D\hr$ and $\langle \cdots
\rangle$ generically denotes a thermal average.

\section{Noninteracting Brownian gas in a Gaussian random potential}
\label{sec:GRF}
\setcounter{equation}{0}

We may now specialize the above equations in accordance with the aim
of the present study, which is to investigate the effect of a
quenched-random environment on the dynamics of colloids.  To this end,
we consider what appears to be the simplest nontrivial case.  First,
in most of this work, we will simply ignore the particle interactions
and set $u(r)=0$ for all $r$, in order to merely focus on the aspect
of quenched disorder.  Second, the one-body potential energy function
$v({\bf r})$, from which the external potential is built, should be
sampled from a convenient functional probability space.  A natural
option is to turn to a homogeneous and isotropic Gaussian random
field, whose statistical properties are fully encoded in its mean,
which can be set to zero without loss of generality, and its
covariance.  Therefore, we shall assume Gaussian statistics for
$v({\bf r})$, with
\begin{equation} \la{eq3.1}
  \overline{v({\bf r})} \equiv 0, \qquad \overline{v({\bf r})v({\bf
      r'})} \equiv w \Phi(|{\bf r}-{\bf r}'|),
\end{equation}
where $\overline{\cdots}$ denotes an average over the random-field
distribution.  The normalized random-field covariance $\Phi(r)$ obeys
$\Phi(0)=1$, so that $w$ appears as a straightforward measure of the
disorder strength.  It will determine the behavior of the system and
should be compared with the typical thermal energy fluctuations, a
purpose readily served by a single dimensionless control parameter
representing the relative disorder strength, $\l \equiv w/T^2$.

Although we choose to introduce Gaussian statistics for the external
potential from the outset, it might be useful to recall that this
represents a common assumption in a number of simple circumstances of
interest.  For instance, a standard argument based on the central
limit theorem and used in a variety of related problems
\cite{HalLax66PR, SimDobStr90JCP, DeeCha94JSP} states that the
one-body potential generated by a statistically homogeneous frozen
matrix of randomly placed interaction sites is expected to develop
Gaussian statistics under suitable conditions, as it is a sum of a
large number of random fluid-matrix pair interactions in the
thermodynamic limit.  It should nevertheless be stressed that,
although this argument can be made rigorous in some special limits
\cite{LifGrePasbook, AkkMonbook}, it can lead to difficulties in more
generic cases \cite{DeaDruHorLef04JPA, DeaDruHor07JSM}.  Other
possible situations expected to yield Gaussian random fields are
associated with linear combinations of random Fourier modes
\cite{Kra70PF, TouDea07JPA} or with coarse-graining of a random field,
be it Gaussian or not, over extended enough regions
\cite{ChuDic98PRB}.  The latter approach is practically relevant to
polarizable colloids in speckle patterns, in the regime where the
effective external potential results from the integrated effect of the
random light intensity field over the whole volume of a particle
\cite{HanDalSchJenEge12SM, EveZunHanBewLadHeuEge13PRE, BewEge16PRA}.

Before considering the dynamics, a few structural properties of the
system should be derived.  From a configurational point of view, one
actually deals with an ideal gas in an external potential.  Its
one-particle configurational integral is readily shown to be
self-averaging, with the nonrandom limit
\begin{equation} \la{eq3.5}
  \lim_{V\to+\infty} \frac{1}{V} \int_V d{\bf r} \, e^{-v({\bf r})/T}
  = \overline{e^{-v({\bf r})/T}} = e^{\lambda/2}.
\end{equation}
Therefore, for any single realization of $v({\bf r})$ in the
thermodynamic limit, one straighforwardly gets
\begin{subequations} \la{eq3.4}
\begin{gather}
  \langle \r ({\bf r}) \rangle = \r_0 e^{-\lambda/2} e^{-v({\bf
      r})/T}, \la{eq3.4a} \\
  \langle \r ({\bf r}) \r ({\bf r'}) \rangle = \r_0 e^{-\lambda/2}
  e^{-v({\bf r})/T} \delta({\bf r}-{\bf r'}) + \r_0^2 e^{-\lambda}
  e^{-[v({\bf r}) + v({\bf r'})]/T}, \la{eq3.4b}
\end{gather}
\end{subequations}
where the normalization factors precisely stem from the one-particle
configurational integral.  Computing now disorder averaged quantities,
one gets $\overline{\langle \r ({\bf r}) \rangle} = \r_0$, as it
should, and
\begin{subequations} \la{eq3.6}
\begin{gather}
  C_\text{st}(|{\bf r}-{\bf r}'|) \equiv \overline{\langle \del\r({\bf
      r}) \del\r({\bf r'}) \rangle} = \r_0 \delta({\bf r}-{\bf r'}) +
  \r_0^2 \left[ e^{\lambda \Phi(|{\bf r}-{\bf r}'|)} - 1 \right], \\
  C_\text{d}(|{\bf r}-{\bf r}'|) \equiv \overline{\langle
    \delta\r({\bf r}) \rangle \langle \delta\r({\bf r'}) \rangle} =
  \r_0^2 \left[ e^{\lambda \Phi(|{\bf r}-{\bf r}'|)} - 1 \right],
  \la{eq3.6b}
\end{gather}
\end{subequations}
where $C_\text{st}(r)$ denotes the static density correlation function
and $C_\text{d}(r)$ the so-called disconnected density correlation
function.  In reciprocal space, the same density correlations are
described in terms of the static and disconnected structure factors,
$S^\text{st}_k$ and $S^\text{d}_k$.  They are obtained by Fourier
transforming $C_\text{st}(r)$ and $C_\text{d}(r)$, respectively, and
normalizing by $\r_0$. Since
\begin{equation} \la{eq3.7}
  C_\text{st}(|{\bf r}-{\bf r}'|) - C_\text{d}(|{\bf r}-{\bf r}'|) =
  \r_0 \delta({\bf r}-{\bf r'}),
\end{equation}
the structure factors obey $S^\text{st}_k=1+S^\text{d}_k$. Note that,
in fact, both equalities generically hold for a noninteracting gas in
any type of homogeneous and isotropic random environment.  The actual
dependence of the above structural quantities on $|{\bf r}-{\bf r}'|$
or on the wavevector modulus $k$ follows from this property of
homogeneity and isotropy.

We now turn to dynamics.  Since $Z[l=0,\hat{l}=0;v]$ is normalized to
1, hence independent from any specific random potential-energy
realization, the noise-averaged dynamical quantities given by
Eq.~\eqref{eq1.21} can be further disorder-averaged as \cite{Dom78PRB}
\begin{equation} \la{eq3.8}
  \overline{\langle A[\r,\hr] \rangle} = \int_{\r, \hr} A[\r,\hr]
  \overline{e^{S[\r, \hr;v]}} \equiv \int_{\r, \hr} A[\r,\hr]
  e^{S_\text{eff}[\r, \hr]} \equiv \langle A[\r,\hr]
  \rangle_\text{eff},
\end{equation}
where the effective action $S_\text{eff}[\r, \hr]$ generically
consists of two terms,
\begin{equation} \la{eq3.9}
  S_\text{eff}[\r, \hr] = S_\text{bulk}[\r, \hr] + S_\text{dis}[\r,
    \hr].
\end{equation}
The first one is that part of $S[\r,\hr;v]$ that does not explicitly
involve $v({\bf r})$ and is therefore left unaffected by the disorder
average.  It generically reads
\begin{equation} \la{eq3.10}
  S_\text{bulk}[\r, \hr] = \int_{{\bf r},t} \left\{ i\hr({\bf r},t)
  \left( \partial_t \r({\bf r},t) - \frac{D_0}{T} \nabla \cdot \left[
    \r({\bf r},t) \nabla \left. \frac{\del {\cal
        F}_\text{bulk}[\r]}{\del \r({\bf r})} \right|_{\rho({\bf
        r},t)} \right] \right) - D_0 \r({\bf r},t) [\nabla \hr({\bf
      r},t)]^2 \right\},
\end{equation}
with ${\cal F}_\text{bulk}[\r] \equiv {\cal F}_\text{id}[\r] + {\cal
  F}_\text{int}[\r]$, and describes the dynamics of a bulk fluid in
the absence of an external field.  In the present noninteracting case
[we have set $u(r)=0$ for all $r$], ${\cal F}_\text{bulk}[\r]$ reduces
to ${\cal F}_\text{id}[\r]$, and $S_\text{bulk}[\r, \hr]$ to
\begin{equation}
    S_\text{free}[\r, \hr] = \int_{{\bf r},t} \left\{ i\hr({\bf r},t)
    \left( \partial_t - D_0 \nabla^2 \right) \r({\bf r},t) - D_0
    \r({\bf r},t) [\nabla \hr({\bf r},t)]^2 \right\},
\end{equation}
which rules ``free'' dynamics in the absence of disorder and
interactions.  Note that $S_\text{free}[\r, \hr]$ is
\emph{non-Gaussian} and possesses a cubic nonlinearity arising from
the multiplicative thermal noise.  Since the quenched randomness has
Gaussian statistics, one can readily perform the disorder average on
the remaining factor in $e^{S[\r, \hr;v]}$,
\begin{equation} \la{eq3.11}
  \overline{\exp \left( -\frac{D_0}{T} \int_{{\bf r},t} \, i\hr({\bf
      r},t) \nabla \cdot \left[ \r({\bf r},t) \nabla v({\bf r})
      \right] \right)} \equiv e^{S_\text{dis}[\r, \hr]},
\end{equation}
and obtain the second term,
\begin{equation} \la{eq3.12}
  S_\text{dis}[\r, \hr] = \frac{1}{2} \l D_0^2 \int_{{\bf r},t}
  \int_{{\bf r}',t'} [ \nabla^{\al} \nabla^{\beta} \Phi(|{\bf r}-{\bf
      r}'|) ] [ \r({\bf r},t) \nabla^{\al} \hr({\bf r},t) ] [ \r({\bf
      r}',t') \nabla'^{\beta} \hr ({\bf r}',t') ],
\end{equation}
where the summation convention is implied for the Cartesian indices
(this will systematically be the case in the following) and the
$\nabla'$ operator acts on ${\bf r}'$.  As is common with
quenched-random systems \cite{Dom78PRB}, the disorder-induced
contribution becomes \emph{nonlocal} in time after disorder averaging,
i.e., it does not only couple the fields at any given time, but also
between different time slices.  In fact, $S_\text{dis}[\r,\hr]$
represents an effective time-persistent dynamical interaction between
the fluid particles induced by the presence of the quenched random
potential.  It displays both cubic and quartic nonlinearities in
$\hr({\bf r},t)$ and $\del \r ({\bf r},t)$.  Through integration by
parts, it can be rewritten as
\begin{equation} \la{eq3.13}
  S_\text{dis}[\r, \hr] = - \frac{1}{2}\l \int_{{\bf r},t} \int_{{\bf
      r}',t'} \Phi(|{\bf r}-{\bf r}'|) \Lam({\bf r},t) \Lam({\bf
    r}',t'),
\end{equation}
where we introduce the composite response field
\begin{equation} \la{eq3.14}
  \Lam({\bf r},t) \equiv D_0 \nabla\cdot \left[ \r({\bf r},t) \nabla
    \hr({\bf r},t) \right].
\end{equation}
The latter leads to the physical response function, as discussed in
the next section.

\section{Physical response function, time-reversal symmetry, and
  fluc\-tuation-dissipation relation} \label{sec:TRS}
\setcounter{equation}{0}

We now define our main quantities of interest and discuss some crucial
relations between them.  For the sake of generality, we retain
interactions between the colloids, as they barely add any additional
complexity.

A fundamental feature of the fluid systems when studied at the level
of the density field is that the physical response function $R({\bf
  r},t;{\bf r}',t')$, pertaining to the change of the local average
density under a small external field coupled to the density
fluctuation, differs from the ordinary response function (to the
thermal noise) $G({\bf r},t;{\bf r}',t')$, because of the
multiplicative nature of the noise in the original Langevin equation,
Eq.~\eqref{eq1.14} \cite{MiyRei05JPA}.  The main quantities of
interest are thus the density correlation function $C({\bf r},t;{\bf
  r}',t')$ and the above response functions, defined as
\begin{subequations} \la{tr1}
\begin{flalign}
  C({\bf r},t;{\bf r}',t') & = \langle \del\r({\bf r},t) \del \r({\bf
    r}',t') \rangle_\text{eff}, \la{tr1c} \\
  G({\bf r},t;{\bf r}',t') & = -i \langle \r({\bf r},t) \hr({\bf
    r}',t') \rangle_\text{eff}, \la{tr1g} \\
  \begin{split}
  R({\bf r},t;{\bf r}',t') & = \frac{i}{T} \langle \r({\bf r},t)
  \Lambda({\bf r}',t') \rangle_\text{eff} \\ & = -\frac{\r_0 D_0}{T}
  \nabla^2 G({\bf r},t;{\bf r}',t') + i \frac{D_0}{T} \langle \r({\bf
    r},t) \nabla' \cdot [ \del \r({\bf r}',t') \nabla' \hr ({\bf
      r}',t') ] \rangle_\text{eff}.
  \end{split} \la{tr1r}
\end{flalign}
\end{subequations}
It is also useful to introduce the so-called connected density
correlation function,
\begin{equation} \label{Fconnected}
  F({\bf r},t;{\bf r}',t') = C({\bf r},t;{\bf r}',t') -
  C_\text{d}(|{\bf r}-{\bf r}'|).
\end{equation}
Note that the physical response function involves the composite
response field, Eq.~\eqref{eq3.14}, hence has two contributions: one
is simply proportional to the noise-response function, while an
additional ``anomalous'' term arises from the multiplicative thermal
noise.  Due to the explicit appearance of the temperature $T$ in the
expression of the physical response function $R({\bf r},t;{\bf
  r}',t')$, it is found convenient to instead use the function
$\overline{R}({\bf r},t;{\bf r}',t')$ defined as
\begin{equation} \la{tr2}
  \overline{R}({\bf r},t;{\bf r}',t') \equiv T R({\bf r},t;{\bf
    r}',t') = i \langle \rho({\bf r},t) \Lam({\bf r}',t')
  \rangle_\text{eff}.
\end{equation}

Causality commands that the response functions obey
\begin{equation} \label{eq:causality}
  G({\bf r},t;{\bf r}',t') = 0, \qquad \overline{R}({\bf r},t;{\bf
    r}',t') = 0, \qquad t\leq t'.
\end{equation}
In terms of the fields, this means
\begin{equation} \label{eq:rhorhohat}
  \langle \r({\bf r},t) \hr({\bf r}',t') \rangle_\text{eff} = 0,
  \qquad \langle \rho({\bf r},t) \Lam({\bf r}',t') \rangle_\text{eff}
  = 0, \qquad t\leq t'.
\end{equation}
Moreover, the normalization condition on the dynamical generating
functional in the MSRJD formalism results in additional causality
constraints, among which
\begin{gather}
  \langle \hr({\bf r},t) \rangle_\text{eff} = 0, \qquad \langle
  \hr({\bf r},t) \hr({\bf r}',t') \rangle_\text{eff} = 0, \qquad
  \langle \Lambda({\bf r},t) \rangle_\text{eff} = 0, \qquad \langle
  \Lambda({\bf r},t) \Lambda({\bf r}',t') \rangle_\text{eff} =
  0. \label{eq:rhohatrhohat}
\end{gather}

As with static quantities, the actual spatial dependence of the above
correlation and response functions is on $|{\bf r}-{\bf r}'|$, because
of the homogeneity and isotropy of the random field.  When
time-translation invariance additionnally holds, we will therefore
write $C({\bf r},t;{\bf r}',t') \equiv C(|{\bf r}-{\bf r}'|, t-t')$,
$G({\bf r},t;{\bf r}',t') \equiv G(|{\bf r}-{\bf r}'|, t-t')$, and
$\overline{R}({\bf r},t;{\bf r}',t') \equiv \overline{R}(|{\bf r}-{\bf
  r}'|, t-t')$.

Equilibrium dynamics is known to possess time-reversal symmetry.  This
symmetry is reflected in the invariance (up to irrelevant boundary
terms) of the effective action, Eq.~\eqref{eq3.9}, under special field
transformations with time reversal \cite{AndBirLef06JSMTE,
  VelChaCugKre08JPA, AroBirCug10JSMTE}.

The approach developed in the present work is motivated by the
invariance of the action under the so-called
$\mathcal{T}$-transformation \cite{AndBirLef06JSMTE,
  VelChaCugKre08JPA, AroBirCug10JSMTE}, as shown in
App.~\ref{app:symm}.  This transformation reads
\begin{equation} \la{tr3}
  \mathcal{T} : \left\{ \begin{aligned} \r({\bf r},t) & \to \r({\bf
      r},-t), \\
  \hr({\bf r},t) & \to \hr({\bf r},-t) + i h({\bf r},-t),
  \end{aligned} \right. 
\end{equation}
with the function $h({\bf r},t)$ defined through the equation
\begin{equation} \la{hTtrans}
  D_0 \nabla\cdot [\r({\bf r},t) \nabla h({\bf r},t)] = \partial_t
  \r({\bf r},t),
\end{equation}
which can be solved in Fourier space \cite{VelChaCugKre08JPA}.

This definition implies the relation
\begin{equation} \la{tr5}
  \mathcal{T}\Lam({\bf r}, t) = \Lam({\bf r}, -t) - i \partial_t
  \r({\bf r},-t)
\end{equation}
for the composite response field.  Thus, with the identification of
the physical response function $\o{R}({\bf r},t;{\bf r}',t')$ in
Eq.~\eqref{tr2}, the FDR is immediately obtained from Eqs.~\eqref{tr3}
and \eqref{tr5}.  Indeed, from the Ward-Takahashi identities
\cite{Tauber14book}
\begin{subequations}
\begin{gather}
  \langle \r({\bf r},t) \r({\bf r}',t') \rangle_\text{eff} = \langle
          [\mathcal{T}\r({\bf r},t)] [\mathcal{T}\r({\bf r}',t')]
          \rangle_\text{eff} = \langle \r({\bf r},-t) \r({\bf r}',-t')
          \rangle_\text{eff}, \\
  \langle \r({\bf r},t) \Lambda({\bf r}',t') \rangle_\text{eff} =
  \langle [\mathcal{T}\r({\bf r},t)] [\mathcal{T}\Lambda({\bf r}',t')]
  \rangle_\text{eff} = \langle \r({\bf r},-t) \Lam({\bf r}', -t')
  \rangle_\text{eff} - i \partial_{t'} \langle \r({\bf r},-t) \r({\bf
    r}',-t') \rangle_\text{eff},
\end{gather}
\end{subequations}
follows the relation
\begin{equation}
  \overline{R}({\bf r},t;{\bf r}',t') = \overline{R}({\bf r},-t;{\bf
    r}',-t') + \partial_{t'} C({\bf r},t;{\bf r}',t'),
\end{equation}
i.e., with time-translation invariance,
\begin{equation} \la{tr7}
\o{R}(|{\bf r}-{\bf r}'|, t-t') - \o{R}(|{\bf r}-{\bf r}'|, t'-t) = -
\partial_{t} C(|{\bf r}-{\bf r}'|, t-t').
\end{equation}
For future reference, we note that causality,
Eq.~\eqref{eq:causality}, and the FDR, Eq.~\eqref{tr7}, imply
\begin{equation} \la{tr11} 
  \int_{-\infty}^{+\infty} d t' \, \o{R} (|{\bf r}-{\bf r}'|, t-t') =
  C_\text{st}(|{\bf r}-{\bf r}'|) - C_\text{d}(|{\bf r}-{\bf r}'|),
\end{equation}
where the equilibrium relations $C(|{\bf r}-{\bf r}'|,0) =
C_\text{st}(|{\bf r}-{\bf r}'|)$ and $C(|{\bf r}-{\bf r}'|,t\to
+\infty) = C_\text{d}(|{\bf r}-{\bf r}'|)$ have been used.

Another field transformation exists, that leaves the effective action
invariant.  It will play a minor role in the present work, but should
be mentioned for completeness and because it might be of general
interest in dynamical studies of random-field systems.  Interestingly,
it does not involve a time reversal in its primary formulation and
therefore holds in generic out-of-equilibrium situations.  However, it
can be usefully specialized to equilibrium dynamics through
composition with the $\mathcal{T}$-transformation.

Thus, guided by Ref.~\cite{AroBirCug10JSMTE}, we show in
App.~\ref{app:symm} that $S_\text{eff}[\r, \hr]$ is invariant under
the $\mathcal{U'}$-transformation defined as
\begin{equation} \label{eq:Uprimetrans}
  \mathcal{U'} : \left\{ \begin{aligned} \r({\bf r}, t) & \to \r({\bf
      r}, t), \\
    \hr({\bf r}, t) & \to -\hr({\bf r},t) + 2 i \int_{{\bf r}',t'}
    K_\lambda^{-1}({\bf r},t;{\bf r}',t') \Det([\r],{\bf r}',t').
\end{aligned} \right.
\end{equation}
The functional $\Det([\r],{\bf r},t)$ represents the deterministic
nonrandom part of the density evolution equation and here reads [see
  Eq.~\eqref{eq1.16}]
\begin{equation} \label{eq:deter}
  \Det([\r],{\bf r},t) = \partial_t \r({\bf r},t) - \frac{D_0}{T}
  \nabla \cdot \left[ \r({\bf r},t) \nabla \left. \frac{\del {\cal
        F}_\text{bulk}[\r]}{\del \r({\bf r})} \right|_{\rho({\bf
        r},t)} \right].
\end{equation}
The kernel $K_\lambda^{-1}({\bf r},t;{\bf r}',t')$ is the inverse of
the density-dependent symmetric kernel characterizing both the
Gaussian noise and disorder in the system,
\begin{equation}\begin{split} \label{eq:klambda}
  K_\lambda({\bf r},t;{\bf r}',t') & \equiv \left\{ \nabla^\alpha
  \nabla'^\beta \left[ 2 D_0 \r({\bf r},t) \delta_{\alpha\beta}
    \delta({\bf r}-{\bf r}') \delta(t-t') + \l D_0^2 \r({\bf r},t)
    \r({\bf r}',t') \nabla^{\al} \nabla'^{\beta} \Phi(|{\bf r}-{\bf
      r}'|) \right] \right\} \\
  & \equiv K_0({\bf r},t;{\bf r}',t') + \lambda \Delta K({\bf
    r},t;{\bf r}',t'),
\end{split}\end{equation}  
and is accordingly defined through
\begin{equation} \label{eq:invklambda}
  \delta({\bf r}-{\bf r}'') \delta(t-t'') = \int_{{\bf r}',t'}
  K_\lambda({\bf r},t;{\bf r}',t') K_\lambda^{-1}({\bf r}',t';{\bf
    r}'',t'').
\end{equation}

The composition of $\mathcal{T}$ and $\mathcal{U'}$ yields the
$\mathcal{U}$-transformation.  It obviously leaves the action
invariant, since $\mathcal{T}$ and $\mathcal{U'}$ separately do, and
involves a time reversal inherited from $\mathcal{T}$.  As shown in
App.~\ref{app:symm}, it reads
\begin{equation} \label{eq:Utrans}
  \mathcal{U} : \left\{ \begin{aligned} \r({\bf r}, t) & \to \r({\bf
      r}, -t), \\
  \hr({\bf r}, t) & \to - \hr({\bf r}, -t) + \frac{i}{T}
  \left. \frac{\del {\cal F}_\text{bulk}[\r]}{\del \r({\bf r})}
  \right|_{\rho({\bf r},-t)} \\
  & + i \lambda D_0 \int_{{\bf r}',t'} K_\lambda^{-1}({\bf r},-t;{\bf
    r}',-t') \nabla' \cdot \left\{ \r({\bf r}',-t') \nabla' \int_{{\bf
      r}'',t''} \Phi(|{\bf r}'-{\bf r}''|) \mathcal{T}\Det([\r],{\bf
    r}'',t'') \right\},
\end{aligned} \right.
\end{equation}
with
\begin{equation} \label{eq:Tdeter}
  \mathcal{T}\Det([\r],{\bf r},t) = \partial_t \r({\bf r},-t) -
  \frac{D_0}{T} \nabla \cdot \left[ \r({\bf r},-t) \nabla
    \left. \frac{\del {\cal F}_\text{bulk}[\r]}{\del \r({\bf r})}
    \right|_{\rho({\bf r},-t)} \right].
\end{equation}
It is clear that, in the absence of a random field ($\lambda=0$), this
transformation reduces to the $\mathcal{U}$-transformation as defined
in Ref.~\cite{AndBirLef06JSMTE} for bulk fluids, hence the shared
naming.  It becomes nonlocal in time in the presence of a random
field.  Note that, in principle, the integral over $t''$ of the total
time derivative $\partial_{t''} \r({\bf r}'',-t'')$ contained in
$\mathcal{T}\Det([\r],{\bf r}'',t'')$ vanishes in an equilibrium
setting, but we found that explicitly keeping such terms makes some
calculations in App.~\ref{app:symm} more straightforward.

As the $\mathcal{T}$-transformation, the $\mathcal{U}$-transformation
can be used to derive relations between response functions and
correlations.  In particular, as shown in App.~\ref{app:symm}, the
Ward-Takahashi identity
\begin{equation} \label{eq:WT_G}
  \langle \r({\bf r},t) \hr({\bf r}',t') \rangle_\text{eff} =
  \langle [\mathcal{U}\r({\bf r},t)] [\mathcal{U}\hr({\bf r}',t')]
  \rangle_\text{eff}
\end{equation}
leads to the following decomposition of the noise-response function,
\begin{multline} \label{genDHMR}
  G(|{\bf r}-{\bf r}'|, t-t') + G(|{\bf r}-{\bf r}'|, t'-t) =
  \\ \int_{{\bf r}''} C(|{\bf r}-{\bf r}''|, t-t') Q^{-1}(|{\bf
    r}''-{\bf r}'|) + \Delta C^\text{nG}(|{\bf r}-{\bf r}'|,t-t') +
  \Delta C^\text{dis}(|{\bf r}-{\bf r}'|,t-t').
\end{multline}
Here, $Q^{-1}(|{\bf r}-{\bf r}'|)$ is the functional inverse of the
static density correlation function of the bulk fluid if its free
energy is restricted to its Gaussian approximation, $\Delta
C^\text{nG}(|{\bf r}-{\bf r}'|,t-t')$ originates in the non-Gaussian
nature of ${\cal F}_\text{bulk}[\r]$ due to ${\cal F}_\text{id}[\r]$,
and $\Delta C^\text{dis}(|{\bf r}-{\bf r}'|,t-t')$ is a
disorder-induced contribution.  Their detailed expressions can be
found in App.~\ref{app:symm}.

For systems with a Gaussian bulk free energy and no random potential,
$Q^{-1}(|{\bf r}-{\bf r}'|) = C^{-1}(|{\bf r}-{\bf r}'|, 0)$, while
$\Delta C^\text{nG}(|{\bf r}-{\bf r}'|,t-t')$ and $\Delta
C^\text{dis}(|{\bf r}-{\bf r}'|,t-t')$ vanish.  One then recovers the
familiar Deker-Haake-Miyazaki-Reichman (DHMR) linear relation between
the noise-response function and the density correlation function
(Deker and Haake first considered the case of additive noise
\cite{DekHaa75PRA}, then Miyazaki and Reichman extended the result to
multiplicative noise \cite{MiyRei05JPA}).

In the following, it will be found convenient to work in reciprocal
space, i.e., with correlation and response functions Fourier
transformed with respect to their spatial variations.  Thus, in
Fourier space, Eqs.~\eqref{tr7} and \eqref{genDHMR} take the form
(setting $t'=0$)
\begin{subequations} \la{tr16}
\begin{gather}
  \o{R}_k(t) - \o{R}_k(-t) = - \partial_{t}  C_k(t), \la{tr16a} \\
  G_k(t) + G_k(-t) = \frac{C_k(t)}{Q_k} + \Delta C^\text{nG}_{k}(t) +
  \Delta C^\text{dis}_k(t).  \la{tr16b}
\end{gather} 
\end{subequations}
For noninteracting colloids, $Q_k=\r_0$. 

\section{Expansion around the disorder-free dynamics} 
\setcounter{equation}{0} \label{sec:expansion}

We now describe the main theoretical development at the heart of the
present work, which is a perturbative expansion dictated by the
$\mathcal{T}$-transformation, Eq.~\eqref{tr3}.  The key point here is
that the two contributions $S_\text{free}[\rho,\hat\rho]$ (to which
$S_\text{bulk}[\rho,\hat\rho]$ reduces in the noninteracting case) and
$S_\text{dis}[\rho,\hat\rho]$ to the effective action
$S_\text{eff}[\r,\hr]$ are separately invariant under this
transformation, as shown in App.~\ref{app:symm}.  Therefore, with a
due account of this property, it should be possible to lay out a
scheme that preserves the FDR, which precisely stems from the
$\mathcal{T}$-transformation, order by order.

The present perturbative approach first involves an expansion in terms
of $S_\text{dis}[\rho,\hat\rho]$ about the free dynamics ruled by
$S_\text{free}[\r,\hr]$, as
\begin{equation} \la{eq5.1}
  \langle A[\r,\hr] \rangle_\text{eff} = \int_{\r,\hr} A[\r,\hr]
  e^{S_\text{free}[\r,\hr]} e^{S_\text{dis}[\r, \hr]}= \int_{\r, \hr}
  A[\r, \hr] e^{S_\text{free}[\r,\hr]} \sum_{n_\text{dis}=0}^\infty
  \frac{S_\text{dis}[\r,\hr]^{n_\text{dis}}}{n_\text{dis}!}.
\end{equation} 
This step can be seen as a weak-disorder or high-temperature
expansion, since $S_\text{dis}[\r,\hr] $ is proportional to the
relative disorder strength $\l=w/T^2$.  Defining the average over the
free part of the action as
\begin{equation} \la{eq5.2}
  \langle A[\r,\hr] \rangle_\text{f} \equiv \int_{\r,\hr} A[\r,\hr]
  e^{S_\text{free}[\r,\hr]},
\end{equation}
we thus have
\begin{equation} \la{eq5.3}
  \langle A[\r,\hr] \rangle_\text{eff} = \sum_{n_\text{dis}=0}^\infty
  \frac{1}{n_\text{dis}!} \left\langle A[\r,\hr]
  S_\text{dis}[\r,\hr]^{n_\text{dis}} \right\rangle_\text{f}.
\end{equation}

Now, the free part of the action has a non-Gaussian cubic nonlinearity
due to the multiplicative thermal noise.  In order to maintain the
invariance of $S_\text{free}[\r,\hr]$ under the
$\mathcal{T}$-transformation and preserve the FDR order by order, this
nonlinearity should be treated \emph{exactly}.  It turns out that this
can be readily achieved thanks to the causality conditions and the
presence of two $\hr$ fields in this cubic contribution.  Indeed,
splitting the free part of the action into its Gaussian and
non-Gaussian components, $S_0[\r,\hr]$ and $S_\text{m}[\r,\hr]$,
respectively, with
\begin{gather}
  S_0[\rho, \hat \rho] \equiv \int_{{\bf r},t} \left\{ i\hat\rho({\bf
    r},t) \left( \partial_t -D_0 \nabla^2 \right) \delta\r({\bf r},t)
  - D_0 \r_0 \left[ \nabla \hr({\bf r},t) \right]^2 \right\},
  \la{eq5.4} \\
  S_\text{m} [\rho, \hat \rho] \equiv -D_0 \int_{{\bf r},t}
  \delta\r({\bf r},t) [ \nabla \hr({\bf r},t) ]^2, \la{eq5.5}
\end{gather}
one can rewrite the averages over the free dynamics as 
\begin{equation} \la{eq5.6}
  \langle B[\rho, \hat \rho] \rangle_\text{f} = \int_{\r, \hr} B[\rho,
    \hat \rho] e^{S_0[\rho, \hat \rho]} e^{S_\text{m}[\rho, \hat
      \rho]} = \langle B[\rho, \hat \rho] e^{S_\text{m}[\rho, \hat
      \rho]} \rangle_0 = \sum_{n_\text{m}=0}^{\infty}
  \frac{1}{n_\text{m}!} \langle B[\rho, \hat \rho] S_\text{m} [\rho,
    \hat \rho]^{n_\text{m}} \rangle_0,
\end{equation}
where $\langle \cdots \rangle_0$ denotes the Gaussian average defined
as
\begin{equation} \la{eq5.7}
  \langle B[\rho, \hat \rho] \rangle_0 \equiv \int_{\r, \hr} B[\rho,
    \hat \rho] e^{S_0[\rho, \hat \rho]}.
\end{equation}
The key observation is that, due to the twice faster increase of the
number of $\hr$ fields with $n_\text{m}$, the summation in
Eq.~\eqref{eq5.6} will be rapidly \emph{terminated} at a low order.
Indeed, consider a generic product of $\delta\r$ and $\hr$ fields or
space derivatives thereof.  If it has an odd number of factors, its
Gaussian average trivially vanishes.  If its number of factors is
even, one can use Wick's theorem to decompose its Gaussian average as
a sum of products of two-point averages.  Then, if the number of
noise-response fields exceeds the number of density fields
(necessarily, by at least two), each term in the sum will unavoidably
have a factor of the form $\langle \hr({\bf r}_i,t_i) \hr({\bf
  r}_j,t_j)\rangle_0$.  Such factors identically vanish due to
causality [Eq.~\eqref{eq:rhohatrhohat} also holds with the Gaussian
  action $S_0$], hence the whole Gaussian average vanishes.  For
instance, one generically gets
\begin{gather} 
  \langle \delta \r({\bf r}_i,t_i) \delta \r({\bf r}_j,t_j) \hr({\bf
    r}_k,t_k) \hr({\bf r}_l,t_l) \rangle_0 \neq 0, \\
  \langle \delta \r({\bf r}_i,t_i) \delta \r({\bf r}_j,t_j) \hr({\bf
    r}_k,t_k) \hr({\bf r}_l,t_l) \hr({\bf r}_m,t_m) \hr({\bf r}_n,t_n)
  \rangle_0 =0.
\end{gather}
Now, if $B[\rho, \hat\rho]$ is such a generic product with $p$ density
fields and $q$ response fields, then the term of order $n_\text{m}$ in
Eq.~\eqref{eq5.6} also involves such a product, with $p+n_\text{m}$
density fields and $q+2n_\text{m}$ response fields.  As just shown,
its Gaussian average vanishes if $q+2n_\text{m} > p+n_\text{m}$, i.e.,
$n_\text{m}>p-q$.  It is precisely this simplification that makes
possible an exact treatment of the cubic nonlinearity due to the
multiplicative thermal noise.  Indeed, if $p\ge q$, the expansion in
Eq.~\eqref{eq5.6} terminates at most at $n_\text{m}=p-q$, while for
$p<q$, the first term of Eq.~\eqref{eq5.6} already vanishes and one
gets $\langle B[\rho, \hat \rho] \rangle_\text{f} = 0$.

Note that the bound on $n_\text{m}$ is the same for all terms in
Eq.~\eqref{eq5.3}.  Indeed, $S_\text{dis}[\r,\hr]$ given by
Eq.~\eqref{eq3.12} can be rewritten as
\begin{multline} \label{eq:Sdis_expan}
  S_\text{dis}[\r,\hr] = \frac{1}{2} \l D_0^2 \int_{{\bf r},t}
  \int_{{\bf r}',t'} [ \nabla^\alpha \nabla^\beta \Phi(|{\bf r}-{\bf
      r}'|) ] \\ \times [ \r_0^2 + 2 \r_0 \del\r({\bf r},t) +
    \del\r({\bf r},t) \del\r({\bf r}',t') ] [\nabla^\alpha \hr({\bf
      r},t)] [\nabla'^\beta \hr({\bf r}',t')],
\end{multline}
where the factor $2$ in the integrand comes from the exchange symmetry
between the dummy indices ${\bf r},t$ and ${\bf r}',t'$.  So, if
$A[\rho, \hat\rho]$ is a product of $p$ density fields and $q$
response fields, then the term of order $n_\text{dis}$ in
Eq.~\eqref{eq5.3} involves products of $p$ to $p+2n_\text{dis}$
density fields and $q+2n_\text{dis}$ response fields.  Following the
above argument, its Gaussian average vanishes if $n_\text{m} >
p+2n_\text{dis}-q-2n_\text{dis} = p-q$ (the bound is imposed by the
product with the largest number of density fields), independent of
$n_\text{dis}$.  Accordingly, for $p<q$, one also gets $\langle
A[\rho, \hat \rho] \rangle_\text{eff} = 0$.

Further simplifications might occur in the computation of
Gaussian-averaged products when space-time points are repeated.
Indeed, through Eq.~\eqref{eq:rhorhohat}, which also holds with the
Gaussian action $S_0$, causality directly sets $\langle \r({\bf
  r}_i,t_i) \hr({\bf r}_i,t_i) \rangle_0 = 0$.  Less directly, the
time ordering in Eq.~\eqref{eq:rhorhohat} also imposes the vanishing
of certain products of two-point averages with loop-like time
dependence.  For instance, one gets
\begin{gather} 
  \langle \r({\bf r}_i,t_i) \hr({\bf r}_j,t_j) \rangle_0 \langle
  \r({\bf r}_j,t_j) \hr({\bf r}_i,t_i) \rangle_0 = 0,
  \label{eq:loop1} \\
  \langle \r({\bf r}_i,t_i) \hr({\bf r}_j,t_j) \rangle_0 \langle
  \r({\bf r}_j,t_j) \hr({\bf r}_k,t_k) \rangle_0 \langle \r({\bf
    r}_k,t_k) \hr({\bf r}_i,t_i) \rangle_0 = 0. \label{eq:loop2}
\end{gather}
These equalities typically lead to a reduction in the number of terms
in the expansion of Gaussian averages.  Occasionally, they result in a
truncation of Eq.~\eqref{eq5.6} below the above-mentioned threshold.

These crucial features of the theory were first pointed out by
Andreanov \emph{et al.} \cite{AndBirLef06JSMTE} and discussed in
detail by Velenich \emph{et al.}  \cite{VelChaCugKre08JPA}, who
demonstrated how they can be used to exactly compute arbitrary
multi-point correlation functions in the noninteracting Brownian gas
without external field.  In this respect, the present work is, to the
best of our knowledge, the first nontrivial extension of this early
study, aiming at including the effect of a Gaussian quenched-random
potential on the gas.

\section{Dynamical equations for the correlation and response functions}
\setcounter{equation}{0} \label{sec:nonpert}

It remains to derive the dynamical equations for the correlation and
response functions, to which the above perturbation scheme will be
applied.  To this end, the following identities can be used, which are
easily proved by functional integration by parts:
\begin{subequations} \la{eq6.1}
  \begin{align} 
    \left\langle \frac{\delta S_\text{eff}}{\delta \hr(1)} \hr(2)
    \right\rangle_\text{eff} & = -\del(12), \la{eq6.1a} \\
    \left\langle \frac{\delta S_\text{eff}}{\delta \hr(1)} \Lam(2)
    \right\rangle_\text{eff} & = -\r_0 D_0 \nabla^2 \delta(12),
    \la{eq6.1b} \\
   \left\langle \frac{\delta S_\text{eff}}{\delta \hr(1)} \r(2)
   \right\rangle_\text{eff} & = 0. \la{eq6.1c}
 \end{align}
\end{subequations}
In the above expressions and in the following, the notation $1$, $2$,
$3$, etc, is used to refer to space-time points, in order to shorten
the equations.  Specifically, we set $1=({\bf r},t)$ and $2=({\bf
  r}',t')$, then $i=({\bf r}_i,t_i)$, $i\ge3$.  Since
\begin{equation} \la{eq6.2}
  \frac{\delta S_\text{eff}}{\delta \hr(1)} = i ( \partial_t -D_0
  \nabla^2 ) \del \r(1) + 2 \Lam(1) -\l D_0 \nabla \cdot \left( \r(1)
  \int_3 [\nabla \Phi(13) ] \Lam(3) \right),
\end{equation}
where $\Phi(13) \equiv \Phi(|{\bf r}-{\bf r}_3|)$, one obtains the
exact equations
\begin{subequations} \la{eq6.3}
\begin{align}
  ( \partial_t -D_0 \nabla^2 ) G(12) & = \del(12) - \l D_0 \nabla
  \cdot \left( \int_3 [ \nabla \Phi(13) ] \langle \r(1)\Lam(3)\hr(2)
  \rangle_\text{eff} \right), \la{eq6.3a} \\
  ( \partial_t -D_0 \nabla^2 ) \o{R}(12) & = -\r_0 D_0 \nabla^2
  \delta(12) +\l D_0 \nabla \cdot \left( \int_3 [ \nabla \Phi(13) ]
  \langle \r(1)\Lam(3)\Lam(2) \rangle_\text{eff} \right), \la{eq6.3b}
  \\
  ( \partial_t -D_0 \nabla^2 ) C(12) & = 2\o{R}(21) -i \l D_0 \nabla
  \cdot \left( \int_3 [ \nabla \Phi(13) ] \langle \r(1)\Lam(3)\r(2)
  \rangle_\text{eff} \right).  \la{eq6.3c}
 \end{align}
\end{subequations}
These equations show an evident hierarchical structure, which calls
for a perturbative study building on an expansion scheme such as the
one developed in the previous section.  In App.~\ref{app:symm}, we
report an alternative derivation of Eq.~\eqref{eq6.3c} based on the
$\mathcal{T}$- and $\mathcal{U}$-transformations.

Substituting $\r(i)=\r_0+\del\r(i)$ and removing terms that vanish due
to the various simple causality conditions, the multi-point averages
in Eqs.~\eqref{eq6.3} can be simplified to (with the summation
convention for the Cartesian indices)
\begin{subequations} \la{eq6.4}
\begin{align}
  \langle \r(1) \Lam(3) \hr(2) \rangle_\text{eff} & = D_0
  \nabla^{\gamma}_3 \langle \del\r(1) \del\r(3) [\nabla^{\gamma}_3
    \hr(3)] \hr(2) \rangle_\text{eff}, \la{eq6.4a} \\
  \langle \r(1) \Lam(3) \Lam(2) \rangle_\text{eff} & = D_0^2
  \nabla_2^{\beta} \nabla_3^{\gamma} \langle \del\r(1) \left[
    \del\r(3) \r_0 + \r_0 \del\r(2) + \del\r(3) \del\r(2) \right]
        [\nabla^{\gamma}_3 \hr(3)] [\nabla^{\beta}_2 \hr(2)]
        \rangle_\text{eff}, \la{eq6.4b} \\
  \langle \r(1) \Lam(3) \r(2) \rangle_\text{eff} & = - i \r_0 \left[
    \o{R}(13) + \o{R}(23) \right] + D_0 \nabla^{\gamma}_3 \langle
  \del\r(1) \left[ \r_0 + \del\r(3) \right] \del\r(2)
        [\nabla^{\gamma}_3 \hr(3)] \rangle_\text{eff}.  \la{eq6.4c}
\end{align}
\end{subequations}

\section{Zeroth-order theory: Disorder-free case}
\setcounter{equation}{0} \label{sec:zeroth}

In the absence of a random potential ($\lambda=0$), the particle
system is a noninteracting Brownian gas, whose properties are very
well known \cite{VelChaCugKre08JPA}.

In Fourier space, the equations of motion simply reduce to (setting
$t'=0$)
\begin{subequations} \la{eq7.1}
\begin{align}
  ( \partial_t +\Gam_k ) G^0_k(t) & = \del(t), \la{eq7.1a} \\
  ( \partial_t +\Gam_k ) \o{R}^0_k(t) & = \r_0 \Gam_k \delta(t),
  \la{eq7.1b} \\
  ( \partial_t +\Gam_k ) C^0_k(t) & = 2\o{R}^0_k(-t), \la{eq7.1c}
\end{align}
\end{subequations}
where $\Gam_k \equiv D_0 k^2$ and the superscript $0$ on the
correlation and response functions denotes the absence of a random
potential.  The solutions are given by
\begin{subequations} \la{eq7.2}
\begin{align}
  G^0_k(t) & = \theta(t) e^{-\Gam_k t}, \la{eq7.2a} \\
  \o{R}^0_k(t) & = \theta(t) \r_0 \Gam_k e^{-\Gam_k t}, \la{eq7.2b} \\
  C^0_k(t) & = \r_0 e^{-\Gam_k |t|}, \la{eq7.2c}
\end{align}
\end{subequations}
where we used the static input for the density correlation function
$C^0_k(0)=\r_0$, since $C(|{\bf r}-{\bf r}'|,0) = \langle \del\r({\bf
  r}) \del\r({\bf r'}) \rangle = \r_0 \delta({\bf r}-{\bf r'})$ for
the noninteracting system in the absence of an external random
potential.  This $\r_0$ factor in $C^0_k(t)$ is also the one required
for consistency with the FDR.

For future use, it is interesting to note that this free dynamics can
be fully characterized through suitable specializations of the
definitions and symmetry-derived relations given in
Sec.~\ref{sec:TRS}.  Indeed, it appears as the equilibrium dynamics
for which Eqs.~\eqref{tr1r}, \eqref{tr16a}, and \eqref{tr16b}, reduce
to
\begin{subequations} \la{eq7.3}
\begin{gather}
  \o{R}^0_k(t) = \r_0 \Gam_k G^0_k(t), \la{eq7.3a} \\
  \o{R}^0_k(t) - \o{R}^0_k(-t) = - \partial_{t} C^0_k(t), \la{eq7.3b}
  \\
  G^0_k(t) + G^0_k(-t) = \frac{C^0_k(t)}{\r_0} , \la{eq7.3c}
\end{gather}
\end{subequations}
thereby demonstrating that the three functions of interest are
directly related in a simple but fundamental way.  In this respect, it
should be fully appreciated that the considered dynamics involves both
multiplicative noise and a non-Gaussian free energy.  Therefore, the
absence of an anomalous contribution to the physical response function
in Eq.~\eqref{eq7.3a} and the validity of the DHMR linear relation
shown by Eq.~\eqref{eq7.3c} are nontrivial observations. They result
from a specific interplay of both aspects and from the cancellation
effects discussed in Sec.~\ref{sec:expansion}.

Regarding this, it might be useful to briefly show how the
field-theoretic calculation unfolds in the present simple case.  This
serves as a preparation for the more complicated random-field
situation and as a confirmation of the identity between the
correlation and response functions of the disorder-free noninteracting
gas and those of the Gaussian theory based on $S_0[\rho, \hat\rho]$.
To this end, we introduce the compact notations
\begin{equation}
  \del\r(i) \equiv i, \qquad \hr(i) \equiv \h{i}, \qquad
  \nabla_i^{\mu} \hr(i) \equiv \h{i}^{\mu},
\end{equation}
to be used for the evaluation of averages here, in the next section,
and in Apps.~\ref{app:G}-\ref{app:C}.  With them, the cubic thermal
noise term, Eq.~\eqref{eq5.5}, can be written as
\begin{equation} \la{eq8.3}
  S_\text{m}[\r,\hr] \equiv - D_0 \int_4 4 \h4^{\delta} \h4^{\delta},
\end{equation}
and $e^{S_\text{m}[\r,\hr]}$ in Eq.~\eqref{eq5.6} expanded
accordingly,
\begin{equation} \la{eq8.4}
  e^{S_\text{m}[\r,\hr]} = 1 - D_0 \int_4 4 \h4^{\delta} \h4^{\delta}
  + \frac12 D_0^2 \int_4 \int_5 4 5 \h4^{\delta} \h4^{\delta}
  \h5^{\varepsilon} \h5^{\varepsilon} + \cdots.
\end{equation}

From Eqs.~\eqref{tr1g}, we get $G^0(12) = -i \langle \r(1) \hr(2)
\rangle_\text{f} = -i \langle \delta\r(1) \hr(2) \rangle_\text{f}
\equiv -i \langle 1 \hat{2} \rangle_\text{f}$.  With one density field
and one noise-response field, the expansion Eq.~\eqref{eq5.6}
terminates at its first term and
\begin{equation}
  \langle 1 \hat{2} \rangle_\text{f} = \langle 1 \hat{2} \rangle_0.
\end{equation}
The anomalous term in $\overline{R}^0(12)$ reads $i D_0
\nabla_2^{\beta} \langle \r(1) \del \r(2) \nabla_2^{\beta} \hr(2)
\rangle_\text{f} = i D_0 \nabla_2^{\beta} \langle \del\r(1) \del \r(2)
\nabla_2^{\beta} \hr(2) \rangle_\text{f} \equiv i D_0 \nabla_2^{\beta}
\langle 12\hat{2}^{\beta} \rangle_\text{f}$, and the Gaussian
expansion of $\langle 12\hat{2}^{\beta} \rangle_\text{f}$ is
\begin{equation}
  \langle 12\hat{2}^{\beta} \rangle_\text{f} = \langle 12
  \hat{2}^{\beta} \rangle_0 - D_0 \int_4 \langle 124 \hat{2}^{\beta}
  \h4^{\delta} \h4^{\delta} \rangle_0 = 0,
\end{equation}
where we used $\langle 2 \h2^{\beta} \rangle_0 = \langle 4
\h4^{\delta} \rangle_0 = 0$ and $\langle 2 \h4^{\delta} \rangle_0
\langle 4 \h2^{\beta} \rangle_0 =0$.  Therefore, $\overline{R}^0(12)=
- \r_0 D_0 \nabla^2 G^0(12)$ as expected.  Finally, $C^0(12) = \langle
\del\r(1) \del\r(2) \rangle_\text{f} \equiv \langle 12
\rangle_\text{f}$ expands to
\begin{equation}
  \langle 12 \rangle_\text{f} = \langle 12 \rangle_0 - D_0 \int_4
  \langle 124 \h4^{\delta} \h4^{\delta} \rangle_0 + \frac12 D_0^2
  \int_4 \int_5 \langle 1245 \h4^{\delta} \h4^{\delta}
  \h5^{\varepsilon} \h5^{\varepsilon} \rangle_0 = \langle 12
  \rangle_0,
\end{equation}
where we used $\langle 4 \h4^{\delta} \rangle_0 = \langle 5
\h5^{\varepsilon} \rangle_0 = 0$ and $\langle 4 \h5^{\varepsilon}
\rangle_0 \langle 5 \h4^{\delta} \rangle_0 =0$.

\section{First-order perturbation calculation} 
\label{sec:firstorder}
\setcounter{equation}{0}

We may now perturbatively compute the three-point averages in
Eqs.~\eqref{eq6.3} and obtain the first-order corrections to the free
dynamics due to the random potential.

Applying Eq.~\eqref{eq5.3} to the different terms in the simplified
Eqs.~\eqref{eq6.4}, one gets
\begin{subequations} \la{eq8.1}
\begin{align}
  \langle \del\r(1) \del\r(3) [\nabla^{\gamma}_3 \hr(3)] \hr(2)
  \rangle_\text{eff} & \equiv \langle 1 3 \h3^{\gamma} \h2
  \rangle_\text{eff} = \langle 1 3 \h3^{\gamma} \h2 \rangle_\text{f} +
  O(\l), \\
  \langle \del\r(1) \del\r(3) [\nabla^{\gamma}_3 \hr(3)] [
    \nabla^{\beta}_2 \hr(2) ] \rangle_\text{eff} & \equiv \langle 1 3
  \h3^{\gamma} \h2^{\beta} \rangle_\text{eff} = \langle 1 3
  \h3^{\gamma} \h2^{\beta} \rangle_\text{f} + O(\l), \\
  \langle \del\r(1) \del\r(2) [\nabla^{\gamma}_3 \hr(3)] [
    \nabla^{\beta}_2 \hr(2)] \rangle_\text{eff} & \equiv \langle 1 2
  \h3^{\gamma} \h2^{\beta} \rangle_\text{eff} = \langle 1 2
  \h3^{\gamma} \h2^{\beta} \rangle_\text{f} + O(\l), \\
  \langle \del\r(1) \del\r(3) \del\r(2) [\nabla^{\gamma}_3 \hr(3)] [
    \nabla^{\beta}_2 \hr(2)] \rangle_\text{eff} & \equiv \langle 1 3 2
  \h3^{\gamma} \h2^{\beta} \rangle_\text{eff} = \langle 1 3 2
  \h3^{\gamma} \h2^{\beta} \rangle_\text{f} + O(\l), \\
  \langle \del\r(1) \del\r(2) [\nabla^{\gamma}_3 \hr(3)]
  \rangle_\text{eff} & \equiv \langle 1 2 \h3^{\gamma}
  \rangle_\text{eff} = \langle 1 2 \h3^{\gamma} \rangle_\text{f} +
  O(\l), \\
  \langle \del\r(1) \del\r(3) \del\r(2) [\nabla^{\gamma}_3 \hr(3)]
  \rangle_\text{eff} & \equiv \langle 1 3 2 \h3^{\gamma}
  \rangle_\text{eff} = \langle 1 3 2 \h3^{\gamma} \rangle_\text{f} +
  O(\l),
\end{align}
\end{subequations}
where we used the compact notations introduced above.  For a
first-order calculation, it is enough to compute the first term in the
right-hand side of each line in Eqs.~\eqref{eq8.1}, since the
contributions in which the three-point averages appear in
Eqs.~\eqref{eq6.3} already involve $\lambda$ as a prefactor.

With Eq.~\eqref{eq5.6}, the free averages are turned into Gaussian
averages defined through Eq.~\eqref{eq5.7}.  As discussed in
Sec.~\ref{sec:expansion}, the number of useful terms in
Eq.~\eqref{eq5.6} is \emph{a priori} determined by the number of
$\del\r$ and $\hr$ fields in the quantity to be averaged, through the
requirements of causality.

The first average in Eqs.~\eqref{eq8.1} is thus obtained as
\begin{subequations} \la{eq8.5}
\begin{equation}
  \langle 1 3 \h3^{\gamma} \h2 \rangle_\text{f} = \langle 1 3
  \h3^{\gamma} \h2 \rangle_0 = \langle 1 \h3^{\gamma} \rangle_0
  \langle 3 \h2 \rangle_0, \la{eq8.5a}
\end{equation}
where we used $\langle \h3^{\gamma} \h2 \rangle_0=0$ and $\langle 3
\h3^{\gamma} \rangle_0 =0$.  Similarly, the second and third are
\begin{gather}  
  \langle 1 3 \h3^{\gamma} \h2^{\beta} \rangle_\text{f} = \langle 1 3
  \h3^{\gamma} \h2^{\beta} \rangle_0 = \langle 1 \h3^{\gamma}
  \rangle_0 \langle 3 \h2^{\beta} \rangle_0, \la{eq8.5b} \\
  \langle 1 2 \h3^{\gamma} \h2^{\beta} \rangle_\text{f} = \langle 1 2
  \h3^{\gamma} \h2^{\beta} \rangle_0 = \langle 1 \h2^{\beta} \rangle_0
  \langle 2 \h3^{\gamma} \rangle_0. \la{eq8.5c}
\end{gather}
The average $\langle 1 3 2 \h3^{\gamma} \h2^{\beta} \rangle_\text{f}$
is shown to vanish,
\begin{equation}
  \langle 1 3 2 \h3^{\gamma} \h2^{\beta} \rangle_\text{f} = \langle 1
  3 2 \h3^{\gamma} \h2^{\beta} \rangle_0 - D_0 \int_4 \langle 1 3 2 4
  \h3^{\gamma} \h2^{\beta} \h4^{\delta} \h4^{\delta} \rangle_0 = 0,
  \la{eq8.5d}
\end{equation}
since $\langle 3 \h4^{\delta} \rangle_0 \langle 4 \h3^{\gamma}
\rangle_0 =0$ and $\langle 3 \h2^{\beta} \rangle_0 \langle 2
\h4^{\delta} \rangle_0 \langle 4 \h3^{\gamma} \rangle_0 =0$.  One
analogously gets
\begin{equation}
  \langle 1 2 \h3^{\gamma} \rangle_\text{f} = \langle 1 2 \h3^{\gamma}
  \rangle_0 - D_0 \int_4 \langle 1 2 4 \h3^{\gamma} \h4^{\delta}
  \h4^{\delta} \rangle_0 = -2 D_0 \int_4 \langle 1 \h4^{\delta}
  \rangle_0 \langle 2 \h4^{\delta} \rangle_0 \langle 4 \h3^{\gamma}
  \rangle_0.  \la{eq8.5e}
\end{equation}
Note that the effect of the multiplicative noise enters in
Eq.~\eqref{eq8.5e}, making a nonperturbative contribution from the
point of view of the free dynamics.  Finally, one computes the
remaining average as
\begin{equation}\begin{split}
  \langle 1 3 2 \h3^{\gamma} \rangle_\text{f} & = \langle 1 3 2
  \h3^{\gamma} \rangle_0 - D_0 \int_4 \langle 1 3 2 4 \h3^{\gamma}
  \h4^{\delta} \h4^{\delta} \rangle_0 + \frac{1}{2} D_0^2 \int_4
  \int_5 \langle 1 3 2 4 5 \h3^{\gamma} \h4^{\delta} \h4^{\delta}
  \h5^{\varepsilon} \h5^{\varepsilon} \rangle_0 \\
  & = \langle 1 \h3^{\gamma} \rangle_0 \langle 3 2 \rangle_0 + \langle
  1 3 \rangle_0 \langle 2 \h3^{\gamma} \rangle_0, \la{eq8.5f}
\end{split}\end{equation}
since one gets $\langle 1 3 2 4 5 \h3^{\gamma} \h4^{\delta}
\h4^{\delta} \h5^{\varepsilon} \h5^{\varepsilon} \rangle_0 = 0$ due to
the repeated space-time points.
\end{subequations}

Taking the necessary spatial derivatives of the nonvanishing terms,
one finally gets
\begin{subequations}
\begin{align}
  \nabla^{\gamma}_3 \langle 1 3 \h3^{\gamma} \h2 \rangle_\text{f} & =
  \nabla^{\gamma}_3 [ \langle 1 \h3^{\gamma} \rangle_0 \langle 3 \h2
    \rangle_0 ] = \nabla^{\beta}_3 [ \langle 1 \h3^{\beta} \rangle_0
    \langle 3 \h2 \rangle_0 ], \\
  \nabla_2^{\beta} \nabla_3^{\gamma} \langle 1 3 \h3^{\gamma}
  \h2^{\beta} \rangle_\text{f} & = \nabla_2^{\beta} \nabla_3^{\gamma}
     [ \langle 1 \h3^{\gamma} \rangle_0 \langle 3 \h2^{\beta}
       \rangle_0 ] = \nabla_3^{\gamma} [ \langle 1 \h3^{\gamma}
       \rangle_0 \nabla_2^2 \langle 3 \h2 \rangle_0 ] =
     \nabla_3^{\beta} [ \langle 1 \h3^{\beta} \rangle_0 \nabla_3^2
       \langle 3 \h2 \rangle_0 ], \\
  \nabla_2^{\beta} \nabla_3^{\gamma} \langle 1 2 \h3^{\gamma}
  \h2^{\beta} \rangle_\text{f} & = \nabla_2^{\beta} \nabla_3^{\gamma}
     [ \langle 1 \h2^{\beta} \rangle_0 \langle 2 \h3^{\gamma}
       \rangle_0 ] = \nabla_2^{\beta} [ \langle 1 \h2^{\beta}
       \rangle_0 \nabla_3^2 \langle 2 \h3 \rangle_0 ] =
     \nabla_2^{\beta} [ \langle 1 \h2^{\beta} \rangle_0 \nabla_2^2
       \langle 2 \h3 \rangle_0 ], \\
  \begin{split}
    \nabla_3^{\gamma} \langle 1 2 \h3^{\gamma} \rangle_\text{f} & =
    \nabla_3^{\gamma} \left[ -2 D_0 \int_4 \langle 1 \h4^{\delta}
      \rangle_0 \langle 2 \h4^{\delta} \rangle_0 \langle 4
      \h3^{\gamma} \rangle_0 \right] = -2 D_0 \int_4 \langle 1
    \h4^{\delta} \rangle_0 \langle 2 \h4^{\delta} \rangle_0 \nabla_3^2
    \langle 4 \h3 \rangle_0 \\
    & = -2 D_0 \int_4 \langle 1 \h4^{\beta} \rangle_0 \langle 2
    \h4^{\beta} \rangle_0 \nabla_4^2 \langle 4 \h3 \rangle_0
  \end{split} \\
  \nabla_3^{\gamma} \langle 1 3 2 \h3^{\gamma} \rangle_\text{f} & =
  \nabla_3^{\gamma} [ \langle 1 \h3^{\gamma} \rangle_0 \langle 3 2
    \rangle_0 + \langle 1 3 \rangle_0 \langle 2 \h3^{\gamma} \rangle_0
  ] = \nabla_3^{\beta} [ \langle 1 \h3^{\beta} \rangle_0 \langle 3 2
    \rangle_0 + \langle 1 3 \rangle_0 \langle 2 \h3^{\beta} \rangle_0
  ].
\end{align}
\end{subequations}
In the final expressions, all dummy Cartesian indices have been
uniformly denoted by $\beta$.

With these results, the dynamical equations can be written down, up to
the first order of the disorder-strength expansion.  Restoring the
explicit field notation, they read :
\begin{subequations} \label{eq8.6}
\begin{gather}
  ( \partial_t -D_0 \nabla^2 ) G(12) = \del(12) - \l D_0^2
  \nabla^\alpha \left( \int_3 [ \nabla^\alpha \Phi(13) ]
  \nabla_3^{\beta} \{ \langle \del\r(1) \nabla_3^{\beta} \hr(3)
  \rangle_0 \langle \del\r(3) \hr(2) \rangle_0 \} \right), \la{eq8.6a}
  \\
  \begin{split}
    ( \partial_t -D_0 \nabla^2 ) \o{R}(12) & = -\r_0 D_0 \nabla^2
    \del(12) \\
    & \quad - i \l D_0^2 \nabla^\alpha \left( \int_3 [ \nabla^\alpha
      \Phi(13) ] \nabla^{\beta}_3 \{ \langle \del\r(1)
    \nabla_3^{\beta} \hr(3) \rangle_0 [ i \r_0 D_0 \nabla^2_3 \langle
      \del\r(3) \hr(2) \rangle_0 ] \} \right) \\
    & \quad - i \l D_0^2 \nabla^\alpha \left( \int_3 [ \nabla^\alpha
      \Phi(13) ] \nabla_2^{\beta} \{ \langle \del\r(1)
    \nabla_2^{\beta} \hr(2) \rangle_0 [ i \r_0 D_0 \nabla_2^2 \langle
      \del\r(2) \hr(3) \rangle_0 ] \} \right),
  \end{split} \label{eq8.6b} \\
  \begin{split}
    ( \partial_t & -D_0 \nabla^2 ) C(12) = 2\o{R}(21) - \l \r_0 D_0
    \nabla^\alpha \left( \int_3 [ \nabla^\alpha \Phi(13) ] [
      \o{R}^0(13) + \o{R}^0(23) ] \right) \\
    & - i \l D_0^2 \nabla^\alpha \left( \int_3 [ \nabla^\alpha
      \Phi(13) ] \nabla_3^{\beta} \{ \langle \del\r(1)
    \nabla_3^{\beta} \hr(3) \rangle_0 \langle \del\r(3) \del\r(2)
    \rangle_0 + \langle \del\r(1) \del\r(3) \rangle_0 \langle
    \del\r(2) \nabla_3^{\beta} \hr(3) \rangle_0 \} \right) \\
    & + 2 \l D_0^2 \nabla^\alpha \left( \int_3 \int_4 [ \nabla^\alpha
      \Phi(13) ] \langle \del\r(1) \nabla_4^{\beta} \hr(4) \rangle_0
    \langle \del\r(2) \nabla_4^{\beta} \hr(4) \rangle_0 [ i \r_0 D_0
      \nabla_4^2 \langle \del\r(4) \hr(3) \rangle_0 ] \right).
  \end{split} \la{eq8.6c}
\end{gather}
\end{subequations} 

In these evolution equations, there are four space-time integrals in
which the time integral can actually be detached from the
corresponding space integral.  We shall refer to these situations as
isolated time integrals, which are due to the nonlocality in time
induced by the quenched randomness.  Indeed, they appear when a
space-time integral acts on a variable which is present both in the
time-independent random-field covariance and in a single
time-dependent response function.  Then, the time integral obviously
acts on the response function only.  We will next focus on these
isolated time integrals to structure our analysis.

In our derivation, two of these isolated time integrals are directly
obtained as $\int dt_3 \o{R}^0(13)$ and $\int dt_3 \o{R}^0(23)$.  They
correspond to the first integral in Eq.~\eqref{eq8.6c} and originate
from the first term in Eq.~\eqref{eq6.4c}.  We have purposefully
arranged the above formulas to make the two others specifically appear
as $\int dt_3 [i \r_0 D_0 \nabla_2^2 \langle \del\r(2) \hr(3)
  \rangle_0]$ and $\int dt_3 [i \r_0 D_0 \nabla_4^2 \langle \del\r(4)
  \hr(3) \rangle_0]$, in the last integrals of Eqs.~\eqref{eq8.6b} and
\eqref{eq8.6c}, respectively.  Indeed, although it might look like
there are two distinct types of isolated time integrals, our claim is
that the difference is only superficial.  To see this, it must be kept
in mind that, within the zeroth-order theory, there is no distinction
between $\o{R}^0(12)$ and $-\r_0 D_0 \nabla^2 G^0(12)$.  Therefore,
one can safely replace $i \r_0 D_0 \nabla_2^2 \langle \del\r(2) \hr(3)
\rangle_0$ and $i \r_0 D_0 \nabla_4^2 \langle \del\r(4) \hr(3)
\rangle_0$ with $\o{R}^0(23)$ and $\o{R}^0(43)$ in the corresponding
integrals.  A direct hint in favor of this substitution is provided by
a third appearance of this specific combination, $i \r_0 D_0
\nabla^2_3 \langle \del\r(3) \hr(2) \rangle_0$, in the first integral
of Eq.~\eqref{eq8.6b}.  Indeed, it is only when it is interpreted as
$\o{R}^0(32)$ that the equations for $G$, $\o{R}$, and $C$, share the
typical structure of the Schwinger-Dyson equation with the same
self-energy.  Accordingly, we translate Eqs.~\eqref{eq8.6} as
\begin{subequations} \label{eq8.7}
\begin{flalign}
  ( \partial_t -D_0 \nabla^2 ) G(12) & = \del(12) + \l D_0^2
  \nabla^\alpha \left( \int_3 [ \nabla^\alpha \Phi(13) ]
  \nabla_3^{\beta} \left\{ [ \nabla_3^{\beta} G^0(13) ] G^0(32)
  \right\} \right), \la{eq8.7a} \\
  \begin{split}
    ( \partial_t -D_0 \nabla^2 ) \o{R}(12) & = -\r_0 D_0 \nabla^2
    \del(12) + \l D_0^2 \nabla^\alpha \left( \int_3 [ \nabla^\alpha
      \Phi(13) ] \nabla_3^{\beta} \left\{ [ \nabla_3^{\beta} G^0(13) ]
    \o{R}^0(32) \right\} \right) \\
    & \quad + \l D_0^2 \nabla^\alpha \left( \int_3 [ \nabla^\alpha
      \Phi(13) ] \nabla_2^{\beta} \left\{ [ \nabla_2^{\beta} G^0(12) ]
    \o{R}^0(23) \right\} \right),
    \end{split} \label{eq8.7b} \\
  \begin{split}
    ( \partial_t -D_0 \nabla^2 ) C(12) & = 2\o{R}(21) - \l \r_0 D_0
    \nabla^\alpha \left( \int_3 [ \nabla^\alpha \Phi(13) ] [
      \o{R}^0(13) + \o{R}^0(23) ] \right) \\
    & \quad + \l D_0^2 \nabla^\alpha \left( \int_3 [ \nabla^\alpha
      \Phi(13) ] \nabla_3^{\beta} \left\{ [ \nabla_3^{\beta} G^0(13) ]
    C^0(32) + C^0(13) [ \nabla_3^{\beta} G^0(23) ] \right\} \right) \\
    & \quad - 2 \l D_0^2 \nabla^\alpha \left( \int_3 \int_4 [
      \nabla^\alpha \Phi(13) ] [ \nabla_4^{\beta} G^0(14) ] [
      \nabla_4^{\beta} G^0(24) ] \o{R}^0(43) \right).
  \end{split} \la{eq8.7c}
\end{flalign}
\end{subequations} 
Note that, when the bare perturbation expansion is pushed to the
second order, one can actually recognize the first-order expansion of
$\o{R}$ precisely at the places where the proposed substitution is
possible, as seen in the derivations of Eqs.~\eqref{d12} and
\eqref{e15} in Apps.~\ref{app:R} and \ref{app:C}.  As a corollary, the
first-order renormalized theory deriving from the second-order bare
theory also features isolated time integrals that are mere integrals
of the now renormalized density response function, as seen in
Eqs.~\eqref{d13} and \eqref{e16}.  These observations clearly lend
further support to the above substitutions.  More broadly, they hint
at the possibility of a generic reduction of the isolated time
integrals to integrals of the physical response function within the
present framework, although a formal proof hereof is currently
lacking.

Finally, once an isolated time integral is expressed as an integral of
the physical response function, any reference to the corresponding
space-time point can be fully eliminated, thanks to Eqs.~\eqref{tr11}
and \eqref{eq3.7} giving
\begin{equation} \label{intFDR}
  \int_{-\infty}^{+\infty} d t'\, \o{R} (|{\bf r}-{\bf r}'|, t-t') =
  \r_0 \delta({\bf r}-{\bf r'}).
\end{equation}
This relation has for sole basic ingredients the exact FDR and the
exact equilibrium statistical mechanics of ideal gases.  It thus holds
nonperturbatively as well as at any order in $\lambda$ of the present
FDR-preserving perturbation scheme.  Although technically unrelated to
the substitutions advocated above, it acts as a natural continuation
thereof, making the structure of the dynamical equations immediately
simpler.  Thus, specializing Eq.~\eqref{intFDR} to the equilibrium
free dynamics with $\o{R}\equiv\o{R}^0$, one eventually gets from
Eqs.~\eqref{eq8.7} [after some rearrangements using integrations by
  parts and space-translation invariance to make all spatial
  derivatives act on $1=({\bf r},t)$],
\begin{subequations} \label{eq8.8}
\begin{flalign}
  ( \partial_t -D_0 \nabla^2 ) G(12) & = \del(12) - \l D_0^2 \int_3
  \nabla^\alpha \left( [ \nabla^\alpha \nabla^\beta \Phi(13) ] [
    \nabla^{\beta} G^0(13) ] \right) G^0(32) , \la{eq8.8a} \\
  \begin{split}
    ( \partial_t -D_0 \nabla^2 ) \o{R}(12) & = -\r_0 D_0 \nabla^2
    \del(12) - \l D_0^2 \int_3 \nabla^\alpha \left( [ \nabla^\alpha
      \nabla^\beta \Phi(13) ] [ \nabla^{\beta} G^0(13) ] \right)
    \o{R}^0(32) \\
    & \quad + \l \r_0 D_0^2 \nabla^\alpha \nabla^\beta \left( [
      \nabla^\alpha \Phi(12) ] [ \nabla^\beta G^0(12) ] \right),
  \end{split} \label{eq8.8b} \\
  \begin{split}
    ( \partial_t -D_0 \nabla^2 ) C(12) & = 2\o{R}(21) - \l \r_0^2 D_0
    \nabla^2 \Phi(12) \\
      & \quad - \l D_0^2 \int_3 \nabla^\alpha \left( [ \nabla^\alpha
      \nabla^\beta \Phi(13) ] [ \nabla^{\beta} G^0(13) ] \right)
    C^0(32) \\
      & \quad + \l D_0^2 \int_3 \nabla^\alpha \nabla^{\beta} \left( [
      \nabla^\alpha \nabla^\beta \Phi(13) ] C^0(13) \right) G^0(23) \\
      & \quad + 2 \l \r_0 D_0^2 \int_3 \nabla^\alpha \nabla^{\beta}
    \left( [ \nabla^\alpha \Phi(13) ] [ \nabla^{\beta} G^0(13) ]
    \right) G^0(23).
    \end{split} \la{eq8.8c}
  \end{flalign}
\end{subequations} 
One sees that the time integral $\int dt_3 \o{R}^0(23)$ has generated
a mere time-persistent term in the equation for the density
correlation function [the contribution from $\int dt_3 \o{R}^0(13)$
  vanishes by isotropy of the random field], while the last term in
Eq.~\eqref{eq8.8b} is now evidently local in time.  The last term in
Eq.~\eqref{eq8.8c} is entirely due to the multiplicative thermal noise
[see the comment about Eq.~\eqref{eq8.5e} above].

The latter equations will be the basis for all developments in the
remainder of this work.

\section{Equilibrium dynamics: First-order bare theory} 
\setcounter{equation}{0}
\label{sec:FOBT}

By themselves, Eqs.~\eqref{eq8.8} form a first-order bare theory
(FOBT) for the equilibrium dynamics of a noninteracting Brownian gas
plunged in a quenched Gaussian random field.  After Fourier
transformation, under the assumption of time-translation invariance,
one gets the following equilibrium dynamical equations (setting $t\geq
t'=0$),
\begin{subequations} \la{eqFOBT1}
  \begin{align}
    \left( \partial_{t} + \Gam_k \right) G_k (t) & = \del(t) -
    \int_{0}^{t} ds \S^0_k(t-s) G^0_k (s), \la{eqFOBT1a} \\
    \left( \partial_{t} + \Gam_k \right) \o{R}_k (t) & = \r_0 \Gam_k
    \del(t) - \int_{0}^{t} ds \S^0_k(t-s) \o{R}^0_k (s) + L^0_k(t),
    \la{eqFOBT1b} \\
    \left( \partial_{t} + \Gam_k \right) C_k (t) & = \l \r_0^2 \Gam_k
    \Phi_k - \int_{0}^{t} ds \S^0_k(t-s) C^0_k (s) + N^0_k(t).
    \la{eqFOBT1c}
  \end{align}
\end{subequations}

There appear three memory kernels.  The memory functions $\S^0_k(t)$
and $L^0_k(t)$ are explicitly given by
\begin{align} 
  \S^0_k(t) & = \l D_0^2 \int_{\bf q} {\b q} \cdot {\b p} [{\b k}
    \cdot {\b q} \Phi_q ] G^0_p(t), \la{eqFOBTsigG} \\
  L^0_k(t) & = \l \r_0 D_0^2 \int_{\bf q} {\b k} \cdot {\b p} [{\b k}
    \cdot {\b q} \Phi_q ] G^0_p(t), \la{eqFOBTellG}
\end{align}
where $\int_{\b q} \equiv \int d{\b q}/(2\pi)^d$ and ${\b p} \equiv
{\b k}-{\b q}$.  Note that the kernel $L^0_k(t)$ would be absent in
the usual case of a Langevin equation with additive thermal noise.  It
is hence associated with the anomalous part of the physical response
function, arising from the multiplicative nature of the basic
stochastic equation for the density variable.

In fact, one can further investigate the origin of $L^0_k(t)$ by going
back to the initial dynamics.  Indeed, within the operator formalism
of Martin, Siggia, and Rose \cite{MarSigRos73PRA,Phy75JPA}, an
evolution equation for $\overline{R}({\bf r},t;{\bf r}',t')$ can be
obtained from Eq.~\eqref{eq1.14}, through multiplication by $i
\Lam({\bf r}',t')$ and double-averaging over thermal fluctuations and
disorder.  The contribution of the random forces deriving from the
external potential then reads ($\nabla$ and $\nabla'$ act on ${\bf r}$
and ${\bf r}'$, respectively)
\begin{equation} \label{eqFOBT4}
  i \frac{D_0}{T} \overline{\langle \nabla \cdot \left[ \r({\bf r},t)
      \nabla v({\bf r}) \right] \Lam({\bf r}',t') \rangle} = i
  \frac{D_0^2}{T} \nabla \cdot \overline{\langle \r({\bf r},t) \nabla'
    \cdot [\r({\bf r}',t') \nabla' \hr({\bf r}',t') ] \rangle \nabla
    v({\bf r})},
\end{equation}
where a realization-dependent physical response function is clearly
visible.  Now, we may split $\r({\bf r}',t')$ as $\langle \r({\bf
  r'}) \rangle + \r({\bf r}',t') - \langle \r({\bf r'}) \rangle$,
where $\langle \r({\bf r'}) \rangle$ corresponds to the static density
profile induced by the random field and $\r({\bf r}',t') - \langle
\r({\bf r'}) \rangle$ to the thermal fluctuations about this profile.
Focusing on the first contribution, one gets
\begin{equation} \label{eqFOBT5}
  i \frac{D_0^2}{T} \nabla \cdot \overline{\langle \r({\bf r},t)
    \nabla' \cdot [ \langle \r({\bf r'}) \rangle \nabla' \hr({\bf
        r}',t') ] \rangle \nabla v({\bf r})} = i \frac{D_0^2}{T}
  \nabla^\alpha \nabla'^\beta \overline{[\nabla'^\beta \langle \r({\bf
        r},t) \hr({\bf r}',t') \rangle] [\nabla^\alpha \{ \langle
      \r({\bf r'}) \rangle v({\bf r}) \} ]},
\end{equation}
where a noise-response function appears.  If the latter is evaluated
with respect to the free dynamics, in the spirit of the present FOBT,
the averages factorize and one obtains
\begin{equation} \label{eqFOBT6}
  \begin{split} 
    i \frac{D_0^2}{T} \nabla^\alpha \nabla'^\beta &
    \overline{[\nabla'^\beta \langle \r({\bf r},t) \hr({\bf r}',t')
        \rangle_\text{f}] [\nabla^\alpha \{ \langle \r({\bf r'})
        \rangle v({\bf r}) \} ]} \\
    & \qquad\qquad = i \frac{D_0^2}{T} \nabla^\alpha
    \nabla'^\beta \left\{ [\nabla'^\beta \langle \r({\bf r},t)
      \hr({\bf r}',t') \rangle_\text{f}] [\nabla^\alpha \overline{
        \langle \r({\bf r'}) \rangle v({\bf r}) } ] \right\} \\
    & \qquad\qquad = \lambda \r_0 D_0^2 \nabla^\alpha
    \nabla^\beta \left( [\nabla^\beta G^0({\bf r},t;{\bf r}',t')]
          [\nabla^\alpha \Phi(|{\bf r}-{\bf r}'|) ] \right) = L^0({\bf
            r},t;{\bf r}',t'),
  \end{split}
\end{equation}
where the real-space expression for $L^0({\bf r},t;{\bf r}',t')$ is
read off Eq.~\eqref{eq8.8b}.  In these final steps, we used
Eq.~\eqref{eq3.4a} to compute the disorder average over the Gaussian
random field, and translational invariance to replace $\nabla'^\beta$
with $-\nabla^\beta$.  Eventually, it thus appears that the kernel
$L^0_k(t)$ arises, at least in part, from the interplay of the
multiplicative nature of the thermal noise and of the static density
heterogeneities imprinted in the fluid by the random external
potential.  Note that, if one repeats all these steps in the case of
the density correlation function, i.e, starting with
Eq.~\eqref{eq1.14} multiplied by $\rho({\bf r}',t')$ and
double-averaged, one obtains the term $- \l \r_0^2 D_0 \nabla^2
\Phi(|{\bf r}-{\bf r}'|)$ of Eq.~\eqref{eq8.8c}, which gives
$\l\r_0^2\Gam_k \Phi_k$ in Eq.~\eqref{eqFOBT1c}.  Indeed, to show
this, one begins with
\begin{equation}
  \frac{D_0}{T} \overline{\langle \nabla \cdot \left[ \r({\bf r},t)
      \nabla v({\bf r}) \right] \r({\bf r}',t') \rangle} =
  \frac{D_0}{T} \nabla \cdot \overline{\langle \r({\bf r},t) \r({\bf
      r}',t') \rangle \nabla v({\bf r})},
\end{equation}
which, after replacing $\r({\bf r}',t')$ with its thermal average,
becomes
\begin{equation}
  \frac{D_0}{T} \nabla \cdot \overline{\langle \r({\bf r},t) \langle
    \r({\bf r'}) \rangle \rangle \nabla v({\bf r})} = \frac{D_0}{T}
  \nabla \cdot \overline{ \langle \r({\bf r},t) \rangle [\nabla \{
      \langle \r({\bf r'}) \rangle v({\bf r}) \} ]}.
\end{equation}
Then, if $\r({\bf r},t)$ is set to evolve according to the free
dynamics, one gets
\begin{equation}
  \frac{D_0}{T} \nabla \cdot \overline{\langle \r({\bf r},t)
    \rangle_\text{f} [\nabla \{ \langle \r({\bf r'}) \rangle v({\bf
        r}) \} ]} = \frac{D_0}{T} \nabla \cdot \left\{ \langle \r({\bf
    r},t) \rangle_\text{f} [\nabla \overline{ \langle \r({\bf r'})
      \rangle v({\bf r}) } ] \right\} = - \lambda \r_0^2 D_0 \nabla^2
  \Phi(|{\bf r}-{\bf r}'|),
\end{equation}
as announced.  As discussed below, this contribution is clearly an
outgrowth of the disorder-induced static density profile.

The kernel $N^0_k(t)$ originally consists of three integrals,
\begin{equation} \la{eqFOBTenn3int}
\begin{split}
  N^0_k(t) & = - \l D_0^2 \int_{-\infty}^{0} ds \int_{\bf q}
  {\b q} \cdot {\b p} [{\b k} \cdot {\b q} \Phi_q ] G^0_p(t-s)
  C^0_k(s) \\
  & \quad + \l D_0^2 \int_{-\infty}^{0} ds \int_{\bf q} {\bf k}\cdot
  {\bf q} [{\bf k}\cdot {\bf q} \Phi_q ] C^0_p(t-s) G^0_k(-s) \\
  & \quad + 2 \l \r_0 D_0^2 \int_{-\infty}^{0} ds \int_{\bf q} {\bf
    k}\cdot {\bf p} [{\bf k}\cdot {\bf q} \Phi_q ] G^0_p(t-s)
  G^0_k(-s),
\end{split}
\end{equation}
but actually reduces to a local function of time if one uses the
identities Eqs.~\eqref{eq7.3} to rearrange this expression.  Indeed,
using Eq.~\eqref{eq7.3c} to distribute the last integral over the
first two, one gets
\begin{equation}
  N^0_k(t) = \l D_0^2 \int_{-\infty}^{0} ds \int_{\bf q} [{\b k} \cdot
    {\b q} \Phi_q ] \left[ p^2 G^0_p(t-s) C^0_k(s) + C^0_p(t-s) k^2
    G^0_k(-s) \right],
\end{equation}
which, with Eq.~\eqref{eq7.3a} followed by Eq.~\eqref{eq7.3b}, leads
to
\begin{equation} \la{eqFOBTenn}
  N^0_k(t) = \frac{\l D_0}{\r_0} \int_{-\infty}^{0} ds \int_{\bf q}
  [{\b k} \cdot {\b q} \Phi_q ] \, \partial_s \left[ C^0_p(t-s)
    C^0_k(s) \right]
  = \l D_0 \int_{\bf q} [{\b k} \cdot {\b q} \Phi_q ] C^0_p(t) .
\end{equation}

The memory functions have to be related with one another, in order for
Eqs.~\eqref{eqFOBT1} to obey the FDR.  Using $\r_0 \Gam_p G^0_p(t) =
\o{R}^0_p(t)$ to rewrite
\begin{align} 
  \S^0_k(t) & = \l \frac{D_0}{\rho_0} \int_{\bf q} \frac{{\b q} \cdot
    {\b p}}{p^2} [{\b k} \cdot {\b q} \Phi_q ] \o{R}^0_p(t),
  \la{eqFOBTsigR} \\
  L^0_k(t) & = \l D_0 \int_{\bf q} \frac{{\b k} \cdot {\b
      p}}{p^2} [{\b k} \cdot {\b q} \Phi_q ] \o{R}^0_p(t),
  \la{eqFOBTellR}
\end{align}
and forming the combination $\r_0 \S^0_k(t) - L^0_k(t)$, one
immediately finds that the kernels obey the FDR-like relation
\begin{equation} \la{eqFOBTkernFDR}
  \r_0 \S^0_k(t) - L^0_k(t) = \partial_{t} N^0_k(t)
\end{equation}
as a mere corollary of the FDR $\o{R}^0_p(t)=-\partial_t C^0_p(t)$.

Another interesting rearrangement of Eq.~\eqref{eqFOBTenn3int} through
$\r_0 \Gam_k G^0_k(t) = \o{R}^0_k(t)$ is
\cite{DekHaa75PRA,MiyRei05JPA}
\begin{equation} \label{eqFOBTennalt}
  N^0_k(t) = - \int_{-\infty}^{0} ds \S^0_k(t-s) C^0_k(s) +
  \int_{-\infty}^{0} ds D^0_k(t-s) \o{R}^0_k(-s),
\end{equation}
where the new kernel $D^0_k(t)$ consists of the two parts:
\begin{align}
  D^0_k(t) & \equiv M^0_k(t) + \frac{2}{\r_0 \Gam_k} L^0_k(t),
  \la{eqFOBTDML} \\
  M^0_k(t) & \equiv \frac{\l D_0}{\r_0} \int_{\bf q} ({\hat {\bf
      k}}\cdot {\bf q})^2 \Phi_q C^0_p(t). \la{eqFOBTM}
\end{align}
Here, ${\hat {\bf k}}$ denotes the unit vector ${\bf k}/k$ and
$M^0_k(t)$ turns out to be the MCT memory kernel \cite{Kra05PRL,
  Kra07PRE, Kra09PRE, KonKra17SM}, albeit in its ``bare'' form (see
below). It is then straightforward to show that $M^0_k(t)$ is related
to $N^0_k(t)$ and $L^0_k(t)$ [after using $\r_0 G^0_p(t) = C^0_p(t)$
  in Eq.~\eqref{eqFOBTellG}] as
\begin{equation} \la{eqFOBTMNL}
  M^0_k(t) = \frac{1}{\r_0} \left( N^0_k(t) - \frac{L^0_k(t)}{\Gam_k}
  \right).
\end{equation}
In combination with Eqs.~\eqref{eqFOBTkernFDR} and \eqref{eqFOBTDML},
this immediately leads to
\begin{align}
  D^0_k(t) & = \frac{1}{\r_0} \left( N^0_k(t) +
  \frac{L^0_k(t)}{\Gam_k} \right), \\
  \partial_t D^0_k(t) & = \S^0_k(t) + \frac{1}{\r_0 \Gam_k} \left(
  \partial_t -\Gam_k \right) L^0_k(t).
\end{align}
Again, one can see that the presence of $L^0_k(t)$ deeply changes the
structure of the dynamics. Indeed, if the kernel $L^0_k(t)$ were
absent, one would simply get $D^0_k(t) = M^0_k(t) = N^0_k(t)/\r_0$
with the familiar relation $\partial_t D^0_k(t) = \S^0_k(t)$, as found
in the case of Langevin dynamics with additive thermal noise
\cite{DekHaa75PRA, BouCugKurMez96PA}.

We now consider some key features of these dynamical equations.

\subsection{Consistency with the FDR.} 
The present perturbation expansion is dictated by the time reversal
invariance of the effective dynamical action. It is hence guaranteed
to preserve the FDR at each order of the expansion. This is confirmed
by explicitly showing that the above first-order dynamical equations
for $\o{R}_k(t)$ and $C_k(t)$ are indeed consistent with the FDR.

Taking the time derivative of the FDR, $\o{R}_k(t) = -\theta(t)
\partial_{t} C_k(t) $, one gets
\begin{equation} \la{eqconsist1}
  \partial_{t} \o{R}_k(t) = -\delta(t) \partial_{t} C_k(0) - \theta(t)
  \partial^2_{t} C_k(t).
\end{equation}
With the second derivative of $C_k(t)$ obtained from
Eq.~\eqref{eqFOBT1c},
\begin{equation} \la{eqconsist2}
  \partial^2_t C_k(t) = -\Gam_k \partial_t C_k(t) -\S^0_k(t) C^0_k(0)
  - \int_0^t ds \, \S^0_k(t-s) \partial_s C^0_k(s) +\partial_{t}
  N^0_k(t),
\end{equation}
Eq.~\eqref{eqconsist1} takes the form
\begin{equation} \la{eqconsist3} %
  \Big(\partial_{t} + \Gam_k \Big) \o{R}_k(t) = - \delta(t) \partial_t
  C_k(0) - \int_0^t ds \, \S^0_k(t-s) \o{R}^0_k(s) + L^0_k(t),
\end{equation}
where Eqs.~\eqref{tr16a} and \eqref{eqFOBTkernFDR} [$C^0_k(0)=\r_0$]
have been used.

Comparing Eq.~\eqref{eqconsist3} with Eq.~\eqref{eqFOBT1b}, we see
that the dynamics obeys the FDR under the condition $\r_0 \Gam_k =
-\partial_t C_k(0) = \Gam_k [C_k(0) - \l \r_0^2 \Phi_k]$, where the
last equality follows from Eq.~\eqref{eqFOBT1c} at $t=0$, knowing that
$N^0_k(0) = \l \r_0 D_0 \int_{\bf q} {\b k} \cdot {\b q} \Phi_q =0$ by
isotropy. This requires that
\begin{equation} \la{eqconsist4} %
  C_k(0)=\r_0+\l\r_0^2 \Phi_k.
\end{equation}

\subsection{Presence of a static nonvanishing component.}
Since the memory terms in Eqs.~\eqref{eqFOBT1} only involve the bare
correlation and response functions that are exponentially relaxing in
time, the present FOBT does not sustain the possibility of a
transition to a kinetically generated nonergodic state driven by the
Gaussian random potential.  This feature is at variance with the
self-consistent MCT predictions \cite{KonKra17SM}.

Yet, it follows from Eq.~\eqref{eqFOBT1c} that the density correlation
function $C_k(t)$ does exhibit a \emph{disorder-induced}
time-persistent component,
\begin{equation} \la{eqconsist5} %
  C_k(t \to +\infty) = \l \r_0^2 \Phi_k.
\end{equation}
This contribution is of a strictly \emph{static} nature and must be
distinguished from a kinetically generated nonergodicity parameter
such as predicted by the MCT, for instance.

\subsection{Disorder-induced static structure factors.}
We are examining the equilibrium dynamics, hence the initial condition
for the density correlation function $C_k(0)$ should yield the
equilibrium static structure factor of the fluid $S^\text{st}_k$,
through the relation $C_k(0)=\r_0 S^\text{st}_k$.  The latter acquires
a \emph{disorder-induced} contribution in the presence of the Gaussian
random potential and Eq.~\eqref{eqconsist4} gives $S^\text{st}_k = 1 +
\l \r_0 \Phi_k$.

The time-persistent component of the density correlation function
$C_k(t \to +\infty)$ should similarly be related to the
\emph{disorder-induced} disconnected static structure factor
$S^\text{d}_k$ through $C_k(t \to +\infty) = \r_0 S^\text{d}_k$, and
one gets $S^\text{d}_k = \l \r_0 \Phi_k$ from Eq.~\eqref{eqconsist5}.

Both expressions for $S^\text{st}_k$ and $S^\text{d}_k$ agree to first
order with the exact static results, Eqs.~\eqref{eq3.6}.  In
particular, the equality $S^\text{st}_k=1+S^\text{d}_k$ is obeyed,
ensuring the validity of the crucial relation Eq.~\eqref{eq3.7}.

In summary, it comes out of these first three points that the present
FOBT is plainly consistent both with the FDR and with the equilibrium
static results at the same level of approximation.  Hence, it
manifestly fulfills all the basic requirements for a \emph{bona fide}
theory of equilibrium dynamics.

Once $C_k(t \to +\infty)$ is linked to the disconnected static
structure factor, it might be subtracted from the density correlation
function to get its connected component, $F_k(t) = C_k(t)-\r_0
S^\text{d}_k$, according to Eq.~\eqref{Fconnected}.  Rewriting
Eq.~\eqref{eqFOBT1c}, $F_k(t)$ obeys
\begin{equation} \la{eqconsist6} %
  \Big( \partial_{t} + \Gam_k \Big) F_k (t) = - \int_0^{t} ds 
  \S^0_k(t-s) C^0_k (s) + N^0_k(t),
\end{equation}
with $F_k(0)=\r_0$, based on Eqs.~\eqref{eqconsist4} and
\eqref{eqconsist5}.  It is clear that $F_k(t)$ can only relax to zero.

\subsection{Bare mode-coupling equations for the connected density
  correlation function.} We may further try and simplify the dynamical
equations for the connected density correlation function $F_k(t)$.
With the FDR for the kernels, Eq.~\eqref{eqFOBTkernFDR}, $\S^0_k(t)$
is straightforwardly eliminated from Eq.~\eqref{eqconsist6}, to get
\begin{equation} \la{eqBMCT1} %
  \Big( \partial_{t} + \Gam_k \Big) F_k (t) = - \frac{1}{\r_0}
  \int_0^{t} ds N^0_k(t-s) \partial_s C^0_k (s) - \frac{1}{\r_0}
  \int_0^{t} ds L^0_k(t-s) C^0_k (s),
\end{equation}
where $N^0_k(0)=0$ is used in an integration by parts. With the
diffusion equation $\partial_s C^0_k(s)=-\Gam_k C^0_k(s)$, which
follows from Eqs.~\eqref{eq7.3}, one can put the above equation into
the form
\begin{equation} \la{eqBMCT2}
  \left( \partial_{t} + \Gam_k \right) F_k(t) = - \int_0^{t} ds
  M^0_k(t-s) \partial_s C^0_k (s),
\end{equation}
where we used Eq.~\eqref{eqFOBTMNL}.  The explicit expression for the
kernel $M^0_k(t)$ is found in Eq.~\eqref{eqFOBTM}.  Note that there is
a significant qualitative difference between the present use of
Eqs.~\eqref{eq7.3} and the previous ones.  Indeed, up to now, these
equations were invoked to make substitutions within the kernels only,
while here, a change in the formal structure of Eq.~\eqref{eqBMCT1},
hence of Eq.~\eqref{eqFOBT1c}, is achieved.

Apart from the bare nature of the memory term, Eqs.~\eqref{eqBMCT2}
and \eqref{eqFOBTM} have the same form as those of the self-consistent
MCT developed by one of us for the study of fluids in random
environments \cite{Kra05PRL, Kra07PRE, Kra09PRE, KonKra17SM}.  Indeed,
using the present notations, the latter read for a noninteracting
Brownian gas:
\begin{subequations} \la{MCT}
  \begin{align}
    \left( \partial_{t} + \Gam_k \right) F_k(t) & = - \int_0^{t} ds
         {\cal M}_k(t-s) \partial_s F_k (s), \la{eqMCTa} \\
   {\cal M}_k(t) & = \frac{D_0}{\r_0^2} \int_{\bf q} ({\hat {\bf
       k}}\cdot {\bf q})^2 S^\text{d}_q F_p(t), \la{eqMCTb}
 \end{align}
\end{subequations}
with $F_k(0)=\r_0$ and $S^\text{d}_q$ the exact disconnected structure
factor.  These equations can immediately be brought forth from the
former through a simple \emph{ad hoc} renormalization scheme in which
the linearized disconnected structure factor $\r_0\l\Phi_q$ is
replaced with its exact value $S^\text{d}_q$ and the bare density
correlation function $C^0_k(t)$ is replaced with the connected density
correlation function $F_k(t)$ [note that both $C_k(t)$ and $F_k(t)$
  reduce to $C^0_k(t)$ in the absence of disorder].
 
\subsection{Mean-squared displacement and related quantities.}
In many studies, the interest mostly revolves around the mean-squared
displacement (MSD) of a particle. Thus, this is a quantity of choice
to investigate here.

Since we are dealing with a noninteracting gas, the connected density
correlation function $F_k(t)$ coincides with the self intermediate
scattering function (with an additional $\r_0$ factor).  The MSD
$\Delta(t)$ can therefore be obtained through the standard low-$k$
expansion
\begin{equation} \la{eqMSD1} %
  F_k(t) = \r_0 \left[ 1 - \frac{k^2 \Delta(t)}{2 d} + O(k^4) \right],
\end{equation} 
where $d$ is the space dimension.

Equation \eqref{eqBMCT2} [since Eqs.~\eqref{eqconsist6},
  \eqref{eqBMCT1}, and \eqref{eqBMCT2}, are fully equivalent, the
  choice of the starting equation is immaterial] can be
straightforwardly integrated to get
\begin{equation} \la{eqMSD2} %
  F_k(t) = C^0_k(t) \left[ 1 - \int_0^t ds C^0_k(s)^{-1} \int_0^{s} du
    M^0_k(s-u) \partial_u C^0_k (u) \right],
\end{equation} 
which, in the low-$k$ limit, yields
\begin{equation} \la{eqMSD3} %
  \frac{\Delta(t)}{2 d D_0} = t - \int_0^{t} ds \int_0^{s} du
  M^0_0(u),
\end{equation} 
with
\begin{equation} \la{eqMSD4} %
  M^0_0(t) = \lim_{k\to 0} M^0_k(t) = \frac{\lambda D_0}{d \r_0}
  \int_{\bf q} q^2 \Phi_q C^0_q(t).
\end{equation}
Equation \eqref{eqFOBTMNL} implies $M^0_0(t) = -\lim_{k \to 0}
[L^0_k(t)/(\r_0 \Gam_k)]$, since $\lim_{k \to 0} N^0_k(t)=0$.  Thus,
we observe that the diffusion of a particle is fully determined by the
small-wavevector behavior of the sole kernels $M^0_k(t)$ or
$L^0_k(t)$.  With a direct integration of Eq.~\eqref{eq1.1} leading to
\begin{equation} \la{eqMSD3bis} %
  \frac{\Delta(t)}{2 d D_0} = t + \frac{D_0}{2 d T^2} \int_0^{t} ds
  \int_0^{t} du \overline{\langle {\bf F}_i(s) {\bf F}_i(u) \rangle} =
  t + \frac{D_0}{d T^2} \int_0^{t} ds \int_0^{s} du \overline{\langle
    {\bf F}_i(s) {\bf F}_i(u) \rangle},
\end{equation} 
these low-$k$ kernels are immediately recognized as approximations for
the force autocorrelation function.

Using the diffusion equation $C^0_q(t) = - \partial_t C^0_q(t)/(D_0
q^2)$ in Eq.~\eqref{eqMSD4} to perform the inner time integration in
Eq.~\eqref{eqMSD3}, one alternatively obtains
\begin{equation} \la{eqMSD5} %
  \frac{\Delta(t)}{2 d D_0} = \left( 1 - \frac{\l}{d} \right) t +
  \int_0^{t} ds \, m^0(s),
\end{equation} 
with 
\begin{equation} \la{eqMSD6} %
  m^0(t) = \frac{\l}{\r_0 d} \int_{\bf q} \Phi_q C^0_q(t) =
  \frac{\l}{\r_0 d} \int_{\bf r} \Phi(r) C^0(r,t).
\end{equation} 
The second expression involving the bare diffusion kernel $C^0(r,t) =
\r_0 (4 \pi D_0 t)^{-d/2} e^{-r^2/(4 D_0 t)}$ results from Parseval's
theorem. 

From these relations, expressions for the time-dependent diffusion
coefficient $D(t) = \dot{\Delta}(t) / (2d)$ and the velocity
autocorrelation function $Z(t) = \ddot{\Delta}(t) / (2d)$ immediately
follow, which read
\begin{gather} 
  \frac{D(t)}{D_0} = 1 - \int_0^{t} ds M^0_0(s) = 1 - \frac{\l}{d} +
  m^0(t), \la{eqMSD7} \\
  \frac{Z(t)}{D_0} = -M^0_0(t) = \dot{m}^0(t). \la{eqMSD8}
\end{gather}

These results show that the present FOBT fully agrees with earlier
perturbative calculations at the same order \cite{DeaDruHor07JSM} in
predicting for the long-time diffusion coefficient
\begin{equation} \la{eqMSD9} %
  \frac{D_\infty}{D_0} = \lim_{t\to+\infty} \frac{D(t)}{D_0} = 1 -
  \frac{\l}{d}.
\end{equation} 
They also unambiguously demonstrate the breakdown of the approach at
strong disorder, since negative values of $D_\infty$, hence of
$\Delta(t)$, are obtained when $\l$ exceeds the space dimension $d$.
Correspondingly, anomalies (nonmonotonicity, overshoot above the
initial value) appear in the density correlation functions at low $k$
when this threshold is approached.

\subsection{Asymptotic analysis and long-time tails.}
Making use of the explicit forms of $C^0_k(t)$ or $C^0(r,t)$ in
Eqs.~\eqref{eqFOBTM}, \eqref{eqMSD4}, and \eqref{eqMSD6}, the presence
of long-time tails in the problem is straightforwardly demonstrated,
since one obtains for the memory kernels
\begin{subequations} \la{eqLTT1}
\begin{align}
  M^0_k(t) & \sim \frac{D_0 k^2 \l \Phi_k}{(4 \pi D_0 t)^{d/2}}, & &
  t\to+\infty, k\neq0, \la{eqLTT1a} \\
  M^0_0(t) & \sim \frac{2 \pi D_0 \l \Phi_0}{(4 \pi D_0 t)^{d/2+1}}, &
  & t\to+\infty, \la{eqLTT1b} \\
  m^0(t) & \sim \frac{\l \Phi_0}{d (4 \pi D_0 t)^{d/2}}, & &
  t\to+\infty. \la{eqLTT1c}
\end{align}
\end{subequations}
The qualitative behavior of the velocity autocorrelation function
$Z(t)$ [see Eq.~\eqref{eqMSD8}], which is thus found negative, linear
in the disorder strength, and relaxing as $-t^{-(d/2+1)}$, is exactly
the same as in the Brownian random Lorentz gas
\cite{FraHofBauFre10CP}.  More generally, these results are in
agreement with previous phenomenological calculations
\cite{ErnMacDorBei84JSP}.

In order to discuss the correlation functions, Eq.~\eqref{eqMSD2} is
first explicitly written as
\begin{equation} \la{eqLTT2} %
  F_k(t) = \r_0 e^{-D_0 k^2 t} \left[ 1 + D_0 k^2 \int_0^t ds
    \int_0^{s} du \, M^0_k(u)\, e^{D_0 k^2 u} \right],
\end{equation} 
then, after an integration by parts,
\begin{equation} \la{eqLTT3} %
  F_k(t) = \r_0 \left[ e^{-D_0 k^2 t} + D_0 k^2 \int_0^t ds \,
 M^0_k(s)\, (t-s) e^{-D_0 k^2 (t-s)} \right].
\end{equation} 
Standard analysis based on Laplace transforms then allows one to
obtain
\begin{equation} \la{eqLTT4} %
  F_k(t) \sim \r_0 \frac{\l \Phi_k}{(4 \pi D_0 t)^{d/2}}, \qquad
  \qquad t\to+\infty, k\neq0.
\end{equation} 

For completeness, we also report the short-time expansions,
\begin{subequations} \la{eqSTE1}
  \begin{align}
    M^0_k(t) & \sim \frac{D_0 \l}{d} \left[ \int_{\bf q} q^2 \Phi_q -
      D_0 t \int_{\bf q} q^2 (k^2+q^2) \Phi_q \right], & t\to 0, \\
    m^0(t) & \sim \frac{\l}{d} \left[ 1 - D_0 t \int_{\bf q} q^2
      \Phi_q \right], & t\to 0, \\
    F_k(t) & \sim \r_0 \left[ 1 - D_0 k^2 t + \frac{(D_0 t)^2}{2} k^2
      \left(k^2 + \frac{\l}{d} \int_{\bf q} q^2 \Phi_q \right)
      \right], & t\to 0.
  \end{align}
\end{subequations}

\subsection{Explicit example.}
In order to report complete solutions of the FOBT, we have to
particularize the covariance of the Gaussian random field.  Since it
allows one to analytically perform the wavevector integrals appearing
in the definitions of $M_k^0(t)$ and $m^0(t)$, a Gaussian covariance,
\begin{equation} \la{eqexplicit1} %
  \Phi(r) = e^{-r^2/(2R^2)}, \qquad \Phi_k = (2\pi R^2)^{d/2}
  e^{-k^2R^2/2},
\end{equation}
where $R$ controls the range of the random-field correlations, appears
as a particularly convenient choice.  One then obtains (see
App.~\ref{appA})
\begin{gather} 
  M_k^0(t) = \frac{\lambda}{2} \cdot \frac{2 D_0}{R^2} \cdot \frac{1 +
    2 D_0 t / R^2 + k^2 R^2 (2 D_0 t / R^2)^2 }{(1 + 2 D_0 t /
    R^2)^{d/2+2}} \cdot \exp \left[ - \frac{k^2 R^2 (2 D_0 t/
      R^2)}{2(1 + 2 D_0 t / R^2)} \right], \la{eqexplicit4} \\
  m^0(t) = \frac{\lambda}{d} \cdot \frac{1}{(1 + 2D_0 t / R^2)^{d/2}}.
\end{gather}
With these formulas, the MSD can be expressed in closed form, and
reads
\begin{equation}
  \frac{\Delta(t)}{d R^2} = \frac{2 D_0 t}{R^2} \left( 1 -
  \frac{\lambda}{d} \right) + \frac{\lambda}{d}
  \begin{cases}
    2 \left( \sqrt{1 + 2 D_0 t/R^2} - 1 \right) & \text{if $d=1$}, \\
    \ln \left( 1 + 2D_0 t /R^2 \right) &
    \text{if $d=2$}, \\
    \dfrac{2}{d-2} \left[ 1 - \dfrac{1}{( 1 + 2D_0 t/ R^2 )^{d/2-1}}
    \right] & \text{if $d\geq 3$}.
  \end{cases}
\end{equation}
In these expressions, the natural units of length and time, $R$ and
$\tau=R^2/(2D_0)$, respectively, have been made evident. The time
$\tau$ merely is the time at which the characteristic lengthscale of
free diffusion $\sqrt{2D_0 t}$ reaches the correlation length of the
disorder.

\begin{figure}
\centering
\includegraphics{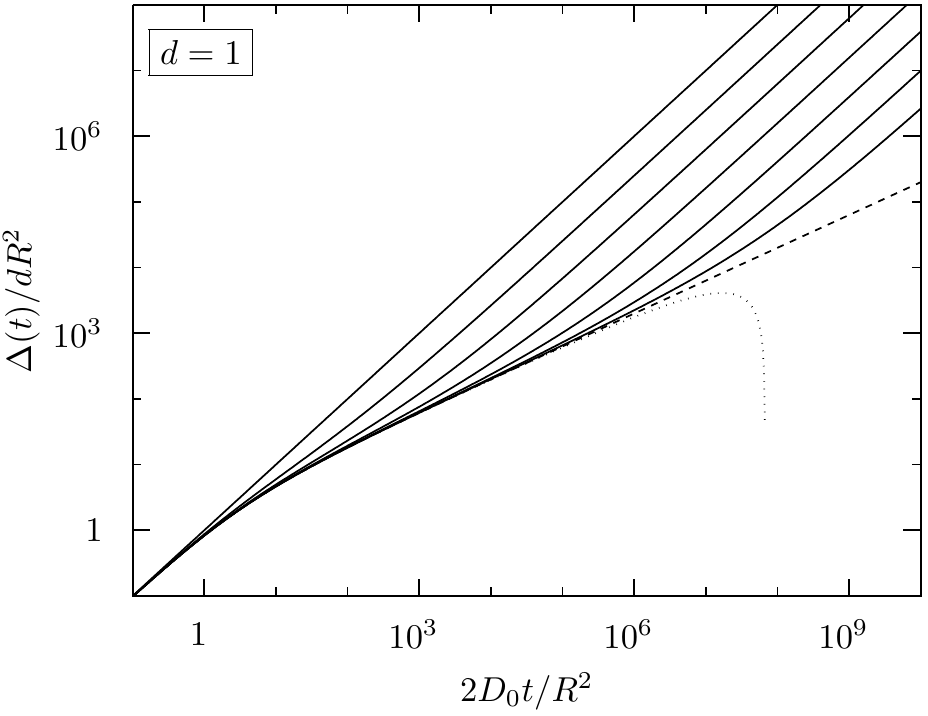}

\includegraphics{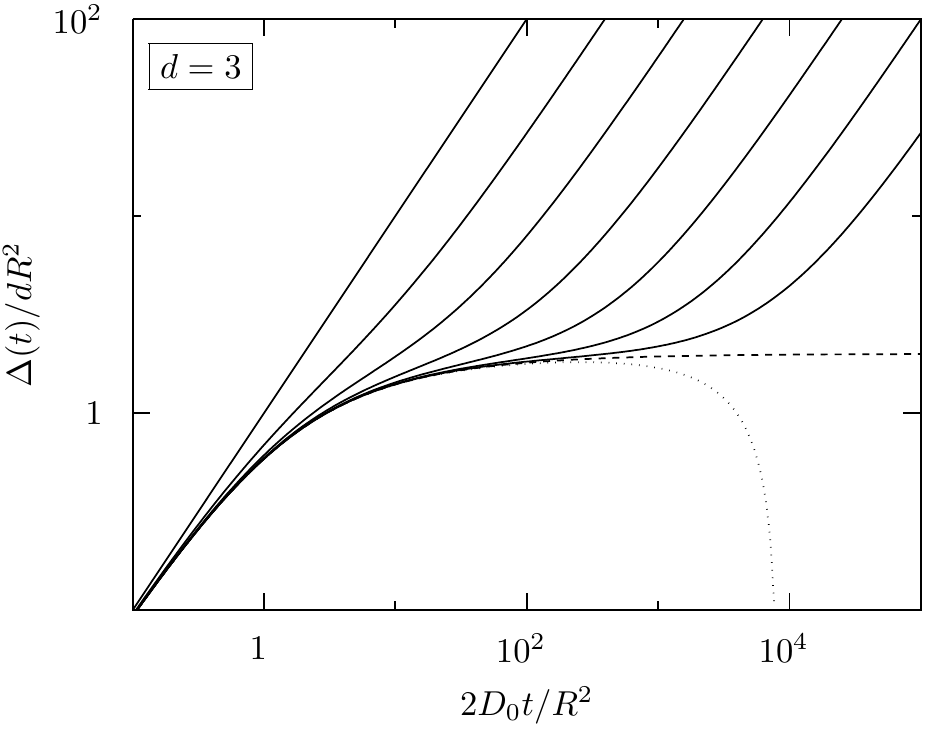}
\caption{\label{fig:bareMSD} Time evolution of the mean-squared
  displacement in a noninteracting Brownian gas exposed to a Gaussian
  random field with Gaussian covariance [specified in
    Eq.~\eqref{eqexplicit1}] in space dimensions $d=1$ (top) and $d=3$
  (bottom), according to the first-order bare theory.  From left to
  right, top to bottom: $\lambda=0$, $\lambda=d(1-1/4^n)$ with
  $n=1,2,\ldots,6$, $\lambda=d$ (dashed line), and
  $\lambda=d(1+1/4^n)$ with $n=6$ (dotted line, unphysically diverging
  to $-\infty$).}
\end{figure}

The effect of the relative disorder strength $\lambda$ on the time
dependence of the MSD is shown in Fig.~\ref{fig:bareMSD} for space
dimensions $d=1$ and $d=3$. Note that the theory is clearly pushed
well beyond its range of validity, since results up to $\lambda=d$ and
slightly above, where its breakdown is obvious, are shown for
completeness. The curve at $\lambda=d$ emphasizes the transient
between the short- and long-time normal diffusive regimes.

\begin{figure}
\centering
\includegraphics{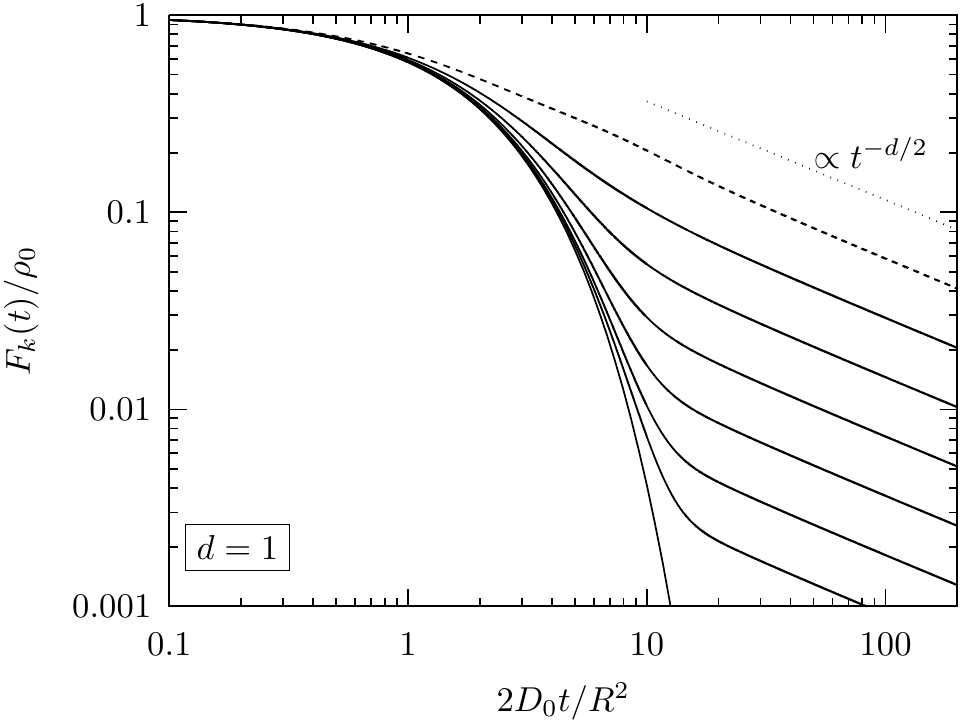}

\includegraphics{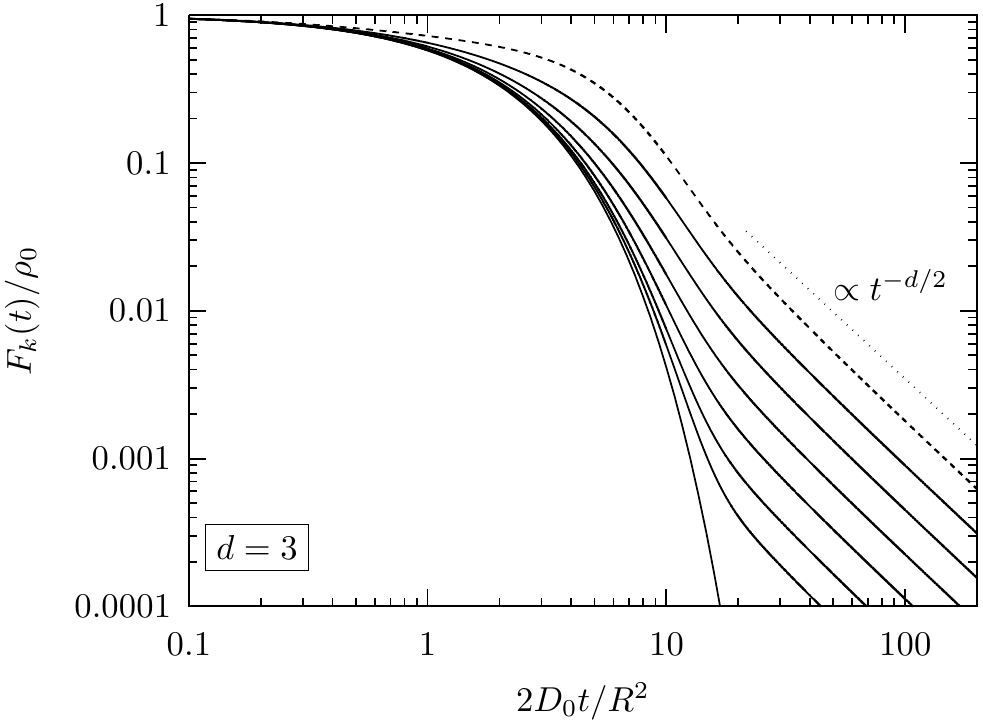}
\caption{\label{fig:barecorrel} Time evolution of the connected
  density correlation function in a noninteracting Brownian gas
  exposed to a Gaussian random field with Gaussian covariance in space
  dimensions $d=1$ (top) and $d=3$ (bottom), according to the
  first-order bare theory.  The wavevector is $k R=\pi/3$. From left
  to right, bottom to top: $\lambda=0$, $\lambda = d/2^{n}$ with
  $n=6,5,\ldots,1$, $\lambda=d$ (dashed line). The dotted line
  illustrates the long-time decay $F_k(t)/\rho_0 \propto t^{-d/2}$.}
\end{figure}

The correlation functions can be computed by a direct numerical
integration of Eq.~\eqref{eqLTT3} with $M_k^0(t)$ given by
Eq.~\eqref{eqexplicit4}.  Figure~\ref{fig:barecorrel} shows the
typical behavior of $F_k(t)/\rho_0$ versus time obtained from this
numerical solution with $d=1$ and $d=3$, $k R=\pi/3$, for different
values of the relative disorder strength $\lambda$.  In this log-log
plots, the algebraic tail $F_k(t)/\rho_0 \sim \lambda [R^2/(2D_0
  t)]^{d/2} e^{-k^2 R^2/2}$ is clearly visible as a linear asymptote
at long times.

\clearpage

\section{Equilibrium dynamics: First-order renormalized theory} 
\setcounter{equation}{0}
\label{sec:FORT}

So far, a bare perturbation theory has been discussed, where the
corrections due to the disorder were expressed in terms of the bare
correlation and response functions.  We now consider renormalized
theories, where the bare correlation and response functions are
replaced with renormalized ones in a self-consistent manner and based
on the exact second order perturbation calculation.

Out of the bare perturbation expansion up to the second order (see
Apps.~\ref{app:G}-\ref{app:C}), one identifies Eqs.~\eqref{c12},
\eqref{d15}, and \eqref{e19}, as a set of first-order renormalized
dynamical equations, which should obey the FDR and reproduce the bare
theory (we set $t\ge t'=0$):
\begin{subequations} \la{eqFORT1}
\begin{align}
  (\partial_{t} + \Gam_k) G_k (t) & = \del(t) - \int_0^{t} ds
  \S_k(t-s) G_k (s), \la{eqFORT1a} \\
  (\partial_{t} + \Gam_k) \o{R}_k (t) & = \r_0 \Gam_k \del(t) -
  \int_0^{t} ds \S_k(t-s) \o{R}_k (s) + L_k(t), \la{eqFORT1b} \\
   (\partial_{t} + \Gam_k) F_k (t) & = - \int_0^{t} ds \S_k(t-s) F_k
  (s) + N_k(t). \la{eqFORT1c}
\end{align}
\end{subequations}
In the latter equation, the static time-persistent part of the density
correlation function has been absorbed into the connected density
correlation function, according to its definition
Eq.~\eqref{Fconnected}.  Clearly, these equations are structurally
similar to the bare Eqs.~\eqref{eqFOBT1a}, \eqref{eqFOBT1b}, and
\eqref{eqconsist6}.  At this stage, the explicit expressions for the
memory kernels $\S_k(t)$, $L_k(t)$, and $N_k(t)$, are left
unspecified.

The relations between these kernels should be constrained by the FDR,
Eq.~\eqref{tr16a}, also expressible as
\begin{equation} \la{eqFORT2}
  \o{R}_k(t) = -\theta(t)\partial_t F_k(t).
\end{equation} 
Taking the time derivative of the above and using
Eq.~\eqref{eqFORT1c}, one gets
\begin{equation} \la{eqFORT3} %
  (\partial_{t} + \Gam_k) \o{R}_k (t) = - \del(t) \partial_{t} F_k(0) -
  \int_0^{t} ds \S_k(t-s) \o{R}_k (s) + \rho_0 \S_k(t) - \partial_t
  N_k(t).
\end{equation}
Comparing with Eq.~\eqref{eqFORT1b}, one sees that the FDR demands the
two relations
\begin{subequations} \la{eqFORT4}
\begin{gather} 
  \partial_{t} F_k(0) = - \r_0 \Gam_k, \la{eqFORT4a} \\
  \partial_t N_k(t) = \rho_0 \S_k(t) - L_k(t). \la{eqFORT4b}
\end{gather}
\end{subequations}
Setting $t=0$ in Eq.~\eqref{eqFORT1c}, one gets the initial condition
$N_k(0)=0$ from Eq.~\eqref{eqFORT4a} and $F_k(0) = \r_0$, to be used
in Eq.~\eqref{eqFORT4b}.  Therefore, one should have
\begin{equation} \la{eqFORT5} %
  N_k(t) = \int_0^t ds [\r_0 \S_k(s) - L_k(s)].
\end{equation}

Using Eq.~\eqref{eqFORT4b} to eliminate $\S_k(t)$ in
Eq.~\eqref{eqFORT1c}, the latter becomes, after an integration by
parts,
\begin{equation} \la{eqFORT6} %
  (\partial_{t} + \Gam_k) F_k (t) = - \frac{1}{\r_0} \int_0^{t} ds
  N_k(t-s) \partial_s F_k (s) - \frac{1}{\r_0} \int_0^{t} ds L_k(t-s)
  F_k (s).
\end{equation}
This equation, a renormalized version of Eq.~\eqref{eqBMCT1}, is
clearly reminiscent of those that can be obtained with standard
projection-operator techniques in the memory-function formalism
\cite{macdohansen3ed}.  However, one can interestingly note that it
mixes two types of convolution integrals which are usually found to be
mutually exclusive and only converted into one another by making use
of special rearrangements \cite{CicHes87PA, GotSjo87ZPB, Kaw95PA}.

Another possibility to eliminate $\S_k(t)$ is through mere Laplace
transforms of Eqs.~\eqref{eqFORT1}. One then obtains nonlinear
relations (once the kernels are specified) expressing the physical
response and correlation functions in terms of the noise-response
function, as
\begin{align} 
  \o{R}_k(t) & = \r_0 \Gam_k G_k(t) + \int_0^t ds L_k(t-s) G_k(s),
  \la{eqFORT7} \\
  F_k(t) & = \r_0 G_k(t) + \int_0^t ds N_k(t-s) G_k(s). \la{eqFORT8}
\end{align}
These expressions will be extremely useful in the following to perform
first-order consistent substitutions, i.e., replacements of one
function with another that entail corrections strictly beyond the
first order.

We might now close the set of dynamical equations with explicit
expressions for the first-order renormalized kernels, which are
self-consistently determined from a second-order bare perturbation
calculation.

\subsection{Native first-order renormalized theory.}
We first consider the first-order renormalized theory (FORT) that
derives in the most literal way from the second-order bare theory.
For this reason, we choose to term it native.

As seen in Apps.~\ref{app:G}-\ref{app:C}, one gets [Eqs.~\eqref{c13}
  and \eqref{d16}]
\begin{align} 
  \S_k(t) & = \l D_0^2 \int_{\bf q} {\b q} \cdot {\b p} [{\b k} \cdot
    {\b q} \Phi_q ] G_p(t), \la{eqFORT9} \\
  L_k(t) & = \l \r_0 D_0^2 \int_{\bf q} {\b k} \cdot {\b p} [{\b k}
    \cdot {\b q} \Phi_q ] G_p(t). \la{eqFORT10}
\end{align}
As for the remaining kernel $N_k(t)$, it is in principle given by
Eq.~\eqref{e20}.  However, as pointed out there, it is very likely
that this expression does not fully comply with the requirements of a
\emph{bona fide} equilibrium dynamics.  Yet, a possible workaround is
to force consistency with the FDR, through the use of
Eq.~\eqref{eqFORT4b}.  One then gets
\begin{equation} \la{eqFORT11} %
  \partial_t N_k(t) = - \l D_0 \int_{\bf q} [{\b k} \cdot {\b q}
    \Phi_q ] [\r_0 D_0 p^2 G_p(t)],
\end{equation}
hence
\begin{equation} \la{eqFORT12} %
  N_k(t) = - \l D_0 \int_0^t ds \int_{\bf q} [{\b k} \cdot {\b q}
    \Phi_q ] [\r_0 D_0 p^2 G_p(s)].
\end{equation}
In App.~\ref{app:C}, we check the suitability of this step, by showing
that, thanks to Eqs.~\eqref{eqFORT7}, \eqref{eqFORT8}, and
\eqref{eqFORT1a}, Eq.~\eqref{e20} can indeed be rewritten within its
order of validity in $\l$, such that it agrees with
Eq.~\eqref{eqFORT12} to first order in $\l$.  Note that, in the
present scheme, the explicit expression of $N_k(t)$ is actually not
needed for the computation of the three functions of interest.
Indeed, with the above form of $\S_k(t)$, the dynamical equation for
the noise-response function $G_k(t)$ is self-closed.  Its solution can
be fed into the dynamical equation for the physical response function,
Eq.~\eqref{eqFORT1b}, or equivalently Eq.~\eqref{eqFORT7}, to obtain
$\o{R}_k(t)$, from which $F_k(t)$ is retrieved by integration of the
FDR, Eq.~\eqref{eqFORT2}.  Self-consistency implies that this solution
for $F_k(t)$ be the same as that from Eqs.~\eqref{eqFORT1c},
\eqref{eqFORT6}, or \eqref{eqFORT8}.

Unfortunately, this theory as it stands does not appear to bring one
very far.  Indeed, our numerical attempts at computing $G_k(t)$, which
is the required first step, faced instabilities that seem to prevent
the application of the theory beyond rather modest disorder strengths
[with a Gaussian random field covariance, Eq.~\eqref{eqexplicit1},
  spurious divergences occur for $\lambda/d > 0.272$ in $d=1$ and
  $\lambda/d > 0.345$ in $d=3$, i.e., significantly below the
  threshold $\lambda/d=1$ beyond which the FOBT produces blatantly
  unphysical results].  Note that these calculations were based on
computing the integrated response function $H_k(t) = \int_0^t ds
G_k(s)$ as an intermediate \cite{KimLat01EL}, whose evolution equation
obtained from Eqs.~\eqref{eqFORT1a} and \eqref{eqFORT9} reads
\begin{equation}
  (\partial_{t} + \Gam_k) H_k (t) = 1 - \int_0^{t} ds S_k(t-s)
  \partial_s H_k (s), \qquad
  S_k(t) = \l D_0^2 \int_{\bf q} {\b q} \cdot {\b p} [{\b k} \cdot {\b
      q} \Phi_q ] H_p(t).
\end{equation}
This is exactly the type of nonlinear integro-differential equation
met within the MCT, for which a well-established and efficient
iterative numerical solution scheme has been developed long ago
\cite{FucGotHofLat91JPCM}.  It usually shows remarkable stability,
provided the underlying equations are themselves stable.  Therefore,
this suggests that the instabilities are intrinsic to the above
renormalized equations, which in particular fail to guarantee that the
kernels $\S_k(t)$ and $S_k(t)$ are nonnegative functions of time,
while this is the case for overdamped dynamics with the standard MCT
kernels.

\subsection{Modified first-order renormalized theory.}
In order to try and overcome these difficulties, one might exploit the
freedom offered by the first-order consistent substitutions to
generate variants of the theory, at the cost of an increased degree of
empiricism in its derivation.  Since $\S_k(t)$, $L_k(t)$, hence
$\partial_t N_k(t)$ via Eq.~\eqref{eqFORT4b}, naturally acquire the
character of response functions within the native FORT, we focused on
the possibilities provided by Eq.~\eqref{eqFORT7} to replace $G_p(t)$
with $\o{R}_p(t)/(\r_0\Gamma_p)$ in Eqs.~\eqref{eqFORT9},
\eqref{eqFORT10}, or \eqref{eqFORT11}.  By separately making one or
the other choice for two kernels, the third one being fixed by
Eq.~\eqref{eqFORT4b}, one obtains eight FDR-consistent theories in
total, including the native one above entirely based on $G_p(t)$.

With respect to the criteria of consistency with the FDR and with the
FOBT, these eight theories are all equally possible and valid by
construction.  Therefore, if one of them is to be favoured, this has
to be based on arguments of a different nature.  Since we identified
difficulties with the native theory through numerical considerations,
we shall pursue this line of reasoning here.  We already know that the
instabilities of the native theory will be present in two other
variants of the FORT, for their $\S_k(t)$ is also given by
Eq.~\eqref{eqFORT9}.

After trying to numerically solve the dynamical equations for the
eight variants of the theory, we find that one of them clearly stands
out.  Indeed, for some relevant choices of parameters, it appears
unique in its ability to deliver physically acceptable numerical
results.  This is particularly the case in the regime of sizeable
disorder strengths, corresponding to $\lambda/d>1$.  The theory in
question, to which we shall refer as the modified FORT, is the one
entirely based on $\o{R}_p(t)$, i.e., with
\begin{align} 
  \S_k(t) & = \l \frac{D_0}{\rho_0} \int_{\bf q} \frac{{\b q} \cdot
    {\b p}}{p^2} [{\b k} \cdot {\b q} \Phi_q ] \o{R}_p(t),
  \la{eqFORT14} \\
  L_k(t) & = \l D_0 \int_{\bf q} \frac{{\b k} \cdot {\b p}}{p^2} [{\b
      k} \cdot {\b q} \Phi_q ] \o{R}_p(t), \la{eqFORT15} \\
  \partial_t N_k(t) & = - \l D_0 \int_{\bf q} [{\b k} \cdot {\b q}
    \Phi_q ] \o{R}_p(t), \la{eqFORT16}
\end{align}
hence
\begin{equation} \la{eqFORT17}
  N_k(t) = \l D_0\int_{\bf q} [{\b k} \cdot {\b q} \Phi_q ] F_p(t),
\end{equation}
where we used the FDR and $\int_{\bf q} {\b k} \cdot {\b q} \Phi_q =
0$ by isotropy.  Note that these expressions achieve consistency with
the FDR in a most natural way, since Eq.~\eqref{eqFORT4b} merely
appears as a trivial corollary of Eq.~\eqref{eqFORT2}.

Beyond the numerical arguments, some aspects of the theory discussed
previously might actually be seen as further hints in favor of these
equations.  For instance, in our physical interpretation of $L_k(t)$
at the bare level [see Eqs.~\eqref{eqFOBT4}-\eqref{eqFOBT6}], the
kernel is proposed to initially involve a composite response field, as
precisely does $\o{R}_k(t)$.  Also, the straightforward appearance of
the combination $\r_0 \Gamma_p G_p(s)$ in Eq.~\eqref{eqFORT11} of the
native theory suggests that a substitution by $\o{R}_p(t)$ might be in
order, as we repeatedly assumed at the bare level [see the transition
  from Eqs.~\eqref{eq8.7} to Eqs.~\eqref{eq8.8}].  On the other hand,
the second-order result of App.~\ref{app:G} does not provide one with
any obvious reason to favour Eq.~\eqref{eqFORT14} over
Eq.~\eqref{eqFORT9}, since both expressions are seen to remain
approximate at this order.

At the level of the response functions, it is now the dynamical
equation for $\o{R}_k(t)$, Eq.~\eqref{eqFORT1b}, which is self-closed.
However, from a physical point of view, the closed coupled set
consisting of Eqs.~\eqref{eqFORT1a} and \eqref{eqFORT7} looks more
telling, as it shows a mixed feedback scheme that might be pictorial
of dynamics with multiplicative noise.  Indeed, on the one hand,
Eq.~\eqref{eqFORT7} formally represents the density response function
as a mere byproduct of the noise-response function, in line with the
fact that fluctuations and dynamics do fundamentally come to the
system precisely through thermal noise.  But, on the other hand, the
couplings and memory effects represented by $\S_k(t)$ and $L_k(t)$ are
ruled by the density response function itself, as a reflection of the
density dependence of the multiplicative thermal noise.

Formally, it is still possible to close Eq.~\eqref{eqFORT1a} and have
the modified theory rest upon the mere determination of $G_k(t)$, as
does the native one.  Indeed, Eqs.~\eqref{eqFORT7} and
\eqref{eqFORT15} can be recursively used to express $\o{R}_k(t)$ as an
infinite sum of integrals of all orders in the disorder strength and
involving $G_k(t)$ only.  A similar series expansion can be derived
for $F_k(t)$, based on Eqs.~\eqref{eqFORT8} and \eqref{eqFORT17}.
When injected into Eqs.~\eqref{eqFORT14}-\eqref{eqFORT17}, these
expressions characterize the present approach as some kind of
resummation scheme beyond the native FORT.

Thanks to the FDR, the dynamical equations for the density correlation
function, Eqs.~\eqref{eqFORT1c} and \eqref{eqFORT6}, are self-closed
as well.  In particular, the latter can be usefully written as
\begin{equation} \la{eqFORT18} 
  ( \partial_{t} + \Gam_k ) F_k (t) = - \frac{1}{\r_0} \int_0^{t} ds
  N_k(t-s) \partial_s F_k (s) - \frac{1}{\r_0} \int_0^{t} ds
  \partial_{t-s} \Lambda_k(t-s) F_k (s) ,
\end{equation}
where $N_k(t)$ is given by Eq.~\eqref{eqFORT17} and $\Lambda_k(t)$
follows from Eq.~\eqref{eqFORT15} and the FDR, Eq.~\eqref{eqFORT2}, as
\begin{equation} \la{eqFORT19} 
  \Lambda_k(t) = - \l D_0 \int_{\bf q} \frac{{\b k} \cdot {\b p}}{p^2}
         [{\b k} \cdot {\b q} \Phi_q ] F_p(t).
\end{equation}
For definiteness, we recall the initial condition $F_k (0) = \r_0$.
Interestingly, these equations are clearly distinct from those
obtained within the MCT, Eqs.~\eqref{MCT}, but they belong to the same
class of self-consistent nonlinear problems and can be analytically
studied \cite{LesHouches,FraGot94JPCM,GotzeBook} and numerically
solved \cite{FucGotHofLat91JPCM} by the same means.

Therefore, we might now discuss the main features of their solutions,
considering again the case of a Gaussian random-field covariance,
Eq.~\eqref{eqexplicit1}, for the purpose of illustration.

\subsection{Numerical solution of the modified first-order
  renormalized theory.}

\begin{figure}
\centering
\includegraphics{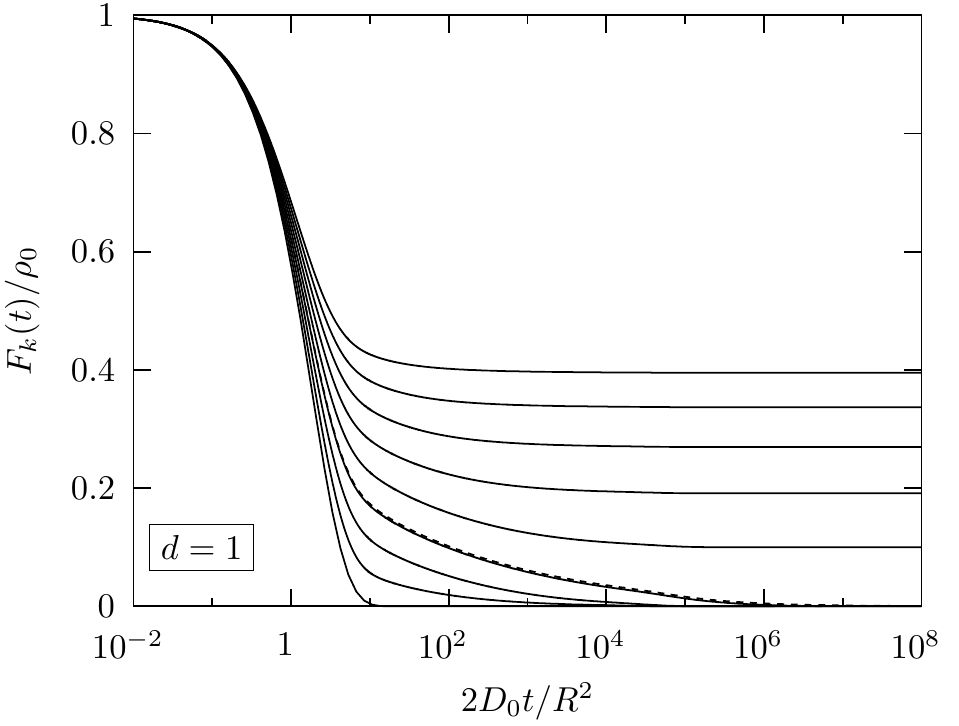}

\includegraphics{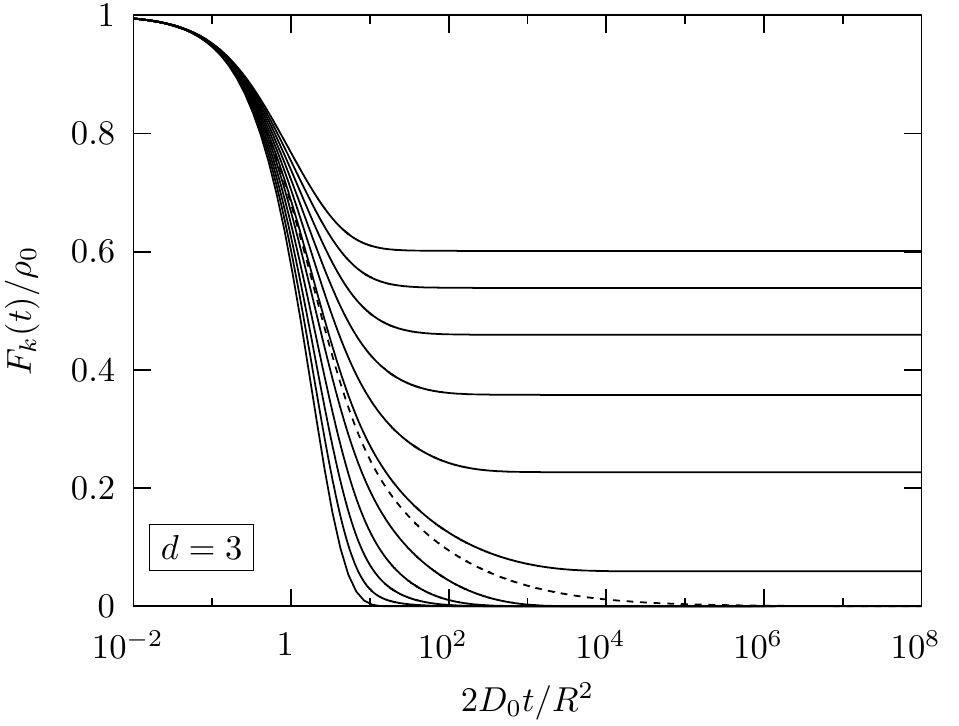}
\caption{\label{fig:equicorrel} Time evolution of the connected
  density correlation function in a noninteracting Brownian gas
  immersed in a Gaussian random field with Gaussian covariance in
  space dimensions $d=1$ (top) and $d=3$ (bottom), according to the
  modified first-order renormalized theory.  The wavevector is $k
  R=\pi/3$. From left to right, bottom to top: $\lambda=0$, $0.25$,
  $0.5$, \dots, $1.75$, $2$, for $d=1$; $\lambda=0$, $0.5$, $1$,
  \dots, $4.5$, $5$, for $d=3$.  Ergodicity is broken for $\lambda$
  larger than $\lambda_\text{c}(d=1)=0.76359$ and
  $\lambda_\text{c}(d=3)=2.34686$.  The corresponding critical density
  correlation functions are reported with dashed lines.}
\end{figure}

\begin{figure}
\centering
\includegraphics{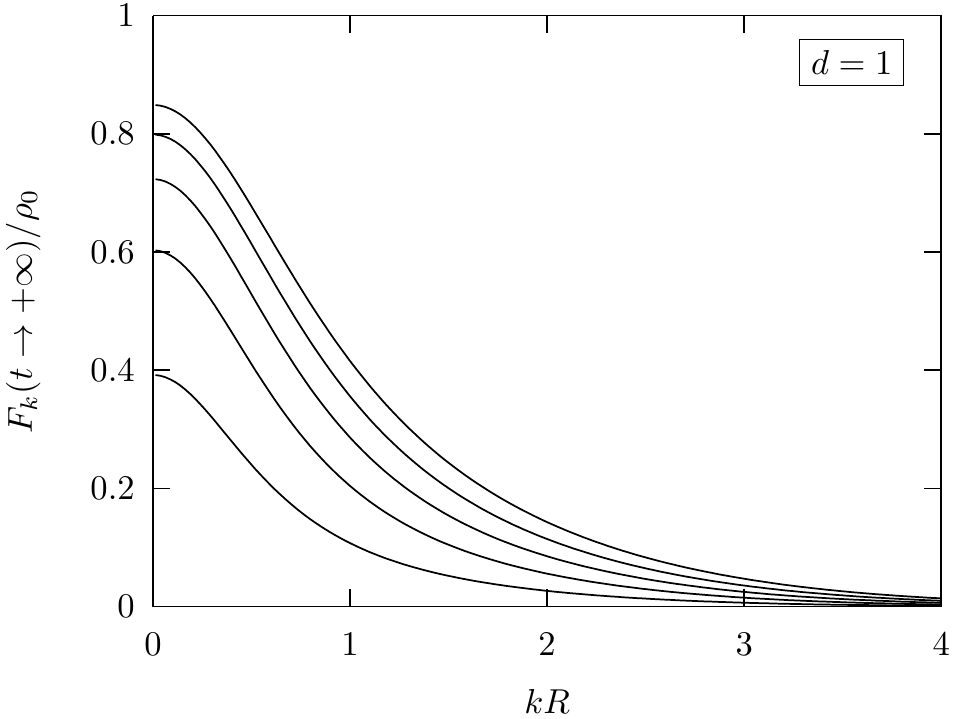}

\includegraphics{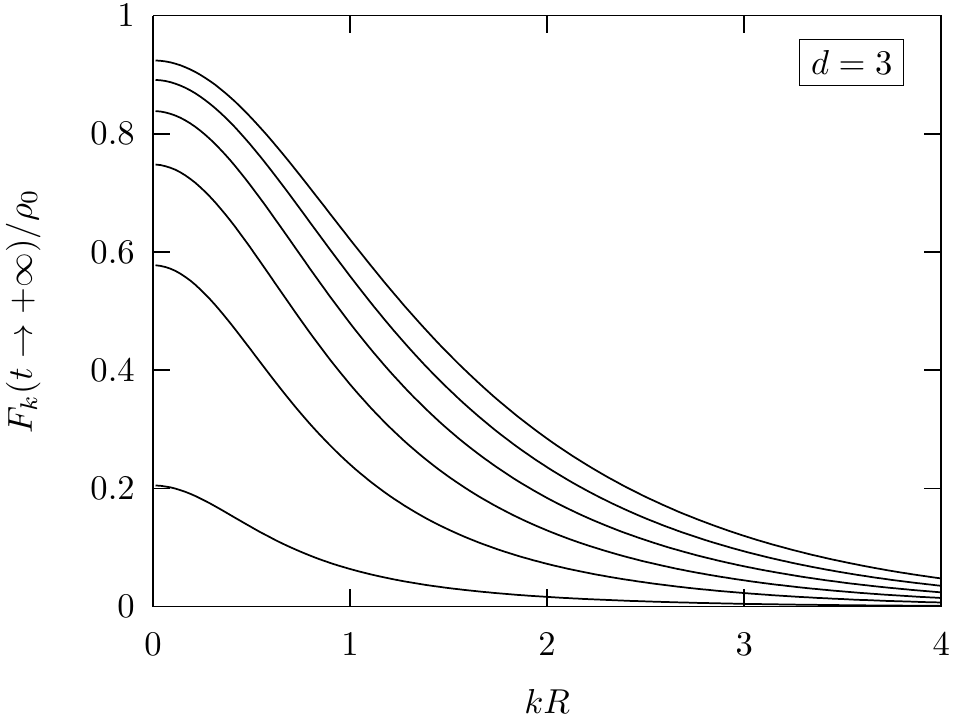}
\caption{\label{fig:nonergparam} Wavevector dependence of the
  nonergodicity parameter of a noninteracting Brownian gas plunged in
  a Gaussian random field with Gaussian covariance in space dimensions
  $d=1$ (top) and $d=3$ (bottom), according to the modified
  first-order renormalized theory.  From bottom to top: $\lambda=1$,
  $1.25$, $1.5$, $1.75$, $2$, for $d=1$; $\lambda=2.5$, $3$, $3.5$,
  $4$, $4.5$, $5$, for $d=3$.}
\end{figure}

The evolution of the correlation function $F_k(t)/\r_0$ with
increasing disorder strength $\lambda$ is displayed in
Fig.~\ref{fig:equicorrel} for a representative wavevector $k R=\pi/3$
in space dimensions $d=1$ and $d=3$.  First, as one would expect, the
dynamics simply slows down as $\lambda$ increases, and a long-time
relaxation tail gradually develops.  Then, at a threshold
$\lambda_\text{c}(d)$, obeying $\lambda_\text{c}(d)<d$ (the importance
of this inequality will be manifest later), the dynamics becomes
nonergodic, i.e., a time-persistent plateau starts to continuously
grow from zero with increasing positive $\lambda-\lambda_\text{c}(d)$,
reflecting a partial arrest of the relaxation of the density
fluctuations.  The so-called nonergodicity parameter
$F_k(t\to+\infty)/\r_0$, corresponding to the height of this plateau,
is solution of the nonlinear equation
\begin{equation} \la{eqFORT20} 
\frac{F_k(t\to+\infty)}{\r_0} = \frac{N_k(t\to+\infty)}{\r_0 \Gamma_k +
  N_k(t\to+\infty) + \Lambda_k(t\to+\infty) - \Lambda_k(0)},
\end{equation}
where $N_k(t\to+\infty)$ and $\Lambda_k(t\to+\infty)$ are linear
functionals of $F_k(t\to+\infty)$, as prescribed by
Eqs.~\eqref{eqFORT17} and \eqref{eqFORT19}.  The wavevector dependence
of $F_k(t\to+\infty)/\r_0$ is shown in Fig.~\ref{fig:nonergparam} for
the values of the disorder strength corresponding to nonergodic
states in Fig.~\ref{fig:equicorrel}.

\begin{figure}
\centering
\includegraphics{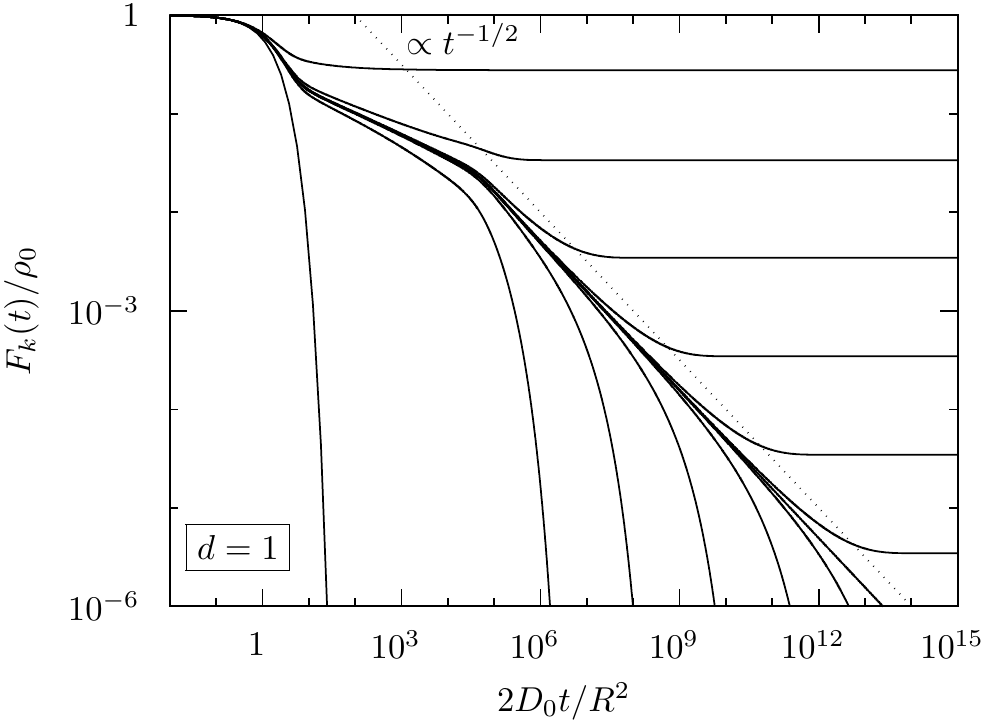}

\includegraphics{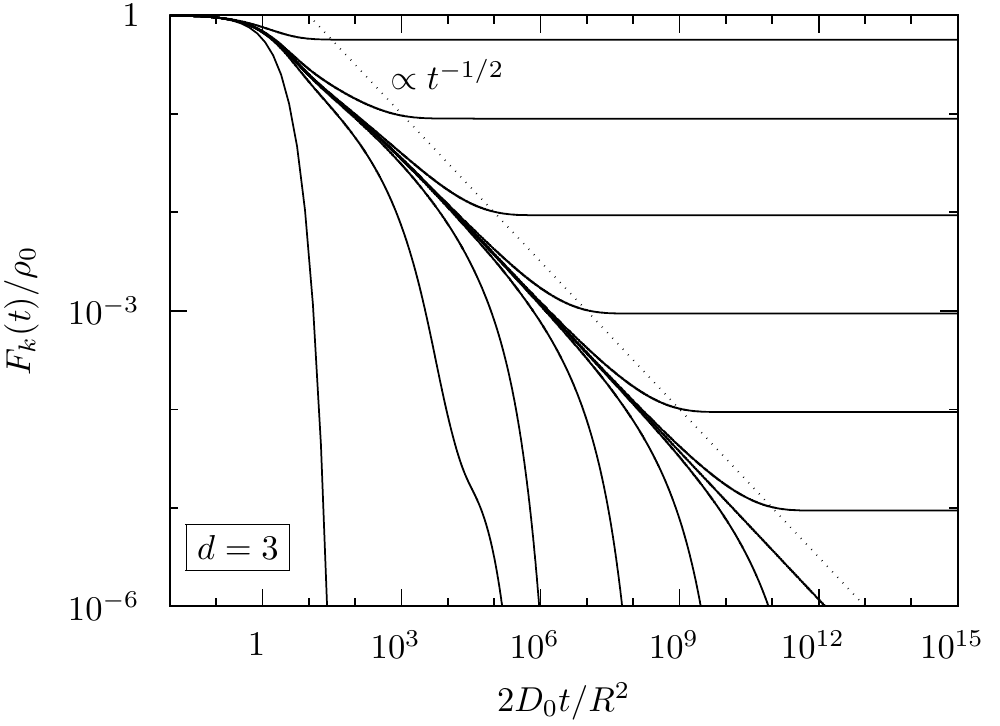}
\caption{\label{fig:critcorrel} Time evolution of the connected
  density correlation function in a noninteracting Brownian gas in a
  Gaussian random field with Gaussian covariance in space dimensions
  $d=1$ (top) and $d=3$ (bottom), according to the modified
  first-order renormalized theory.  The wavevector is $k
  R=\pi/3$. From left to right, bottom to top: $\lambda=0$,
  $0.9\lambda_\text{c}$, $0.99\lambda_\text{c}$,
  $0.999\lambda_\text{c}$, $0.9999\lambda_\text{c}$,
  $0.99999\lambda_\text{c}$, $\lambda_\text{c}$,
  $1.00001\lambda_\text{c}$, $1.0001\lambda_\text{c}$,
  $1.001\lambda_\text{c}$, $1.01\lambda_\text{c}$,
  $1.1\lambda_\text{c}$, $2\lambda_\text{c}$, with
  $\lambda_\text{c}(d=1)=0.76359$ and $\lambda_\text{c}(d=3)=2.34686$.
  The dotted line illustrates the long-time critical decay
  $F_k(t)/\rho_0 \propto t^{-1/2}$.}
\end{figure}

The details of the critical dynamics near the threshold are
illustrated by Fig.~\ref{fig:critcorrel}.  The long-time relaxation
tail is seen to be algebraic, $F_k(t)/\r_0 \propto t^{-1/2}$,
independently of the space dimension.  It lasts longer and longer as
$\lambda_\text{c}(d)$ is approached from below, and gradually recedes,
giving way to the time-persistent plateau, as $\lambda_\text{c}(d)$ is
left from above.  These evolutions are symmetric on both sides of
$\lambda_\text{c}(d)$, with a diverging characteristic timescale
$\propto [\lambda - \lambda_\text{c}(d)]^{-2}$ .  In the partially
arrested state, the nonergodicity parameter grows $\propto [\lambda -
  \lambda_\text{c}(d)]$ to leading order.

In most respects, this scenario is the same as the one found within
the MCT \cite{KonKra17SM}.  This similarity can be traced back to the
linearity of the kernels with the density correlation functions, which
generically enforces continuous ergodicity-breaking transitions, if
any \cite{LesHouches, GotzeBook}.  Such a linearity is an expected
generic feature of MCT-like approaches to fluids in random fields,
which has been found in all previous studies, either strictly
\cite{Got78SSC, Got79JPC, Got81PMB, GotLeuYip81aPRA, GotLeuYip81bPRA,
  Leu83aPRA} or to leading order in the strong disorder regime
\cite{Kra05PRL, Kra07PRE, Kra09PRE, KonKra17SM}.  There is however one
important difference with regard to the behavior of the nonergodicity
parameter.  Indeed, within the MCT, the evolution of the latter with
increasing disorder strength mainly consists of the continuous
broadening of a low-wavevector peak with maximum
$F_0(t\to+\infty)/\r_0=1$, which appears with a vanishing width at the
ergodicity-breaking transition (this behavior is illustrated for the
case of a fluid in a random porous solid in Refs.~\cite{Kra09PRE} and
\cite{SchHofFraVoi11JPCM}).  This implies the existence of a
localization length in the nonergodic phase, which diverges as the
transition is approached from above.  There is no such thing in the
present theory, as readily seen in Fig.~\ref{fig:nonergparam}.  This
difference can be traced back to the contrasting low-wavevector
behaviors of the kernels in the two theories.  Here, both $N_k(t)$ and
$\Lambda_k(t)$ are $O(k^2)$, so that $F_k(t\to+\infty)/\r_0$ in
Eq.~\eqref{eqFORT20} does not have to go to one as $k\to 0$, while it
does have to in the MCT, where ${\cal M}_k(t)$ is $O(k^0)$ [see
  Eq.~\eqref{eqMCTb}] and $F_k(t\to+\infty)/\r_0 = {\cal
  M}_k(t\to+\infty) / [\Gamma_k + {\cal M}_k(t\to+\infty)]$.

\begin{figure}
\centering
\includegraphics{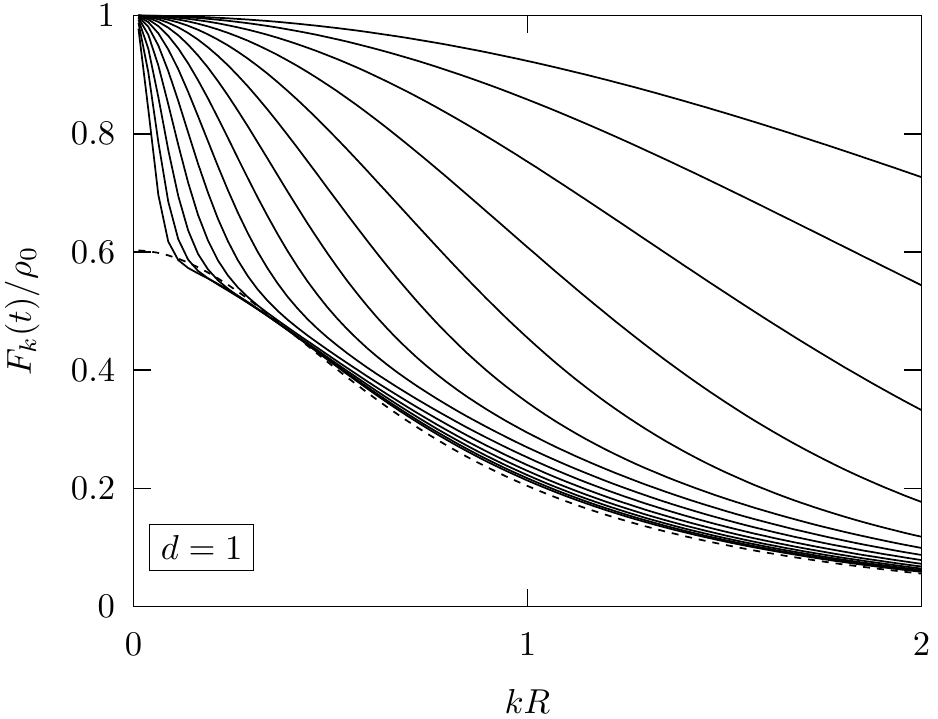}

\includegraphics{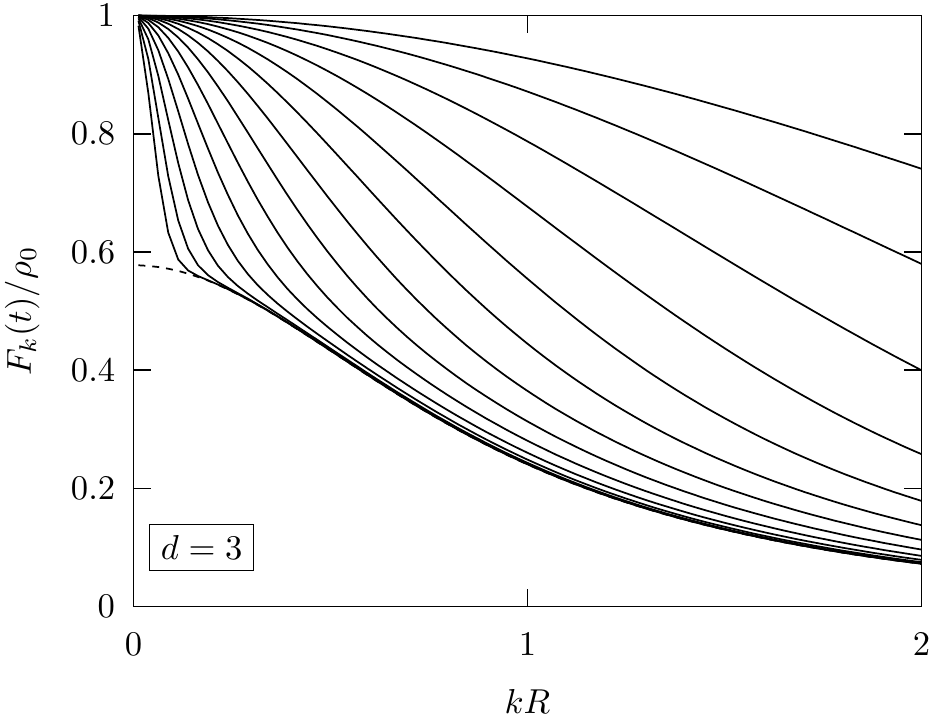}
\caption{\label{fig:correlvsq} Wavevector dependence of the connected
  density correlation function at fixed times in a noninteracting
  Brownian gas exposed to a Gaussian random field with Gaussian
  covariance in space dimensions $d=1$ (top) and $d=3$ (bottom),
  according to the modified first-order renormalized theory.  For
  $d=1$, $\lambda=1.25$; for $d=3$, $\lambda=3$.  In both cases,
  $\lambda > \lambda_\text{c}(d)$ and the system is nonergodic.  From
  top to bottom: $2 D_0 t / R^2 = 2^n\times 10^{-8}$,
  $n=24,\ldots,37$.  The nonergodicity parameter is shown as a dashed
  line.}
\end{figure}

The absence of a localized state in the nonergodic phase is readily
seen in the full wavevector dependence of the dynamics, as reported in
Fig.~\ref{fig:correlvsq}.  Indeed, as the density correlation
functions relax toward their infinite-time limits, a peak forms on top
of the nonergodicity parameter curve at low wavevectors, which becomes
narrower and narrower with time.  From Eq.~\eqref{eqMSD1}, it is clear
that this peak relates to the diffusional properties of the fluid and
that its vanishing width with increasing time implies a diverging
mean-squared displacement (MSD), hence a delocalized state.  In
passing, note that an occasional slight inaccuracy of the theory can
be spotted in the top panel of Fig.~\ref{fig:correlvsq}.  Indeed, at
low wavevectors (below $k R\simeq 0.4$), the nonergodicity parameter
is reached from below, meaning a slightly nonmonotonic behavior of the
density correlation function.  Quantitatively, the phenomenon is very
small, but, in principle, it violates the property that
autocorrelation functions be completely monotone functions of time for
overdamped dynamics.

\begin{figure}
\centering
\includegraphics{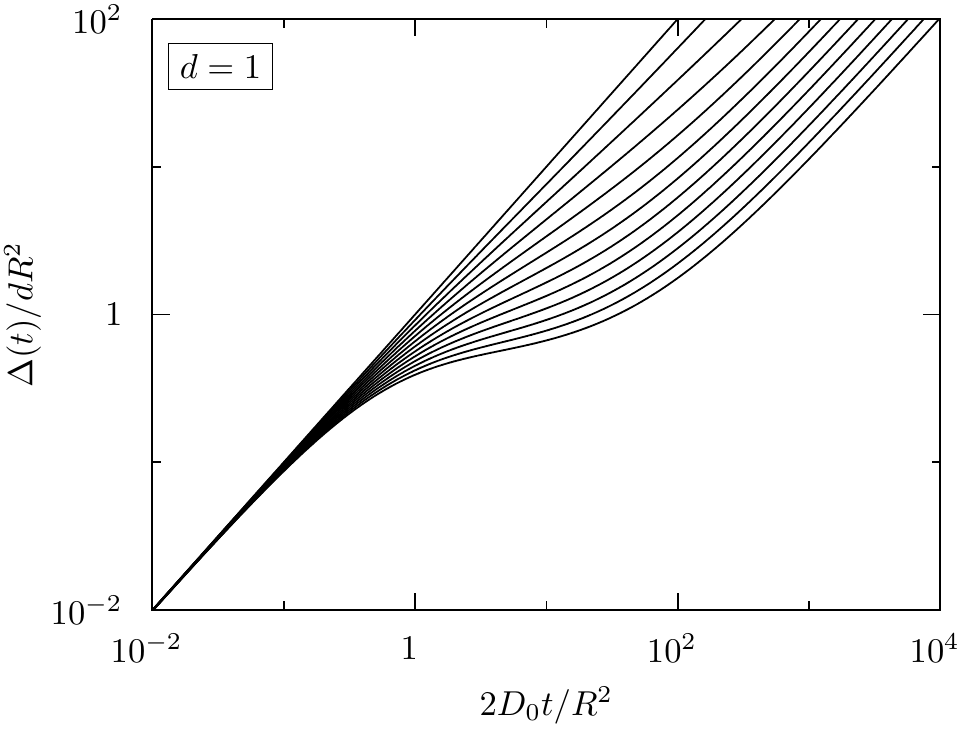}

\includegraphics{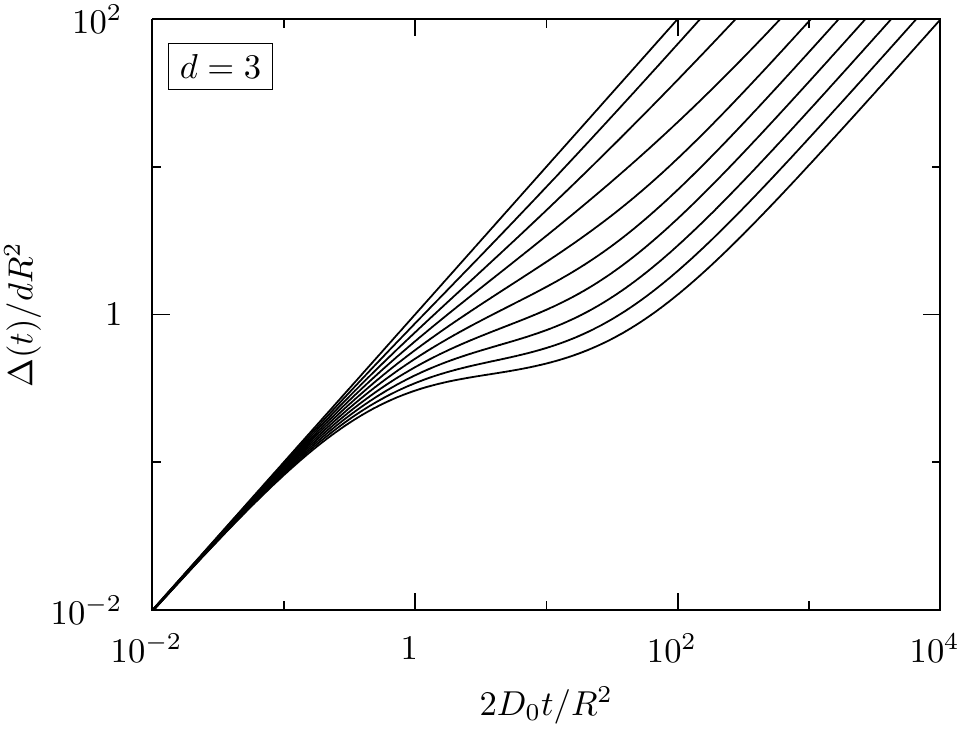}
\caption{\label{fig:renormMSD} Time evolution of the mean-squared
  displacement in a noninteracting Brownian gas in a Gaussian random
  field with Gaussian covariance in space dimensions $d=1$ (top) and
  $d=3$ (bottom), according to the modified first-order renormalized
  theory.  From left to right, top to bottom: $\lambda=0$, $0.5$, $1$,
  \dots, $5.5$, $6$, for $d=1$; $\lambda=0$, $1$, $2$, \dots, $8$,
  $9$, for $d=3$.}
\end{figure}

The above reasoning is confirmed by a direct computation of the MSD.
Using the low-$k$ expansion Eq.~\eqref{eqMSD1} in Eq.~\eqref{eqFORT6},
knowing that $\lim_{k\to 0} N_k(t)=0$ and $L_k(t)=O(k^2)$, one
generically obtains
\begin{equation} \la{eqFORT21} %
  \frac{\Delta(t)}{2 d D_0} = t + \int_0^t ds \int_0^s du \lim_{k\to0}
  \frac{L_k(u)}{\r_0 \Gamma_k},
\end{equation} 
which again connects the low-wavevector limit of $L_k(t)$ to the force
autocorrelation function through Eq.~\eqref{eqMSD3bis}.  Then, within
the modified FORT, where $L_k(t)=\partial_{t} \Lambda_k(t)$ and
$\Lambda_k(t)$ is given by Eq.~\eqref{eqFORT19}, this can be rewritten
as
\begin{equation} \la{eqFORT22} %
  \frac{\Delta(t)}{2 d D_0} = \left( 1 - \frac{\l}{d} \right) t +
  \int_0^t ds \, m(s),
\end{equation} 
with 
\begin{equation} \la{eqFORT23} %
  m(t) = \frac{\l}{\r_0 d} \int_{\bf q} \Phi_q F_q(t) = \frac{\l}{\r_0
    d} \int_{\bf r} \Phi(r) F(r,t),
\end{equation} 
which is an obvious renormalized version of Eq.~\eqref{eqMSD6}.  The
corresponding results for the influence of the relative disorder
strength on the time dependence of the MSD are shown in
Fig.~\ref{fig:renormMSD}.  Remarkably, it is found that a normal
diffusive behavior is reached at long times for all disorder
strengths, even those leading to nonergodic states.  This feature is
definitely at variance with the MCT predictions, where the
ergodicity-breaking transition is also a diffusion-localization
transition \cite{KonKra17SM}, and in complete agreement with the known
rigorous results \cite{MasFerGolWic89JSP}.

\begin{figure}
\centering
\includegraphics{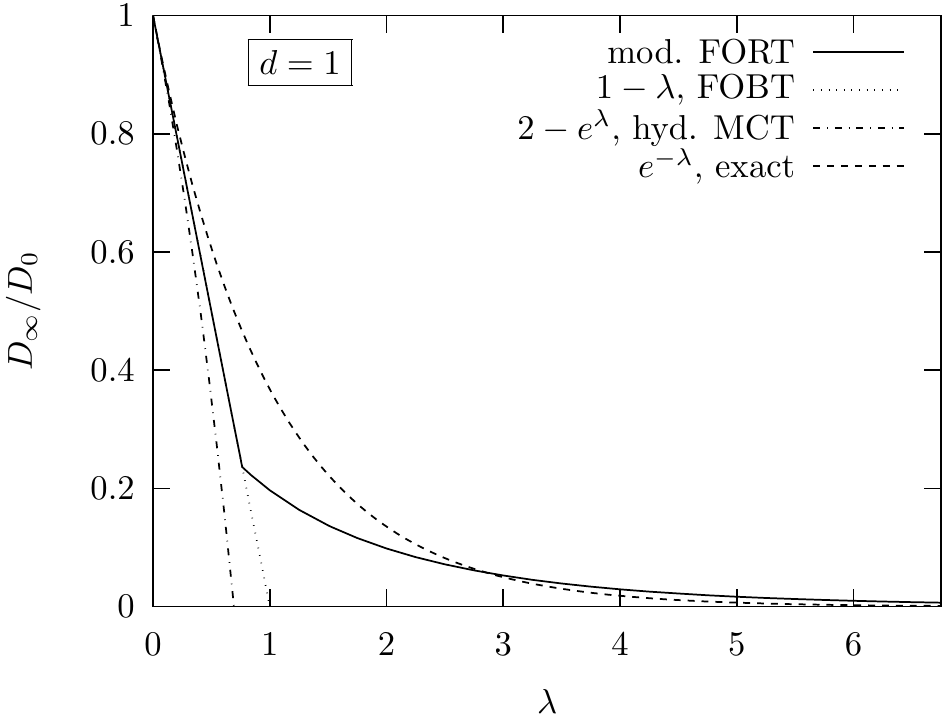}

\includegraphics{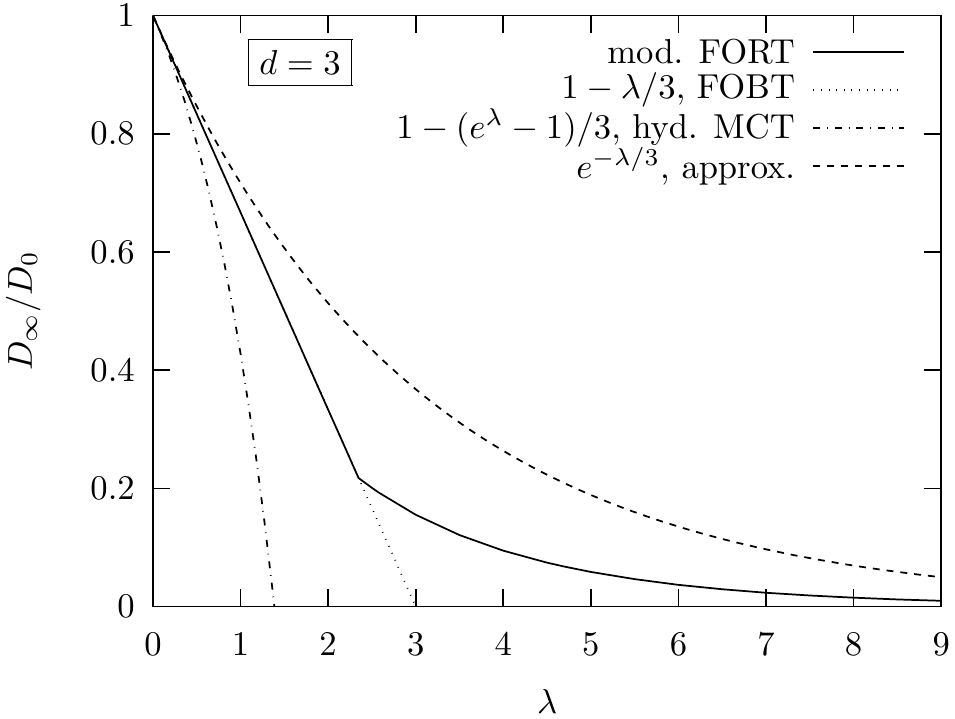}
\caption{\label{fig:diffcoeff} Disorder-strength dependence of the
  long-time diffusion coefficient of a noninteracting Brownian gas
  plunged in a Gaussian random field with Gaussian covariance in space
  dimensions $d=1$ (top) and $d=3$ (bottom). Continuous line: modified
  first-order renormalized theory; the ergodicity-breaking transition
  at $\lambda_\text{c}(d)$ is signalled by a corner singularity where
  the slope of $D_\infty/D_0$ is discontinuous.  Dotted line:
  first-order bare theory, $D_\infty/D_0=1-\lambda/d$. Dashed line:
  $D_\infty/D_0=e^{-\lambda/d}$; this expression is exact in $d=1$ and
  a good approximation in $d=3$. Dash-dotted line: mode-coupling
  theory with hydrodynamic approximation,
  $D_\infty/D_0=1-(e^\lambda-1)/d$.}
\end{figure}

Further insight into this finding can be gained by considering the
time-dependent diffusion coefficient $D(t)$, given by
\begin{equation} \la{eqFORT24} %
  \frac{D(t)}{D_0} = \frac{\dot\Delta(t)}{2d D_0} = 1 - \frac{\l}{d} +
  m(t),
\end{equation}
or, more specifically, its long-time limit $D_\infty =
\lim_{t\to+\infty} D(t)$.  It is plotted in Fig.~\ref{fig:diffcoeff}
for $d=1$ and $d=3$.  For $\lambda<\lambda_\text{c}(d)$, the system is
ergodic, $m(t)$ vanishes at long times because $F_q(t)$ does for all
$q$, and Eq.~\eqref{eqMSD9} from the FOBT is recovered.  On the other
hand, for $\lambda>\lambda_\text{c}(d)$, ergodicity is broken and one
gets
\begin{equation} \la{eqFORT25} %
  \frac{D_\infty}{D_0} = 1 - \frac{\l}{d} + \frac{\l}{\r_0 d}
  \int_{\bf q} \Phi_q F_q(t\to+\infty) > 1 - \frac{\l}{d}.
\end{equation}
As shown by the numerical results, the additional nonergodic
contribution strongly restricts the decrease of $D_\infty$ with the
disorder strength compared to the ergodic regime.  The breakdown of
Eq.~\eqref{eqMSD9} at $\lambda=d$ is therefore avoided, so that
$D_\infty$ remains strictly positive.  Note that this obviously
requires the condition $\lambda_\text{c}(d)<d$.  Unfortunately, this
mechanism generates a corner singularity in $D_\infty$ at
$\lambda_\text{c}(d)$, as a result of the leading linear growth of
$F_q(t\to+\infty)$ above $\lambda_\text{c}(d)$.  This is clearly a
spurious feature of the present theory, as no such corner exists in
the known exact results for $D_\infty$ in $d=1$ and $d=2$
\cite{DeaDruHor07JSM} and there is no obvious reason why this should
be different in other space dimensions.  As for the aspect of
quantitative accuracy, comparison with the law $D_\infty/D_0 =
e^{-\lambda/d}$, which is known to be exact in $d=1$ and a good
approximation in $d=3$ \cite{DeaDruHor07JSM}, immediately shows that
there is room for improvement.  For completeness, we also report an
analytic result from the MCT with an additional hydrodynamic
approximation \cite{Leu83aPRA,Kra09PRE}, $D_\infty/D_0 =
1-(e^\lambda-1)/d$, which shows its predicted vanishing of the
diffusion coefficient.  Note that this expression is based on an exact
treatment of the static correlations.  If the structure factors are
truncated to linear order in $\lambda$, this version of the MCT simply
reproduces $D_\infty/D_0$ from the FOBT in the ergodic phase.

\subsection{Relation between the present theory and the MCT.} 
We close this section by considering how the present FORT can be
related to the MCT.  Indeed, as mentioned in introduction, a major
motivation for the development of field-theoretic approaches to
particle dynamics came from the search of an improved derivation of
the MCT, with better controlled approximations.  It is thus
interesting to see where the present results stand from this
perspective.

With Eqs.~\eqref{eqFORT6} or \eqref{eqFORT18}, which are evocative of
the memory-function formalism, and the closures Eqs.~\eqref{eqFORT15},
\eqref{eqFORT17}, and \eqref{eqFORT19}, the modified FORT manifestly
appears as a FDR-consistent MCT-like theory, in the sense that it
relies on closed self-consistent dynamical equations for the density
correlation function only.

In order to actually get the MCT equations from the present framework,
one needs additional manipulations.  In particular, the derivation of
the bare MCT from the FOBT in the previous section shows that the use
of first-order consistent substitutions has to be pushed further.
Thus, keeping Eq.~\eqref{eqFORT17} for $N_k(t)$, Eq.~\eqref{eqFORT8}
is invoked to set
\begin{equation} \la{eqFORT26}
  L_k(t) = \l D_0^2 \int_{\bf q} {\b k} \cdot {\b p} [{\b k} \cdot {\b
      q} \Phi_q ] F_p(t),
\end{equation}
instead of Eqs.~\eqref{eqFORT10} or \eqref{eqFORT15}.  One then has
the equality
\begin{equation} \la{eqFORT27} 
  \frac{1}{\r_0} \left[ N_k(t) - \frac{L_k(t)}{\Gam_k} \right] =
  \frac{\l D_0}{\r_0} \int_{\bf q} (\h{\b k} \cdot {\b q})^2 \Phi_q
  F_p(t) \equiv M_k(t),
\end{equation}
which reproduces the MCT kernel ${\cal M}_k(t)$, Eq.~\eqref{eqMCTb},
with the linearized disconnected structure factor $S^\text{d}_k =
\lambda \r_0 \Phi_k$.  Therefore, it suffices to eventually replace
$F_k(s)$ with $-\partial_{s} F_k(s) / \Gamma_k$ in the second
convolution integral of Eq.~\eqref{eqFORT6} to get the MCT equations,
Eq.~\eqref{MCT}.  The first-order compatibility of the latter
substitution follows from the combination of the FDR, of
Eq.~\eqref{eqFORT7}, and of Eq.~\eqref{eqFORT8}, or, more directly,
from Eqs.~\eqref{eqFORT1c} or \eqref{eqFORT6}.  With this last step,
however, the structure of the theory is changed and not only the
details of the kernels.  The equivalence of Eqs.~\eqref{eqFORT1c} and
\eqref{eqFORT6} through the FDR corollary for the kernels,
Eq.~\eqref{eqFORT4b}, is broken, with issues for the consistency of
the theory.  For instance, changing Eq.~\eqref{eqFORT4b} to restore
this equivalence would in turn compromise the consistency of
Eqs.~\eqref{eqFORT1b} and \eqref{eqFORT1c} with the FDR.

There would be no such difficulty if the relations
\begin{align} 
  \o{R}_k(t) & = \r_0 \Gam_k G_k(t), \la{eqFORT28} \\
  F_k(t) & = \r_0 G_k(t), \la{eqFORT29} 
\end{align}
complemented by the FDR, held exactly.  They amount to a mere
truncation of both Eqs.~\eqref{eqFORT7} and \eqref{eqFORT8} to their
first term and actually coincide with the defining equations of the
dynamics of the noninteracting Brownian gas without field,
Eqs.~\eqref{eq7.3}.  Generically, Eq.~\eqref{eqFORT28} holds for
dynamics with additive noise, and Eq.~\eqref{eqFORT29}, the DHMR
relation, for nondisordered systems with Gaussian free energy and
either additive or multiplicative noise \cite{DekHaa75PRA,
  MiyRei05JPA}.  Their simultaneous validity for the noninteracting
Brownian gas without field, a non-Gaussian system with multiplicative
noise, stems from special circumstances described in
Sec.~\ref{sec:expansion}.  Those obviously do not survive in the
presence of a Gaussian random field (otherwise, the dynamics should be
the same with and without field), as shown by Eq.~\eqref{tr16b}, in
particular.

We thus conclude that the modified FORT is of a fundamentally distinct
nature from the MCT.

\section{Summary and outlook}
\setcounter{equation}{0}
\label{sec:conc}

The time evolution of the density fluctuations in a system of
colloidal (Brownian) particles is characterized by a Langevin equation
with multiplicative thermal noise, which drives the system into an
equilibrium state governed by a highly non-Gaussian free-energy
density functional.  The multiplicative nature of the time-evolution
equation at the density level generates unique dynamical features
compared to the usual cases of Langevin equations with additive noise.
Indeed, the corresponding free action is a non-Gaussian cubic field
theory, and the physical response function is not the same as the
usual noise-response function, but is given by a three-point function.
It results that the direct loop expansion for the action fails to
satisfy the FDR at each order \cite{MiyRei05JPA}.  These features pose
a theoretical challenge as to how one can develop a FDR-compatible
perturbation theory for the equilibrium dynamics.  A profound
resolution of this issue has recently been proposed, based on the TR
symmetry of the action, i.e., its invariance properties under certain
field transformations when time is reversed \cite{AndBirLef06JSMTE,
  KimKawJacWij14PRE}.  This TR symmetry can indeed dictate
perturbation theories that preserve the FDR.

In the present work, we have developed one such a FDR-preserving
perturbation theory to study the equilibrium dynamics of the density
fluctuations of a noninteracting Brownian gas embedded in a frozen
random potential-energy landscape with Gaussian statistics.
Technically, it is quite different from previous work on bulk
interacting liquids by one of us and others \cite{AndBirLef06JSMTE,
  KimKaw07JPA, KimKaw08JSMTE, KimKawJacWij14PRE}, as it is motivated
by the $\mathcal{T}$- rather than the $\mathcal{U}$-transformation,
does not require the introduction of extra fields into the problem,
and does not rely on a loop expansion.  In practice, the present
perturbation theory involves a double expansion: (i) an expansion
about the dynamics of the pure system, in terms of the
disorder-induced contribution to the dynamical action, then (ii) an
expansion in terms of the cubic contribution generated by the
multiplicative thermal noise in the free dynamics.  The first
expansion can be seen as a weak-disorder or high-temperature
expansion, since the disorder-induced part of the action is
proportional to $\l \equiv w/T^2$, $w$ being the strength of the
Gaussian random potential.  An essential and novel aspect of the
present perturbation theory is the {\it nonperturbative} (exact)
nature of the second expansion.  Indeed, the TR symmetry requires that
the second expansion be carried out exactly.  This is made possible by
the form of the cubic term (containing two noise-response fields as
factors) and by the causality requirements on the vanishing of
averages involving hatted variables.  The latter lead to a quick
termination of the second expansion at each order of the first one.

We carried out a first-order calculation within this FDR-preserving
perturbation scheme.  The corresponding results, the first-order bare
theory, consist of a set of dynamical equations for the correlation
and response functions, which was explicitly checked to be consistent
with the FDR, as intended.  Using the properties of the dynamics of
the pure noninteracting Brownian gas, the equation for the density
correlation function can be rearranged as a MCT equation,
\begin{equation} \la{eqsumm1}
  \left( \partial_{t} + \Gam_k \right) F_k(t) = - \int_0^{t} ds
  M^0_k(t-s) \partial_s C^0_k (s), \quad 
  M^0_k(t) = \frac{\l D_0}{\r_0} \int_{\bf q} ({\hat {\bf
      k}}\cdot {\bf q})^2 \Phi_q C^0_p(t),
\end{equation}
albeit with the memory integral expressed in terms of the bare density
correlation function.  Apart from this, the equation is the same as in
the self-consistent MCT developed by one of us \cite{Kra05PRL,
  Kra07PRE, Kra09PRE, KonKra17SM}.  The bare theory allows one to
compute the MSD, for which we recover results from earlier
calculations at the same order \cite{DeaDruHor07JSM}, and to
characterize the disorder-induced tails that develop in the long-time
dynamics.  The latter reproduce in detail the behavior found in the
Brownian random Lorentz gas, thereby confirming the universal behavior
of the persistent correlations induced by quenched disorder
\cite{FraHofBauFre10CP, ErnMacDorBei84JSP}.  Finally, the bare theory
is clearly found to break down at too strong disorder, when $\lambda$
exceeds the space dimension $d$.  Below this threshold, the dynamics
always remains ergodic.

From the second-order bare perturbation expansion, we also developed a
first-order renormalized theory, constrained to obey the FDR.  Out of
different candidates, all consistent to first order, it is singled out
as the only one delivering useful numerical results (without response
functions that blow up, for instance) over a significant range of
disorder strengths.  It turns out that this theory is distinct from
the MCT, but might be described as MCT-like, in the sense that the
dynamical equation for the density correlation function is also
self-closed:
\begin{subequations} \la{eqsumm2} 
\begin{gather} 
  ( \partial_{t} + \Gam_k ) F_k (t) = - \frac{1}{\r_0} \int_0^{t} ds
  N_k(t-s) \partial_s F_k (s) - \frac{1}{\r_0} \int_0^{t} ds
  \partial_{t-s} \Lambda_k(t-s) F_k (s), \\
  N_k(t) = \l D_0\int_{\bf q} [{\b k} \cdot {\b q} \Phi_q ] F_p(t),
  \qquad
  \Lambda_k(t) = - \l D_0 \int_{\bf q} \frac{{\b k} \cdot {\b p}}{p^2}
         [{\b k} \cdot {\b q} \Phi_q ] F_p(t).
\end{gather}
\end{subequations}
Interestingly, its predictions somewhat improve upon
those of the MCT.  Indeed, in both cases, an ergodicity-breaking
transition occurs in the dynamics of the density fluctuations at
strong enough disorder, but, in the present theory, it does not lead
to a diffusion-localization transition in the MSD, at variance with
the MCT.  This is in agreement with known rigorous results, stating
that normal diffusion is always obtained at long time for Brownian
dynamics \cite{MasFerGolWic89JSP}.  The reason for these contrasting
predictions can be traced back to the distinct low-wavevector
asymptotics of the two theories.  Actually, the low-wavevector
behavior of the single-particle MCT kernel has repeatedly been found
to be a source of difficulties in the theory and is usually considered
as spurious \cite{Got78SSC, Got79JPC, GotPreWol79SSC, Got81PMB,
  Leu83bPRA, SchHofFraVoi11JPCM}.  It is therefore promising that the
present approach seems to naturally circumvent this issue.  It remains
that the sharp ergodicity-breaking transition and the corresponding
singularity in the long-time diffusion coefficient certainly are
artifacts of the self-consistent theory.  Indeed, the exact
expressions of $D_\infty$ are known in $d=1$ and $d=2$
\cite{DeaDruHor07JSM}.  They are infinitely differentiable functions
of the relative disorder strength, and the same can naturally be
expected in other space dimensions.  An ergodicity-breaking transition
would have to display quite unusual characteristics to be consistent
with such a behavior.  However, recent computer simulations in $d=1$
have evidenced strong transient, but long-lived, nonergodic effects in
the system at hand \cite{GoyKhaMet17PRE}.  In this respect, the
theoretical predictions do not appear as an unreasonable first
approximation.

In the present work, we took the initiative of developing a
perturbative expansion method about the highly non-Gaussian pure
noninteracting state.  Compared to the maturity of the fully
renormalized theories such as the loop expansion, such approaches are
still at an early stage.  It would be important for the future to gain
a better understanding of their working principles.  For instance, it
would be useful to put the somewhat ad hoc arguments used in the
derivation of the first-order renormalized theory on firm theoretical
grounds.  This would allow one to further investigate non-equilibrium
phenomena, where by definition the equilibrium theorems cannot be used
as guides.  In this respect, we note that, in principle, the
prediction of an ergodicity-breaking transition in the equilibrium
theory calls for a reassessment within an out-of-equilibrium two-time
formalism.  Finally, it would be most interesting to apply the present
perturbation scheme to the interacting Dean-Kawasaki equation (with or
without the random potential).  This would certainly enrich our
current perspective on the use of field theory in particle-system
dynamics, its relation with the MCT, and the possibilities to go
beyond the latter.

\acknowledgments

We gratefully acknowledge discussions with Kyozi Kawasaki, Thomas
Franosch, J{\"u}rgen Horbach, and Rolf Schilling.  M.F. acknowledges
support from the Deutsche Forschungsgemeinschaft (DFG) through FOR
1394 projects P3 and Z.  B.K. and V.K. warmly acknowledge financial
support from the research unit FOR 1394 'Nonlinear response to probe
vitrification' funded by the DFG, and the hospitality during their
stays at the University of Konstanz.  B.K. was also supported by the
Basic Science Research Program through the National Research
Foundation (NRF) funded by the Ministry of Education, Science, and
Technology (grant No. 2016R1D1A1B03935548), and V.K. by a grant from
R{\'e}gion Rh{\^o}ne-Alpes (grant Explora Pro No. 13 006394 01).

\clearpage
\appendix

\section{Symmetries of the effective dynamical action for colloids in
  a Gaussian random field}
\setcounter{equation}{0} \label{app:symm}

In this Appendix, we provide the technical proofs for the invariance
properties quoted in Sec.~\ref{sec:TRS}, together with some of their
implications.

\subsection{The $\mathcal{T}$-transformation}

We first show the invariance of $S_\text{bulk}[\r, \hr]$,
$S_\text{dis}[\r, \hr]$, and $S_\text{eff}[\r, \hr]$, under the
$\mathcal{T}$-transformation, Eq.~\eqref{tr3}.

With integrations by parts and the definition of the composite
response field, Eq.~\eqref{eq3.14}, Eq.~\eqref{eq3.10} is easily
rewritten as
\begin{equation}
  S_\text{bulk}[\r, \hr] = \int_{{\bf r},t} \hr({\bf r},t) [ i
    \partial_t \r({\bf r},t) + \Lambda({\bf r},t) ] - \int_{{\bf r},t}
  \Lambda({\bf r},t) \frac{i}{T} \left. \frac{\del {\cal
      F}_\text{bulk}[\r]}{\del \r({\bf r})} \right|_{\rho({\bf r},t)}.
\end{equation}
The structure of the first term clearly calls for a field
transformation of the form of Eq.~\eqref{tr5}, requiring
Eq.~\eqref{hTtrans}.  With the explicit application of the field
transformation, one indeed finds
\begin{equation}\begin{split}
  S_\text{bulk}[\mathcal{T}\r, \mathcal{T}\hr] & =
  \int_{{\bf r},t} [ \hr({\bf r},-t) + i h({\bf r},-t) ] \Lambda({\bf
    r},-t) - \int_{{\bf r},t} [ \Lambda({\bf r},-t) - i \partial_{t}
    \r({\bf r},-t) ] \frac{i}{T} \left. \frac{\del {\cal
      F}_\text{bulk}[\r]}{\del \r({\bf r})} \right|_{\rho({\bf r},-t)}
  \\
  & = \int_{{\bf r},t} [ \hr({\bf r},t) + i h({\bf r},t) ]
  \Lambda({\bf r},t) - \int_{{\bf r},t} [ \Lambda({\bf r},t) + i
    \partial_{t} \r({\bf r},t) ] \frac{i}{T} \left. \frac{\del {\cal
      F}_\text{bulk}[\r]}{\del \r({\bf r})} \right|_{\rho({\bf r},t)},
\end{split}\end{equation}
where the second line merely follows from the change of variable
$t\to-t$ in the integrals.  Integrations by parts restore the initial
form of the first integral and, recognizing the chain rule in the
second one, one gets for now
\begin{equation}
  S_\text{bulk}[\mathcal{T}\r, \mathcal{T}\hr] = S_\text{bulk}[\r,
    \hr] + \frac{1}{T} \int_{t} \partial_{t} {\cal F}[\rho({\bf
      r},t)].
\end{equation}

We may repeat the calculation for $S_\text{dis}[\r,\hr]$ as given by
Eq.~\eqref{eq3.13}.  One first gets
\begin{equation}\begin{split}
  S_\text{dis}[\mathcal{T}\r, \mathcal{T}\hr] & = - \frac{1}{2}\l
  \int_{{\bf r},t} \int_{{\bf r}',t'} \Phi(|{\bf r}-{\bf r}'|) [
    \Lambda({\bf r},-t) - i \partial_{t} \r({\bf r},-t) ] [
    \Lambda({\bf r}',-t') - i \partial_{t'} \r({\bf r}',-t') ] \\
  & = - \frac{1}{2}\l \int_{{\bf r},t} \int_{{\bf r}',t'} \Phi(|{\bf
    r}-{\bf r}'|) [ \Lambda({\bf r},t) + i \partial_{t} \r({\bf r},t)
  ] [ \Lambda({\bf r}',t') + i \partial_{t'} \r({\bf r}',t') ],
\end{split}\end{equation}
with again the change of variables $t\to-t$, $t'\to-t'$ in the
integrals to obtain the second line.  Then, the result can be
rearranged as
\begin{multline}
  S_\text{dis}[\mathcal{T}\r, \mathcal{T}\hr] = S_\text{dis}[\r, \hr]
  - i \l \int_{t} \partial_{t} \left[ \int_{{\bf r}} \int_{{\bf
        r}',t'} \Phi(|{\bf r}-{\bf r}'|) \r({\bf r},t) \Lambda({\bf
      r}',t') \right] \\ + \frac{1}{2}\l \int_{t} \partial_{t}
  \int_{t'} \partial_{t'} \left[ \int_{{\bf r}} \int_{{\bf r}'}
    \Phi(|{\bf r}-{\bf r}'|) \r({\bf r},t) \r({\bf r}',t') \right].
\end{multline}

Since the differences $S_\text{bulk}[\mathcal{T}\r, \mathcal{T}\hr] -
S_\text{bulk}[\r, \hr]$ and $S_\text{dis}[\mathcal{T}\r,
  \mathcal{T}\hr] - S_\text{dis}[\r, \hr]$ are mere integrals of total
time derivatives, both $S_\text{bulk}[\r, \hr]$ and $S_\text{dis}[\r,
  \hr]$ are invariant under the $\mathcal{T}$-transformation at
equilibrium.  This obviously implies the invariance of
$S_\text{eff}[\r, \hr]$.

\subsection{The $\mathcal{U}$- and $\mathcal{U'}$-transformations}

A shared feature of the present theory and of the theory of Langevin
processes with colored noise developed in Ref.~\cite{AroBirCug10JSMTE}
is that the dynamical action is a sum of quadratic and linear terms in
the hatted variables, as a consequence of the Gaussianity of the noise
and/or disorder.  In the latter work, a symmetry of the action was
unveiled, which can actually be related to this observation.  We show
that a similar one holds in the present case as well.

Denoting the thermal-noise contribution to the effective dynamical
action as
\begin{equation}
  S_\text{noise}[\r, \hr] = - D_0 \int_{{\bf r},t} \r({\bf r},t)
  [\nabla \hr({\bf r},t)]^2,
\end{equation}
and adding it to the random-field term $S_\text{dis}[\r, \hr]$,
restoration of the noise variance, Eq.~\eqref{eq1.15}, and
integrations by parts can be used to get
\begin{equation}
  S_\text{noise}[\r, \hr] + S_\text{dis}[\r, \hr] = - \frac{1}{2}
  \int_{{\bf r},t} \int_{{\bf r}',t'} K_\lambda({\bf r},t;{\bf r}',t')
  \hr({\bf r},t) \hr({\bf r}',t'),
\end{equation}
where the density-dependent symmetric kernel $K_\lambda({\bf r},t;{\bf
  r}',t')$ is given by Eq.~\eqref{eq:klambda}.  Now, the remaining
part of the action, which only involves the deterministic nonrandom
part of the density evolution equation defined in Eq.~\eqref{eq:deter}
and thus reads
\begin{equation}
  S_\text{bulk}[\r, \hr] - S_\text{noise}[\r, \hr] = \int_{{\bf r},t}
  i\hr({\bf r},t) \Det([\r],{\bf r},t),
\end{equation}
can be rewritten as
\begin{equation}
  S_\text{bulk}[\r, \hr] - S_\text{noise}[\r, \hr] = i \int_{{\bf
      r},t} \int_{{\bf r}',t'} K_\lambda({\bf r},t;{\bf r}',t')
  \hr({\bf r},t) \int_{{\bf r}'',t''} K_\lambda^{-1}({\bf r}',t';{\bf
    r}'',t'') \Det([\r],{\bf r}'',t''),
\end{equation}
through injection of Eq.~\eqref{eq:invklambda} and minor
reorganizations.  It results that
\begin{equation}
  S_\text{eff}[\r, \hr] = - \frac{1}{2} \int_{{\bf r},t} \int_{{\bf
      r}',t'} K_\lambda({\bf r},t;{\bf r}',t') \hr({\bf r},t) \left\{
  \hr({\bf r}',t') -2 i \int_{{\bf r}'',t''} K_\lambda^{-1}({\bf
    r}',t';{\bf r}'',t'') \Det([\r],{\bf r}'',t'') \right\}.
\end{equation}
This expression is manifestly invariant under the
$\mathcal{U'}$-transformation, Eq.~\eqref{eq:Uprimetrans}, thanks to
the symmetry of $K_\lambda({\bf r},t;{\bf r}',t')$.

Although one can directly use Eq.~\eqref{eq:Uprimetrans} to compose
$\mathcal{U'}$ with $\mathcal{T}$, we find it useful to first
reorganize $\mathcal{U'}\hr({\bf r},t)$.  Indeed, this allows one to
isolate contributions with distinct physical origins and facilitates
comparisons with previous results.  Once the explicit expression of
$\Det([\r],{\bf r},t)$ is restored, Eq.~\eqref{hTtrans} and a single
integration by parts lead to
\begin{equation}
  \mathcal{U'}\hr({\bf r}, t) = -\hr({\bf r},t) - 2 i D_0 \int_{{\bf
      r}',t'} \r({\bf r}',t') [ \nabla' K_\lambda^{-1}({\bf r},t;{\bf
      r}',t') ] \cdot \left( \nabla' \left[ h({\bf r}',t') -
    \frac{1}{T} \left. \frac{\del {\cal F}_\text{bulk}[\r]}{\del
      \r({\bf r}')} \right|_{\rho({\bf r}',t')} \right] \right).
\end{equation}  
Treating the integral in the same way as $S_\text{noise}[\r, \hr]$,
one then gets
\begin{equation}
  \mathcal{U'}\hr({\bf r}, t) = -\hr({\bf r},t) - i \int_{{\bf r}',t'}
  \int_{{\bf r}'',t''} K_\lambda^{-1}({\bf r},t;{\bf r}',t') K_0({\bf
    r}',t';{\bf r}'',t'') \left[ h({\bf r}'',t'') - \frac{1}{T}
    \left. \frac{\del {\cal F}_\text{bulk}[\r]}{\del \r({\bf r}'')}
    \right|_{\rho({\bf r}'',t'')} \right],
\end{equation}  
where $K_0({\bf r},t;{\bf r}',t')$ is nothing but $K_\lambda({\bf
  r},t;{\bf r}',t')$ at $\lambda=0$.  Using Eq.~\eqref{eq:klambda}, it
can be replaced with $K_\lambda({\bf r},t;{\bf r}',t') - \lambda
\Delta K({\bf r},t;{\bf r}',t')$ to obtain
\begin{multline}
  \mathcal{U'}\hr({\bf r}, t) = -\hr({\bf r},t) - i h({\bf r},t) +
  \frac{i}{T} \left. \frac{\del {\cal F}_\text{bulk}[\r]}{\del \r({\bf
      r})} \right|_{\rho({\bf r},t)} \\ + i \lambda \int_{{\bf r}',t'}
  \int_{{\bf r}'',t''} K_\lambda^{-1}({\bf r},t;{\bf r}',t') \Delta
  K({\bf r}',t';{\bf r}'',t'') \left[ h({\bf r}'',t'') - \frac{1}{T}
    \left. \frac{\del {\cal F}_\text{bulk}[\r]}{\del \r({\bf r}'')}
    \right|_{\rho({\bf r}'',t'')} \right].
\end{multline}  
It remains to use the expression of $\Delta K({\bf r},t;{\bf r}',t')$
to eventually get
\begin{multline}
  \mathcal{U'}\hr({\bf r}, t) = -\hr({\bf r},t) - i h({\bf r},t) +
  \frac{i}{T} \left. \frac{\del {\cal F}_\text{bulk}[\r]}{\del \r({\bf
      r})} \right|_{\rho({\bf r},t)} \\
  + i \lambda D_0 \int_{{\bf r}',t'} K_\lambda^{-1}({\bf r},t;{\bf
    r}',t') \nabla' \cdot \left\{ \r({\bf r}',t') \nabla' \int_{{\bf
      r}'',t''} \Phi(|{\bf r}'-{\bf r}''|) \Det([\r],{\bf r}'',t'')
  \right\}
\end{multline}  
after a last pair of integrations by parts.  This formula can be used
as an alternative to the second line of Eq.~\eqref{eq:Uprimetrans}.

We may now compose $\mathcal{U'}$ and $\mathcal{T}$ to get the
$\mathcal{U}$-transformation with time reversal.  Since the
application of $\mathcal{T}$ to Eq.~\eqref{hTtrans} gives $D_0
\nabla\cdot [\r({\bf r},-t) \nabla \mathcal{T}h({\bf r},t)] =
\partial_t \r({\bf r},-t)$, hence $\mathcal{T}h({\bf r},t) = - h({\bf
  r},-t)$, the function $h({\bf r},t)$ disappears when $\mathcal{T}$
is applied to $\mathcal{U'}\hr({\bf r}, t)$, giving
Eq.~\eqref{eq:Utrans} as the final result.

Obviously, if $\mathcal{U'}\hr({\bf r},t)$ from
Eq.~\eqref{eq:Uprimetrans} is left untouched, the expression
\begin{equation} \label{eq:altU}
  \mathcal{U} \hr({\bf r}, t) = - \hr({\bf r},-t) - i h({\bf r},-t) +
  2 i \int_{{\bf r}',t'} K_\lambda^{-1}({\bf r},-t;{\bf r}',-t')
  \mathcal{T}\Det([\r],{\bf r}',t')
\end{equation}
is a valid replacement for the second line of Eq.~\eqref{eq:Utrans}.

\subsection{Implications of the $\mathcal{U}$-transformation}

As the $\mathcal{T}$-transformation, the $\mathcal{U}$-transformation
can be used to derive equilibrium relations between correlations and
responses.

In particular, a generalized form of the FDR can be obtained for the
noise-response function.  Indeed, expanding the Ward-Takahashi
identity Eq.~\eqref{eq:WT_G}, one gets
\begin{multline} \label{genDHMRdet}
  G(|{\bf r}-{\bf r}'|, t-t') + G(|{\bf r}-{\bf r}'|, t'-t) =
  \left\langle \r({\bf r},t) \frac{1}{T} \left. \frac{\del {\cal
      F}_\text{bulk}[\r]}{\del \r({\bf r}')} \right|_{\rho({\bf
      r}',t')} \right\rangle_\text{eff} \\
  + \lambda D_0 \left\langle \r({\bf r},t) \int_{{\bf r}'',t''}
  K_\lambda^{-1}({\bf r}',t';{\bf r}'',t'') \nabla'' \cdot \left\{
  \r({\bf r}'',t'') \nabla'' \int_{{\bf r}''',t'''} \Phi(|{\bf
    r}''-{\bf r}'''|) \Det([\r],{\bf r}''',t''') \right\}
  \right\rangle_\text{eff},
\end{multline}
where we used time-translation invariance and the time-reversal
symmetry of the correlations.

Using the explicit expression of ${\cal F}_\text{bulk}[\r]$, the first
average in the right-hand side of Eq.~\eqref{genDHMRdet} can be
rewritten as
\begin{multline}
  \left\langle \r({\bf r},t) \frac{1}{T} \left. \frac{\del {\cal
      F}_\text{bulk}[\r]}{\del \r({\bf r}')} \right|_{\rho({\bf
      r}',t')} \right\rangle_\text{eff} =
  \left\langle \r({\bf r},t) \left[ \frac{\del\r({\bf r}',t')}{\r_0} +
    \frac{1}{T} \int_{{\bf r}''} u(|{\bf r}'-{\bf r}''|) \delta\r({\bf
      r}'',t') \right] \right\rangle_\text{eff} \\
  + \left\langle \r({\bf r},t) \left[ \ln \left( 1+ \frac{\del \r({\bf
        r}',t')}{\r_0} \right) - \frac{\del\r({\bf r}',t')}{\r_0}
    \right] \right\rangle_\text{eff}.
\end{multline} 
The first term is due to the Gaussian part of the free energy,
\begin{equation}
  {\cal F}_\text{bulk,G}[\r] = \frac{T}{2} \int_{\bf r} \int_{{\bf
      r}'} Q^{-1}(|{\bf r}-{\bf r}'|) \del\r({\bf r}) \del\r({\bf
    r}'),
\end{equation}
where
\begin{equation}
  Q^{-1}(|{\bf r}-{\bf r}'|) = \frac{\delta({\bf r}-{\bf r'})}{\r_0} +
  \frac{u(|{\bf r}-{\bf r}'|)}{T}
\end{equation}
is the functional inverse of the static density correlation function
in the Gaussian theory defined by ${\cal F}_\text{bulk,G}[\r]$.  One
can thus write
\begin{equation}
  \left\langle \r({\bf r},t) \left[ \frac{\del\r({\bf r}',t')}{\r_0} +
    \frac{1}{T} \int_{{\bf r}''} u(|{\bf r}'-{\bf r}''|) \delta\r({\bf
      r}'',t') \right] \right\rangle_\text{eff} =
  \int_{{\bf r}''} C(|{\bf r}-{\bf r}''|, t-t') Q^{-1}(|{\bf r}''-{\bf
    r}'|).
\end{equation} 
The second term, which we shall denote by $\Delta C^\text{nG}(|{\bf
  r}-{\bf r}'|,t-t')$, arises from the non-Gaussian nature of ${\cal
  F}_\text{id}[\r]$.  As such, it already appears in the absence of a
random field.

The second average in the right-hand side of Eq.~\eqref{genDHMRdet}
manifestly arises from the presence of the quenched random potential
(it has $\lambda$ as a prefactor).  Accordingly, we shall denote it by
$\Delta C^\text{dis}(|{\bf r}-{\bf r}'|,t-t')$, for which we could not
find any obvious simpler expression.

Combining these notations, Eq.~\eqref{genDHMR} is finally obtained.

As an interesting consistency check, it is also possible to get the
dynamical equations for the density correlation function,
Eqs.~\eqref{eq6.1c} and \eqref{eq6.3c}, directly from the
$\mathcal{T}$- and $\mathcal{U}$-transformations.  Indeed, consider
the Ward-Takahashi identity
\begin{equation} \label{eq:WTdenscorr}
  \left\langle \left[ \int_{{\bf r}'',t''} K_\lambda({\bf r},t;{\bf
      r}'',t'') \hr({\bf r}'', t'') \right] \r({\bf r}',t')
  \right\rangle_\text{eff} = \left\langle \left[ \mathcal{U}
    \int_{{\bf r}'',t''} K_\lambda({\bf r},t;{\bf r}'',t'') \hr({\bf
      r}'', t'') \right] [ \mathcal{U} \r({\bf r}',t')]
  \right\rangle_\text{eff}.
\end{equation}
The direct application of the $\mathcal{U}$-transformation,
Eq.~\eqref{eq:altU}, gives
\begin{multline}
  \mathcal{U} \int_{{\bf r}'',t''} K_\lambda({\bf r},t;{\bf r}'',t'')
  \hr({\bf r}'', t'') = \\
  \int_{{\bf r}'',t''} K_\lambda({\bf r},-t;{\bf r}'',-t'') \left[ -
    \hr({\bf r}'',-t'') -i h({\bf r}'',-t'') \right]
  + 2 i \mathcal{T}\Det([\r],{\bf r},t).
\end{multline}
Using
\begin{equation}
  \int_{{\bf r}'',t''} K_\lambda({\bf r},t;{\bf r}'',t'') \hr({\bf
    r}'', t'') = - 2 \Lambda({\bf r},t) + \l D_0 \nabla \cdot \left[
    \r({\bf r},t) \nabla \int_{{\bf r}'',t''} \Phi(|{\bf r}-{\bf
      r}''|) \Lambda({\bf r}'',t'') \right]
\end{equation}
and
\begin{equation}
  \int_{{\bf r}'',t''} K_\lambda({\bf r},t;{\bf r}'',t'') h({\bf
    r}'',t'') = - 2 \partial_{t} \r({\bf r},t) + \l D_0 \nabla \cdot
  \left[ \r({\bf r},t) \nabla \int_{{\bf r}'',t''} \Phi(|{\bf r}-{\bf
      r}''|) \partial_{t''} \r({\bf r}'',t'') \right],
\end{equation}
this becomes
\begin{multline}
  \mathcal{U} \int_{{\bf r}'',t''} K_\lambda({\bf r},t;{\bf r}'',t'')
  \hr({\bf r}'', t'')
   = 2 i \mathcal{T}\Det([\r],{\bf r},t) + 2 \Lambda({\bf r},-t) - 2 i
   \partial_{t} \r({\bf r},-t) \\
   - \l D_0 \nabla \cdot \left[ \r({\bf r},-t) \nabla \int_{{\bf
         r}'',t''} \Phi(|{\bf r}-{\bf r}''|) \left\{ \Lambda({\bf
       r}'',-t'') - i \partial_{t''} \r({\bf r}'',-t'') \right\}
     \right].
\end{multline}
The Ward-Takahashi identity, Eq.~\eqref{eq:WTdenscorr}, now explicitly
reads
\begin{multline}
  - 2 \langle \Lambda({\bf r},t) \r({\bf r}',t') \rangle_\text{eff} +
  \l D_0 \left\langle \nabla \cdot \left[ \r({\bf r},t) \nabla
    \int_{{\bf r}'',t''} \Phi(|{\bf r}-{\bf r}''|) \Lambda({\bf
      r}'',t'') \right] \r({\bf r}',t') \right\rangle_\text{eff} = \\
  2 i \langle \mathcal{T}\Det([\r],{\bf r},t) \r({\bf r}',-t')
  \rangle_\text{eff} + 2 \langle \Lambda({\bf r},-t) \r({\bf r}',-t')
  \rangle_\text{eff} - 2 i \left\langle \partial_{t} \r({\bf r},-t)
  \r({\bf r}',-t') \right\rangle_\text{eff} \\
  - \l D_0 \left\langle \nabla \cdot \left[ \r({\bf r},-t) \nabla
    \int_{{\bf r}'',t''} \Phi(|{\bf r}-{\bf r}''|) \left\{
    \Lambda({\bf r}'',-t'') - i \partial_{t''} \r({\bf r}'',-t'')
    \right\} \right] \r({\bf r}',-t') \right\rangle_\text{eff}.
\end{multline}
The $\mathcal{T}$-transformation gives
\begin{gather}
  \langle \mathcal{T}\Det([\r],{\bf r},t) \r({\bf r}',-t')
  \rangle_\text{eff} = \langle \Det([\r],{\bf r},t) \r({\bf r}',t')
  \rangle_\text{eff}, \\
  \langle \Lambda({\bf r},t) \r({\bf
    r}',t') \rangle_\text{eff} = \langle \Lambda({\bf r},-t) \r({\bf
    r}',-t') \rangle_\text{eff} - i \langle \partial_{t} \r({\bf
    r},-t) \r({\bf r}',-t') \rangle_\text{eff},
\end{gather}  
and
\begin{multline}
  \left\langle \nabla \cdot \left[ \r({\bf r},t) \nabla \int_{{\bf
        r}'',t''} \Phi(|{\bf r}-{\bf r}''|) \Lambda({\bf r}'',t'')
    \right] \r({\bf r}',t') \right\rangle_\text{eff} = \\ \left\langle
  \nabla \cdot \left[ \r({\bf r},-t) \nabla \int_{{\bf r}'',t''}
    \Phi(|{\bf r}-{\bf r}''|) \{ \Lambda({\bf r}'',-t'') - i
    \partial_{t''} \r({\bf r}'',-t'') \} \right] \r({\bf r}',-t')
  \right\rangle_\text{eff}.
\end{multline}
Therefore, one gets
\begin{equation}
  \left\langle \left\{ i \Det([\r],{\bf r},t) + 2 \Lambda({\bf r},t) -
  \l D_0 \nabla \cdot \left[ \r({\bf r},t) \nabla \int_{{\bf r}'',t''}
    \Phi(|{\bf r}-{\bf r}''|) \Lambda({\bf r}'',t'') \right] \right\}
  \r({\bf r}',t') \right\rangle_\text{eff} = 0,
\end{equation}
which is nothing but
\begin{equation} 
  \left\langle \frac{\delta S_\text{eff}}{\delta \hr({\bf r},t)}
  \r({\bf r}',t') \right\rangle_\text{eff} = 0.
 \end{equation}

\section{Calculation of the  memory kernel $M_k^0(t)$ for a Gaussian
covariance} \setcounter{equation}{0} \label{appA}

We analytically compute the memory kernel $M_k^0(t)$ given in
Eq.~\eqref{eqFOBTM} for the Gaussian random potential with Gaussian
covariance:
\begin{equation} \la{a1} %
  M_k^0(t) = A_k \int d{\bf q} ({\bf k}\cdot {\bf q})^2 e^{-q^2 R^2/2}
  e^{-D_0 ({\bf k}-{\bf q})^2 t}, \qquad A_k \equiv \frac{\l D_0 R^d}
  {k^2 (2\pi)^{d/2}}.
\end{equation}
The integral can be arranged as
\begin{equation}  \la{a2}
  M_k^0(t) = A_k e^{-D_0k^2 t} \int d{\bf q} ({\bf k}\cdot {\bf q})^2
  e^{- (R^2/2+D_0t) q^2 + 2 D_0 t {\bf k}\cdot {\bf q}}.
\end{equation}
Completing the square in the argument of the exponential, we have
\begin{equation} \la{a3}
  M_k^0(t) = A_k e^{-\frac{D_0 k^2 R^2 t }{R^2+2 D_0t}} \int d{\bf q}
  ({\bf k}\cdot {\bf q})^2 e^{- (R^2/2+D_0t) \left( {\bf q}-\frac{D_0
      t}{R^2/2+D_0t} {\bf k} \right)^2}.
\end{equation}
Now, shifting the integration variable via ${\bf u}\equiv {\bf
  q}-\frac{D_0 t}{R^2/2+D_0t} {\bf k}$, we get
\begin{equation} \la{a4} %
  M_k^0(t) = A_k e^{-\frac{D_0 k^2 R^2 t }{R^2+2 D_0t}} \int d{\bf u}
  \left[ ({\bf k}\cdot {\bf u})^2 + 2 {\bf k}\cdot {\bf u} \frac{k^2
      D_0 t}{R^2/2+D_0t} + \left( \frac{k^2 D_0 t}{R^2/2+D_0t}
    \right)^2 \right] e^{- (R^2/2+D_0t) {\bf u}^2}.
\end{equation}
By isotropy, the first term $({\bf k}\cdot {\bf u})^2$ can be replaced
with $k^2 {\bf u}^2/d$ and the second term involving $ {\bf k}\cdot
{\bf u}$ vanishes.  We thus have
\begin{equation} \la{a5} %
  M_k^0(t) = A_k e^{-\frac{D_0 k^2 R^2 t }{R^2+2 D_0t}} \int d{\bf u}
  \left[ \frac{k^2 {\bf u}^2}{d} + \left( \frac{2 k^2 D_0 t}{R^2 + 2
      D_0t} \right)^2 \right] e^{- (R^2/2+D_0t){\bf u}^2}.
\end{equation}
Using the integration formulas
\begin{equation} \la{a6} %
  \int d{\bf u} e^{-\al {\bf u}^2} = \left( \frac{\pi}{\al}
  \right)^{d/2}, \quad \int d{\bf u} {\bf u}^2 e^{-\al {\bf u}^2} =
  \frac{d}{2 \al} \left( \frac{\pi}{\al} \right)^{d/2},
\end{equation}
we obtain
\begin{equation} \la{a7} %
  M_k^0(t) = A_k e^{-\frac{D_0 k^2 R^2 t }{R^2+2 D_0t}} \left[
    \frac{k^2}{R^2+2D_0t} + \left( \frac{2k^2 D_0 t}{R^2+2D_0t}
    \right)^2 \right] \left(\frac{ 2 \pi}{R^2+2D_0t} \right)^{d/2}.
\end{equation}
Putting the explicit expression for $A_k$, we have the final
expression for the memory kernel,
\begin{equation} \la{a8} %
  M_k^0(t) = \l D_0 R^d e^{-\frac{D_0 k^2 R^2 t }{R^2+2 D_0t}}
  \frac{R^2 + 2 D_0 t + (2D_0 t)^2 k^2 }{ \left(R^2 + 2D_0t \right)^2}
  \left(\frac{ 1}{R^2+2D_0t} \right)^{d/2},
\end{equation}
which is Eq.~\eqref{eqexplicit4}.

\section{Renormalized equation for the noise-response function} 
\setcounter{equation}{0} \label{app:G}

The full dynamical equation for the noise-response function $G(12)$ is
given by Eq.~\eqref{eq6.3a} and, after simplification, reads
\begin{equation} \la{c1}
  (\partial_t -D_0 \nabla^2 ) G(12) = \del(12) - \l D_0^2 \nabla^{\al}
  \left( \int_3 [ \nabla^{\al} \nabla^{\gamma} \Phi(13) ] \langle 13
  \h3^{\gamma} \h2 \rangle \right) .
\end{equation}
Since there is no risk of confusion in these appendices, we shall here
denote the averages over the effective action simply as $\langle
\ldots \rangle$.

It is straightforward to calculate the multi-point average up to the
first order, as
\begin{equation} \la{c2}
  \langle 13 \h3^{\gamma} \h2 \rangle = \langle 13 \h3^{\gamma} \h2
  \rangle_\text{f} + \langle 13 \h3^{\gamma} \h2 S_\text{dis}
  \rangle_\text{f} + O(\l^2).
\end{equation}
The first term corresponds to Eq.~\eqref{eq8.5a}, and the first-order
average involves $S_\text{dis}[\r,\hr]$, given by
Eq.~\eqref{eq:Sdis_expan} and rewritten as
\begin{equation} \la{c3}
  S_\text{dis}[\r,\hr] = \frac{1}{2} \l D_0^2 \int_6 \int_9 [
    \nabla_6^a \nabla_6^b \Phi(69) ] [ \r_0^2 + 2 \r_0 \del\r(6) +
    \del\r(6) \del\r(9) ] [\nabla_6^a \hr(6)] [\nabla_9^b \hr(9)],
\end{equation}
where $a$ and $b$ denote summed-upon Cartesian indices.  With these
expressions, one readily obtains
\begin{equation}
  \langle 13 \h3^{\gamma} \h2 \rangle = \langle 1 \h3^{\gamma}
  \rangle_0 \langle 3 \h2 \rangle_0 + \frac{1}{2} \l D_0^2 \int_6
  \int_9 [ \nabla_6^a \nabla_6^b \Phi(69) ] \langle 1369 \h3^{\gamma}
  \h6^a \h9^b \h2 \rangle_0 + O(\l^2),
\end{equation}
hence
\begin{multline} \la{c4}
  \langle 13 \h3^{\gamma} \h2 \rangle = \langle 1 \h3^{\gamma}
  \rangle_0 \langle 3 \h2 \rangle_0 \\
  + \l D_0^2 \int_6 \int_9 [ \nabla_6^a \nabla_6^b \Phi(69) ]
     [ \langle 1 \h3^{\gamma} \rangle_0 \langle 3 \h6^a \rangle_0
       \langle 6 \h9^b \rangle_0 \langle 9 \h2 \rangle_0 + \langle 1
       \h6^a \rangle_0 \langle 6 \h9^b \rangle_0 \langle 9
       \h3^{\gamma} \rangle_0 \langle 3 \h2 \rangle_0 \\
  + \langle 1 \h6^a \rangle_0 \langle 6 \h3^{\gamma} \rangle_0 \langle
  3 \h9^b \rangle_0 \langle 9 \h2 \rangle_0 ] + O(\l^2).
\end{multline}

On the other hand, one has the following (first-order) result for the
noise-response function itself,
\begin{equation}\begin{split} \la{c5}
  \langle 1 \h3^{\gamma} \rangle & = \langle 1 \h3^{\gamma}
  \rangle_\text{f} + \langle 1 \h3^{\gamma} S_\text{dis}
  \rangle_\text{f} + O(\l^2) \\
  & = \langle 1 \h3^{\gamma} \rangle_0 + \frac{1}{2} \l D_0^2 \int_6
  \int_9 [ \nabla_6^a \nabla_6^b \Phi(69) ] \langle 169 \h6^a \h9^b
  \h3^{\gamma} \rangle_0 + O(\l^2) \\
  & = \langle 1 \h3^{\gamma} \rangle_0 + \l D_0^2 \int_6 \int_9 [
    \nabla_6^a \nabla_6^b \Phi(69) ] \langle 1 \h6^a \rangle_0 \langle
  6 \h9^b \rangle_0 \langle 9 \h3^{\gamma} \rangle_0 + O(\l^2).
\end{split}\end{equation}
Equivalently, one can express the bare response in terms of the
renormalized one as
\begin{equation} \la{c6} 
  \langle 1 \h3^{\gamma} \rangle_0 = \langle 1 \h3^{\gamma} \rangle -
  \l D_0^2 \int_6 \int_9 [ \nabla_6^a \nabla_6^b \Phi(69) ] \langle 1
  \h6^a \rangle \langle 6 \h9^b \rangle \langle 9 \h3^{\gamma} \rangle
  + O(\l^2).
\end{equation}
Likewise, one has
\begin{equation} \la{c7} 
  \langle 3 \h2 \rangle_0 = \langle 3 \h2 \rangle - \l D_0^2 \int_6
  \int_9 [ \nabla_6^a \nabla_6^b \Phi(69) ] \langle 3 \h6^a \rangle
  \langle 6 \h9^b \rangle \langle 9\h2 \rangle + O(\l^2).
\end{equation}
Substituting these expressions into Eq.~\eqref{c4}, one
straightforwardly obtains
\begin{equation} \la{c8} 
  \langle 13 \h3^{\gamma} \h2 \rangle = \langle 1 \h3^{\gamma} \rangle
  \langle 3 \h2 \rangle + \l D_0^2 \int_6 \int_9 [ \nabla_6^a
    \nabla_6^b \Phi(69) ] \langle 1 \h6^a \rangle \langle 6
  \h3^{\gamma} \rangle \langle 3 \h9^b \rangle \langle 9 \h2 \rangle +
  O(\l^2).
\end{equation}

The dynamical equation for $G(12)$ is then given by (up to second
order in $\l$)
\begin{multline} \la{c9}
  (\partial_t -D_0 \nabla^2) G(12) = \del(12) - \l D_0^2 \nabla^{\al}
  \left( \int_3 [ \nabla^{\al} \nabla^{\gamma} \Phi(13) ] \langle 1
  \h3^{\gamma} \rangle \langle 3 \h2 \rangle \right) \\
  - \l^2 D_0^4 \nabla^{\al} \left( \int_3 \int_6 \int_9 [\nabla^{\al}
    \nabla^{\gamma} \Phi(13) ] [ \nabla_6^a \nabla_6^b \Phi(69) ]
  \langle 1 \h6^a \rangle \langle 6\h3^{\gamma} \rangle \langle 3
  \h9^b \rangle \langle 9 \h2 \rangle \right) .
\end{multline}
Note that Eq.~\eqref{c5} takes the form of the Schwinger-Dyson
equation,
\begin{equation} \la{c10}
  G = G_0 + G_0 \cdot \S[G] \cdot G = G_0 + G_0 \cdot \S^{0}[G_0]
  \cdot G_0 + \ldots,
\end{equation}
and Eq.~\eqref{c9} would equivalently take the form
\begin{equation} \la{c11} 
  G_0^{-1} \cdot G = I + \S[G] \cdot G,
\end{equation}
where $G_0^{-1}$ is given by $G_0^{-1}(12) = ( \partial_t -D_0
\nabla^2 ) \delta(12)$.

The Fourier-transformed dynamical equation for $G$ is eventually given
by (up to first order)
\begin{gather} 
 (\partial_t + \Gam_k ) G_k(t-t') = \del(t-t') - \int_{t'}^t ds
  \S_k(t-s) G_k(s-t'), \la{c12} \\
  \S_k(t) = \l D_0^2 \int_{\bf q} {\b q} \cdot {\b p} [{\b k} \cdot
    {\b q} \Phi_q ] G_p(t). \la{c13}
\end{gather}

\section{Renormalized equation for the physical response function} 
\setcounter{equation}{0} \label{app:R}

The full dynamical equation for the physical response function is
given by Eq.~\eqref{eq6.3b}, leading to
\begin{multline} \la{d1}
  (\partial_t -D_0 \nabla^2) \o{R}(12) = -\r_0 D_0 \nabla^2 \delta(12)
  \\
 + \l D_0^3 \nabla^{\al} \left( \int_3
 [\nabla^{\al} \Phi(13)] \nabla_2^{\beta} \nabla_3^{\gamma} [ \r_0
   \langle 13 \h3^{\gamma} \h2^{\beta} \rangle + \r_0 \langle 1 2
   \h3^{\gamma} \h2^{\beta} \rangle + \langle 1 3 2 \h3^{\gamma}
   \h2^{\beta} \rangle ] \right) .
\end{multline}

The first two multi-point averages have already been computed [see
  Eq.~\eqref{c8}]:
\begin{align} 
  \langle 13 \h3^{\gamma} \h2^{\beta} \rangle & = \langle 1
  \h3^{\gamma} \rangle \langle 3 \h2^{\beta} \rangle + \l D_0^2 \int_6
  \int_9 [ \nabla_6^a \nabla_6^b \Phi(69) ] \langle 1 \h6^a \rangle
  \langle 6 \h3^{\gamma} \rangle \langle 3 \h9^b \rangle \langle 9
  \h2^{\beta} \rangle + O(\l^2), \la{d2} \\
  \langle 12 \h3^{\gamma} \h2^{\beta} \rangle & = \langle 1
  \h2^{\beta} \rangle \langle 2 \h3^{\gamma} \rangle + \l D_0^2 \int_6
  \int_9 [ \nabla_6^a \nabla_6^b \Phi(69) ] \langle 1 \h6^a \rangle
  \langle 6 \h2^{\beta} \rangle \langle 2 \h9^b \rangle \langle 9
  \h3^{\gamma} \rangle + O(\l^2). \la{d3}
\end{align}
The last one is obtained up to the first order as [see
  Eq.~\eqref{eq8.5d} for the first term]
\begin{equation}\begin{split} \la{d4}
  \langle 1 3 2 \h3^{\gamma} \h2^{\beta} \rangle & = \langle 1 3 2
  \h3^{\gamma} \h2^{\beta} \rangle_\text{f} + \langle 1 3 2
  \h3^{\gamma} \h2^{\beta} S_\text{dis} \rangle_\text{f} + O(\l^2) \\
  & = \l \r_0 D_0^2 \int_6 \int_9 [ \nabla_6^a \nabla_6^b \Phi(69) ]
  \langle 1236 \h2^{\beta} \h3^{\gamma} \h6^a \h9^b \rangle_0 +
  O(\l^2) \\
  & = \l \r_0 D_0^2 \int_6 \int_9 [ \nabla_6^a \nabla_6^b \Phi(69) ] [
    \langle 1 \h2^{\beta} \rangle_0 \langle 2 \h6^a \rangle_0 \langle
    6 \h3^{\gamma} \rangle_0 \langle 3 \h9^b \rangle_0 + \langle 1
    \h3^{\gamma} \rangle_0 \langle 3 \h6^a \rangle_0 \langle 6
    \h2^{\beta} \rangle_0 \langle 2 \h9^b \rangle_0 \\
  & \qquad\qquad\qquad\qquad + \langle 1 \h6^a \rangle_0 \langle 6
    \h2^{\beta} \rangle_0 \langle 2 \h3^{\gamma} \rangle_0 \langle 3
    \h9^b \rangle_0 + \langle 1 \h6^a \rangle_0 \langle 6 \h3^{\gamma}
    \rangle_0 \langle 3 \h2^{\beta} \rangle_0 \langle 2 \h9^b
    \rangle_0 ] + O(\l^2),
\end{split}\end{equation}
where we used $\int_9 [ \nabla_6^a \nabla_6^b \Phi(69) ] \langle 6
\h9^b \rangle_0 = 0$ by isotropy.

Now, recall that the physical response function $\o{R}(12)$ is related
to the noise-response function $G(12)$ as shown by Eq.~\eqref{tr1r},
hence
\begin{equation} \la{d5}
  \o{R}(12) = i \r_0 D_0 \nabla^2 \langle 1 \h2 \rangle + i D_0
  \nabla_2^{\beta}\langle 1 2 \h2^{\beta} \rangle.
\end{equation}
The three-point average $\langle 1 2 \h2^{\beta} \rangle$ is evaluated
up to the first order as
\begin{equation}\begin{split}
  \langle 1 2 \h2^{\beta} \rangle & = \langle 1 2 \h2^{\beta}
  \rangle_\text{f} + \langle 1 2 \h2^{\beta} S_\text{dis}
  \rangle_\text{f} + O(\l^2) \\
  & = \l \r_0 D_0^2 \int_6 \int_9 [ \nabla_6^a \nabla_6^b \Phi(69) ]
  \langle 1 2 6 \h2^{\beta} \h6^a \h9^b \rangle_0 + O(\l^2) \\
  & = \l \r_0 D_0^2 \int_6 \int_9 [ \nabla_6^a \nabla_6^b \Phi(69) ]
  \langle 1 \h6^a \rangle_0 \langle 6 \h2^{\beta} \rangle_0 \langle 2
  \h9^b \rangle_0 + O(\l^2).  \la{d6}
\end{split}\end{equation}
One thus has 
\begin{gather}
  \o{R}(23) = i \r_0 D_0 \nabla_2^2 \langle 2 \h3 \rangle + i D_0
  \nabla_3^{\gamma} \langle 2 3 \h3^{\gamma} \rangle, \la{d7} \\
  \o{R}(32) = i \r_0 D_0 \nabla_3^2 \langle 3 \h2 \rangle + i D_0
  \nabla_2^{\beta} \langle 3 2 \h2^{\beta} \rangle, \la{d8}
\end{gather}
with
\begin{gather}
  \langle 2 3 \h3^{\gamma} \rangle = \l \r_0 D_0^2 \int_6 \int_9 [
    \nabla_6^a \nabla_6^b \Phi(69) ] \langle 2 \h6^a \rangle_0 \langle
  6 \h3^{\gamma} \rangle_0 \langle 3 \h9^b \rangle_0 + O(\l^2),
  \la{d9} \\
  \langle 3 2 \h2^{\beta} \rangle = \l \r_0 D_0^2 \int_6 \int_9 [
    \nabla_6^a \nabla_6^b \Phi(69) ] \langle 3 \h6^a \rangle_0 \langle
  6 \h2^{\beta} \rangle_0 \langle 2 \h9^b \rangle_0 + O(\l^2).
  \la{d10}
\end{gather}
Crucially, the above integrals can be straightforwardly recognized in
the first two terms of the right-hand side of Eq.~\eqref{d4}.  These
terms are thus associated with the first-order expansion of $\o{R}$
and should be accounted for accordingly in the renormalization
process.

Hence, from Eqs.~\eqref{d2}, \eqref{d3}, \eqref{d4}, \eqref{d9}, and
\eqref{d10}, the following first-order renormalized expressions
result:
\begin{equation}
  \langle 13 \h3^{\gamma} \h2^{\beta} \rangle = \langle 1 \h3^{\gamma}
  \rangle \langle 3 \h2^{\beta} \rangle, \ \langle 12 \h3^{\gamma}
  \h2^{\beta} \rangle = \langle 1 \h2^{\beta} \rangle \langle 2
  \h3^{\gamma} \rangle, \ \langle 1 3 2 \h3^{\gamma} \h2^{\beta}
  \rangle = \langle 1 \h3^{\gamma} \rangle \langle 3 2 \h2^{\beta}
  \rangle + \langle 1 \h2^{\beta} \rangle \langle 2 3 \h3^{\gamma}
  \rangle. \la{d11}
\end{equation}
They provide one with the first-order renormalization
\begin{equation}\begin{split} \la{d12}
  D_0^3 & \nabla_2^{\beta} \nabla_3^{\gamma} [ \r_0 \langle 13
    \h3^{\gamma} \h2^{\beta} \rangle + \r_0 \langle 1 2 \h3^{\gamma}
    \h2^{\beta} \rangle + \langle 1 3 2 \h3^{\gamma} \h2^{\beta}
    \rangle ] \\
  & = D_0^2 \nabla_3^{\gamma} [ \langle 1 \h3^{\gamma} \rangle ( \r_0
    D_0\nabla_2^{\beta} \langle 3 \h2^{\beta} \rangle + D_0
    \nabla_2^{\beta} \langle 3 2 \h2^{\beta} \rangle ) ] + D_0^2
  \nabla_2^{\beta} [ \langle 1 \h2^{\beta} \rangle ( \r_0 D_0
    \nabla_3^{\gamma} \langle 2 \h3^{\gamma} \rangle + D_0
    \nabla_3^{\gamma} \langle 2 3 \h3^{\gamma} \rangle ) ] \\
  & = D_0^2 \nabla_3^{\gamma} [ \langle 1 \h3^{\gamma} \rangle ( \r_0
    D_0\nabla_2^{2} \langle 3 \h2 \rangle + D_0 \nabla_2^{\beta}
    \langle 3 2 \h2^{\beta} \rangle ) ] + D_0^2 \nabla_2^{\beta} [
    \langle 1 \h2^{\beta} \rangle ( \r_0 D_0 \nabla_3^{2} \langle 2
    \h3 \rangle + D_0 \nabla_3^{\gamma} \langle 2 3 \h3^{\gamma}
    \rangle ) ] \\
  & = D_0^2 \nabla_3^{\beta} [ \langle 1 \h3^{\beta} \rangle ( \r_0
    D_0\nabla_3^{2} \langle 3 \h2 \rangle + D_0 \nabla_2^{\gamma}
    \langle 3 2 \h2^{\gamma} \rangle ) ] + D_0^2 \nabla_2^{\beta} [
    \langle 1 \h2^{\beta} \rangle ( \r_0 D_0 \nabla_2^{2} \langle 2
    \h3 \rangle + D_0 \nabla_3^{\gamma} \langle 2 3 \h3^{\gamma}
    \rangle ) ] \\
  & = -i D_0^2 \{ \nabla_3^{\beta} [ \langle 1 \h3^{\beta} \rangle
    \o{R}(32) ] + \nabla_2^{\beta} [ \langle 1 \h2^{\beta} \rangle
    \o{R}(23) ] \}.
\end{split}\end{equation}

Therefore, one obtains the first-order renormalized dynamical equation
\begin{multline} \la{d13}
  (\partial_t -D_0 \nabla^2 ) \o{R}(12) = -\r_0 D_0 \nabla^2
  \delta(12) \\
  - i \l D_0^2 \nabla^{\alpha} \left( \int_3 [\nabla^{\alpha} \Phi(13)
  ] \{ \nabla_3^{\beta} [ \langle 1 \h3^{\beta} \rangle \o{R}(32) ] +
  \nabla_2^{\beta} [ \langle 1 \h2^{\beta} \rangle \o{R}(23) ] \}
  \right),
\end{multline}
or, through elimination of the isolated time integral thanks to the
FDR,
\begin{multline} \la{d14}
  (\partial_t -D_0 \nabla^2 ) \o{R}(12) = - \r_0 D_0 \nabla^2
  \delta(12) \\
  - i \l D_0^2 \nabla^{\alpha} \left( \int_3 [\nabla^{\alpha}
    \nabla^{\beta} \Phi(13) ] \langle 1 \h3^{\beta} \rangle \o{R}(32)
  \right) + i \l \r_0 D_0^2 \nabla^{\alpha} \nabla^{\beta} (
        [\nabla^{\alpha} \Phi(12) ] \langle 1 \h2^{\beta} \rangle ).
\end{multline}

In Fourier space, this equation takes the form
\begin{equation}
  (\partial_t +\Gam_k ) \o{R}_k(t-t') = \r_0 \Gam_k \del(t-t') -
  \int_{t'}^t ds \S_k(t-s) \o{R}_k(s-t') + L_k(t-t'), \la{d15}
\end{equation}
where the kernel $\S_k(t)$ is given in Eq.~\eqref{c13}.  As for the
new kernel $L_k(t)$ arising from the composite nature of the physical
response function, it can be obtained from Eq.~\eqref{d14} as
\begin{equation} \la{d16}
  L_k(t) = \l \r_0 D_0^2 \int_{\bf q} {\b k} \cdot {\b p} [{\b k}
    \cdot {\b q} \Phi_q ] G_p(t).
\end{equation}

\section{Renormalized equation for the correlation function} 
\setcounter{equation}{0} \label{app:C}

From Eq.~\eqref{eq6.3c}, the full dynamical equation for the
correlation function is given by
\begin{multline} \la{e1}
  ( \partial_t -D_0 \nabla^2 ) C(12) = 2\o{R}(21) - \l \r_0 D_0
  \nabla^{\al} \left( \int_3 [\nabla^{\al} \Phi(13)] [ \o{R}(13) +
    \o{R}(23) ] \right) \\
  - i \l D_0^2 \nabla^{\al} \left( \int_3 [\nabla^{\al} \Phi(13)]
  \nabla_3^{\gamma} [ \r_0 \langle 1 2 \h3^{\gamma} \rangle + \langle
    123 \h3^{\gamma} \rangle ] \right).
\end{multline}

We calculate the multi-point averages up to the first order in $\l$
with the bare perturbation expansion.  The three-point average is
given by
\begin{equation} \la{e2}
  \langle 12 \h3^{\gamma} \rangle = \langle 12 \h3^{\gamma}
  \rangle_\text{f} + \langle 12 \h3^{\gamma} S_\text{dis}
  \rangle_\text{f} + O(\l^2).
\end{equation}
The first term corresponds to Eq.~\eqref{eq8.5e}. The first-order
contribution reads
\begin{equation}\begin{split} \la{e3}
  \langle 12 \h3^{\gamma} S_\text{dis} \rangle_\text{f} & =
  \frac{1}{2} \l D_0^2 \int_6 \int_9 [ \nabla_6^a \nabla_6^b \Phi(69)
  ] [ 2 \r_0 \langle 126 \h3^{\gamma} \h6^a \h9^b \rangle_0 + \langle
    1269 \h3^{\gamma} \h6^a \h9^b e^{-D_0 \int_4 4 \h4^\delta
      \h4^\delta} \rangle_0 ] \\
  & = \frac{1}{2} \l D_0^2 \int_6 \int_9 [ \nabla_6^a \nabla_6^b
    \Phi(69) ] \left[ 2 \r_0 \langle 126 \h3^{\gamma} \h6^a \h9^b
    \rangle_0 - D_0 \int_4 \langle 12469 \h3^{\gamma} \h4^\delta
    \h4^\delta \h6^a \h9^b \rangle_0 \right] \\
  & = \l \r_0 D_0^2 \int_6 \int_9 [ \nabla_6^a \nabla_6^b \Phi(69) ] [
    \langle 1 \h6^a \rangle_0 \langle 2 \h9^b \rangle_0 + \langle 1
    \h9^b \rangle_0 \langle 2 \h6^a \rangle_0 ] \langle 6 \h3^{\gamma}
  \rangle_0 \\
  & \quad -2 \l D_0^3 \int_4 \int_6 \int_9 [ \nabla_6^a \nabla_6^b
    \Phi(69) ] \{ \langle 1 \h4^\delta \rangle_0 [ \langle 4 \h6^a
    \rangle_0 \langle 6 \h9^b \rangle_0 \langle 9 \h3^{\gamma}
    \rangle_0 ] \langle 2 \h4^\delta \rangle_0 \\
  & \qquad\qquad + \langle 1 \h4^\delta \rangle_0 \langle 4
  \h3^{\gamma} \rangle_0 [ \langle 2 \h6^a \rangle_0 \langle 6 \h9^b
    \rangle_0 \langle 9 \h4^\delta \rangle_0 ] + [ \langle 1 \h6^a
    \rangle_0 \langle 6 \h9^b \rangle_0 \langle 9 \h4^\delta \rangle_0
  ] \langle 4 \h3^{\gamma} \rangle_0 \langle 2 \h4^\delta \rangle_0 \}
  \\
  & \quad - 2 \l D_0^3 \int_4 \int_6 \int_9 [ \nabla_6^a \nabla_6^b
    \Phi(69) ] \{ \langle 1 \h4^\delta \rangle_0 [ \langle 4 \h9^b
    \rangle_0 \langle 9 \h3^{\gamma} \rangle_0 ] [ \langle 2 \h6^a
    \rangle_0 \langle 6 \h4^\delta \rangle_0 ] \\
  & \qquad\qquad + [\langle 1 \h6^a \rangle_0 \langle 6 \h4^\delta
    \rangle_0 ] [ \langle 4 \h9^b \rangle_0 \langle 9 \h3^{\gamma}
    \rangle_0 ] \langle 2 \h4^\delta \rangle_0 + [\langle 1 \h6^a
    \rangle_0 \langle 6 \h4^\delta \rangle_0 ] \langle 4 \h3^{\gamma}
  \rangle_0 [ \langle 2 \h9^b \rangle_0 \langle 9 \h4^\delta \rangle_0
  ] \}.
\end{split}\end{equation}
In the right-hand side of this equation, the first line is part of the
second-order contribution to the time-persistent term [whose
  first-order expression is $- \l \r_0^2 D_0 \nabla^2 \Phi(12)$, see
  Eq.~\eqref{eq8.8c}].  The next three terms are those that contribute
to renormalize Eq.~\eqref{eq8.5e}, whereas the last contributions
belong to the second-order renormalization.  Therefore, apart from the
time-persistent terms, one has the renormalized expression
\begin{equation} \la{e4}
  \langle 12 \h3^{\gamma} \rangle = - 2 D_0 \int_4 \langle 1
  \h4^\delta \rangle \langle 4 \h3^{\gamma} \rangle \langle 2
  \h4^\delta \rangle + O(\l).
\end{equation}

Now, the four-point average in Eq.~\eqref{e1} is given by
\begin{equation} \la{e5}
  \langle 12 3 \h3^{\gamma} \rangle = \langle 123 \h3^{\gamma}
  \rangle_\text{f} + \langle 12 3\h3^{\gamma} S_\text{dis}
  \rangle_\text{f} + O(\l^2).
\end{equation}
The first term is already computed in Eq.~\eqref{eq8.5f}.  The
first-order contribution consists of three Gaussian averages:
\begin{multline} \la{e6}
  \langle 12 3\h3^{\gamma} S_\text{dis} \rangle_\text{f} =
  \frac{1}{2} \l D_0^2 \int_6 \int_9 [ \nabla_6^a \nabla_6^b \Phi(69)
  ] \\
  \times \{ \r_0^2 \langle 123 \h3^{\gamma} \h6^a \h9^b \rangle_0 + 2
  \r_0 \langle 1236 \h3^{\gamma} \h6^a \h9^b e^{-D_0 \int_4 4
    \h4^\delta \h4^\delta} \rangle_0 + \langle 1236 9 \h3^{\gamma}
  \h6^a \h9^b e^{-D_0 \int_4 4 \h4^\delta \h4^\delta} \rangle_0 \}.
\end{multline}
These averages are straightforward to compute.  The first one is given
by
\begin{equation} \la{e7}
  \frac{1}{2} \r_0^2 \langle 123 \h3^{\gamma} \h6^a \h9^b \rangle_0 =
  \r_0^2 ( \langle 1 \h3^{\gamma} \rangle_0 \langle 3 \h6^a \rangle_0
  \langle 2 \h9^b \rangle_0 + \langle 1 \h6^a \rangle_0 \langle 3
  \h9^b \rangle_0 \langle 2 \h3^{\gamma} \rangle_0 ).
\end{equation}
It is also part of the time-persistent contribution at second order.
Combining this term with the previous one of the same nature in
Eq.~\eqref{e3} and using the procedure introduced in
Sec.~\ref{sec:firstorder} to eliminate the isolated time integrals,
one gets
\begin{multline} \la{e8}
  - i \l D_0^2 \nabla^{\al} \Bigg( \int_3 [\nabla^{\al} \Phi(13)]
  \nabla_3^{\gamma} \Bigg[ \l \r_0^2 D_0^2 \int_6 \int_9 [ \nabla_6^a
      \nabla_6^b \Phi(69) ] \\
  \times \{ [ \langle 1 \h6^a \rangle_0 \langle 2 \h9^b \rangle_0 +
    \langle 1 \h9^b \rangle_0 \langle 2 \h6^a \rangle_0 ] \langle 6
  \h3^{\gamma} \rangle_0 + \langle 1 \h3^{\gamma} \rangle_0 \langle 3
  \h6^a \rangle_0 \langle 2 \h9^b \rangle_0 + \langle 1 \h6^a
  \rangle_0 \langle 3 \h9^b \rangle_0 \langle 2 \h3^{\gamma} \rangle_0
  \} \Bigg] \Bigg) = \\
  - \r_0^2 D_0 \nabla^2 \left[\frac{\l^2\Phi(12)^2}2 \right].
\end{multline}
Adding the first-order term $-\r_0^2 D_0 \nabla^2 [ \l \Phi(12)]$, one
recognizes the second-order expansion of $- D_0 \nabla^2
C_\text{d}(12)$, with $C_\text{d}(12) \equiv C_\text{d}(|{\bf r}-{\bf
  r}'|)$, according to the exact static equilibrium calculation,
Eq.~\eqref{eq3.6b}.  Such an identification is actually required to
insure consistency between statics and dynamics.

The remaining Gaussian averages in Eq.~\eqref{e6} are given by
\begin{equation}\begin{split} \la{e9}
  \r_0 & \langle 1236 \h3^{\gamma} \h6^a \h9^b e^{-D_0 \int_4 4
    \h4^\delta \h4^\delta} \rangle_0 = - \r_0 D_0 \int_4 \langle 12346
  \h3^{\gamma} \h4^\delta \h4^\delta \h6^a \h9^b \rangle_0 \\
   & = -2\r_0 D_0 \int_4 \{ \langle 1 \h3^{\gamma} \rangle_0 [
      \langle 3 \h4^\delta \rangle_0 \langle 4 \h6^a \rangle_0 \langle
      6 \h9^b \rangle_0 \langle 2 \h4^\delta \rangle_0 + \langle 3
      \h4^\delta \rangle_0 \langle 4 \h9^b \rangle_0 \langle 2 \h6^a
      \rangle_0 \langle 6 \h4^\delta \rangle_0 + \langle 3 \h6^a
      \rangle_0 \langle 6 \h4^\delta \rangle_0 \langle 4 \h9^b
      \rangle_0 \langle 2 \h4^\delta \rangle_0 ] \\
  & \qquad\qquad\quad + \langle 1 \h4^\delta \rangle_0 [ \langle 4
      \h6^a \rangle_0 \langle 6 \h9^b \rangle_0 \langle 3 \h4^\delta
      \rangle_0 + \langle 4 \h9^b \rangle_0 \langle 3 \h6^a \rangle_0
      \langle 6 \h4^\delta \rangle_0 ] \langle 2 \h3^{\gamma}
    \rangle_0 +\langle 1 \h6^a \rangle_0 \langle 6 \h4^\delta
    \rangle_0 \langle 4 \h9^b \rangle_0 \langle 3 \h4^\delta \rangle_0
    \langle 2 \h3^{\gamma} \rangle_0 \\
  & \qquad\qquad\quad + \langle 1 \h4^\delta \rangle_0 [ \langle 4
      \h6^a \rangle_0 \langle 6 \h3^{\gamma} \rangle_0 \langle 3 \h9^b
      \rangle_0 ] \langle 2 \h4^\delta \rangle_0 \\
  & \qquad\qquad\quad + \langle 1 \h4^\delta \rangle_0 [ \langle 4
      \h3^{\gamma} \rangle_0 \langle 3 \h9^b \rangle_0 \langle 6
      \h4^\delta \rangle_0 + \langle 4 \h9^b \rangle_0 \langle 6
      \h3^{\gamma} \rangle_0 \langle 3 \h4^\delta \rangle_0 ] \langle
    2 \h6^a \rangle_0 \\
  & \qquad\qquad\quad + \langle 1 \h6^a \rangle_0 [ \langle 6
      \h3^{\gamma} \rangle_0 \langle 3 \h4^\delta \rangle_0 \langle 4
      \h9^b \rangle_0 + \langle 6 \h4^\delta \rangle_0 \langle 4
      \h3^{\gamma} \rangle_0 \langle 3 \h9^b \rangle_0 ] \langle 2
    \h4^\delta \rangle_0 \}
\end{split}\end{equation}
and
\begin{equation}\begin{split} \la{e10}
  \frac{1}{2} & \langle 1236 9 \h3^{\gamma} \h6^a \h9^b e^{-D_0 \int_4
    4 \h4^\delta \h4^\delta} \rangle_0 = \frac{1}{2} \langle 1236 9
  \h3^{\gamma} \h6^a \h9^b \rangle_0 \\
  & = \langle 1 3 \rangle_0 [\langle 2 \h6^a \rangle_0 \langle 6 \h9^b
    \rangle_0 \langle 9 \h3^{\gamma} \rangle_0 ] + \langle 1 6
  \rangle_0 \langle 3 \h9^b \rangle_0 \langle 9 \h6^a \rangle_0
  \langle 2 \h3^{\gamma} \rangle_0 \\
  & \quad + \langle 1 \h3^{\gamma} \rangle_0 [ \langle 3 \h9^b
    \rangle_0 \langle 9 \h6^a \rangle_0 \langle 6 2 \rangle_0 +
    \langle 2 \h6^a \rangle_0 \langle 3 9 \rangle_0 \langle 6 \h9^b
    \rangle_0 + \langle 2 \h6^a \rangle_0 \langle 3 \h9^b \rangle_0
    \langle 6 9 \rangle_0 ] \\
  & \quad + [ \langle 1 \h6^a \rangle_0 \langle 6 \h9^b \rangle_0
    \langle 9 \h3^{\gamma} \rangle_0 ] \langle 3 2 \rangle_0 + \langle
  1 \h6^a \rangle_0 [ \langle 3 \h9^b \rangle_0 \langle 6 9 \rangle_0
    \langle 2 \h3^{\gamma} \rangle_0 + \langle 3 9 \rangle_0 \langle 6
    \h9^b \rangle_0 \langle 2 \h3^{\gamma} \rangle_0 ] \\
  & \quad + [ \langle 1 6 \rangle_0 \langle 3 \h6^a \rangle_0 +
    \langle 1 \h6^a \rangle_0 \langle 6 3 \rangle_0 ] \langle 2 \h9^b
  \rangle_0 \langle 9 \h3^{\gamma} \rangle_0 + \langle 1 \h6^a
  \rangle_0 \langle 6 \h3^{\gamma} \rangle_0 [ \langle 3 \h9^b
    \rangle_0 \langle 9 2 \rangle_0 + \langle 2 \h9^b \rangle_0
    \langle 9 3 \rangle_0 ].
\end{split}\end{equation}

We now need to calculate the correlation function itself up to the
first order of the bare perturbation expansion:
\begin{equation}\begin{split} \la{e11}
  \langle 13 \rangle & = \langle 13 \rangle_\text{f} + \langle 13
  S_\text{dis} \rangle_\text{f} + O(\l^2) \\
  & = \langle 13 \rangle_0 + \frac{1}{2} \l D_0^2 \int_6 \int_9 [
    \nabla_6^a \nabla_6^b \Phi(69) ] \left\{ \r_0^2 \langle 13 \h6^a
  \h9^b \rangle_0 - 2 \r_0 D_0 \int_4 \langle 1346 \h4^\delta
  \h4^\delta \h6^a \h9^b \rangle_0 + \langle 13 6 9 \h6^a \h9^b
  \rangle_0 \right\} + O(\l^2) \\
  & = \langle 13 \rangle_0 + \l D_0^2 \int_6 \int_9 [ \nabla_6^a
    \nabla_6^b \Phi(69) ] \left\{ \r_0^2 \langle 1 \h6^a \rangle_0
  \langle 3 \h9^b \rangle_0 \right. \\
  & \qquad\qquad - 2 \r_0 D_0 \int_4 ( \langle 1 \h4^\delta \rangle_0
          [ \langle 4 \h6^a \rangle_0 \langle 6 \h9^b \rangle_0
            \langle 3 \h4^\delta \rangle_0 + \langle 4 \h9^b \rangle_0
            \langle 3 \h6^a \rangle_0 \langle 6 \h4^\delta \rangle_0 ]
          + \langle 1 \h6^a \rangle_0 \langle 6 \h4^\delta \rangle_0
          \langle 4 \h9^b \rangle_0 \langle 3 \h4^\delta \rangle_0 )
          \\
  & \qquad\qquad \left. + \langle 1 6 \rangle_0 \langle 3 \h9^b
          \rangle_0 \langle 9 \h6^a \rangle_0 + \langle 1 \h6^a
          \rangle_0 \langle 3 9 \rangle_0 \langle 6 \h9^b \rangle_0 +
          \langle 1 \h6^a \rangle_0 \langle 3 \h9^b \rangle_0 \langle
          6 9 \rangle_0 \right\} + O(\l^2)
\end{split}\end{equation}
and
\begin{equation}\begin{split} \la{e12}
  \langle 32 \rangle & = \langle 32 \rangle_0 + \l D_0^2 \int_6 \int_9
          [ \nabla_6^a \nabla_6^b \Phi(69) ] \left\{ \r_0^2 \langle 3
          \h6^a \rangle_0 \langle 2 \h9^b \rangle_0 \right. \\
  & \qquad\qquad - 2 \r_0 D_0 \int_4 ( \langle 3 \h4^\delta \rangle_0
             [ \langle 4 \h6^a \rangle_0 \langle 6 \h9^b \rangle_0
               \langle 2 \h4^\delta \rangle_0 + \langle 4 \h9^b
               \rangle_0 \langle 2 \h6^a \rangle_0 \langle 6
               \h4^\delta \rangle_0 ] + \langle 3 \h6^a \rangle_0
             \langle 6 \h4^\delta \rangle_0 \langle 4 \h9^b \rangle_0
             \langle 2 \h4^\delta \rangle_0 ) \\
  & \qquad\qquad \left. + \langle 2 6 \rangle_0 \langle 3 \h9^b
             \rangle_0 \langle 9 \h6^a \rangle_0 + \langle 2 \h6^a
             \rangle_0 \langle 3 9 \rangle_0 \langle 6 \h9^b \rangle_0
             + \langle 2 \h6^a \rangle_0 \langle 3 \h9^b \rangle_0
             \langle 6 9 \rangle_0 \right\} + O(\l^2).
\end{split}\end{equation}
The first term in each integral corresponds to the first-order
contribution to the time-persistent part of the correlation function,
$\l \r_0^2 \Phi(13)$ in $\langle 13 \rangle$ and $\l \r_0^2 \Phi(32)$
in $\langle 32 \rangle$. It should be discarded to avoid
double-counting with Eq.~\eqref{e8} and we accordingly define $\langle
13 \rangle_\text{c}$ and $\langle 32 \rangle_\text{c}$, where c stands
for connected, from Eqs.~\eqref{e11} and \eqref{e12} without this
term.

Therefore, we identify the first-order renormalization for the average
$\langle 12 3 \h3^{\gamma} \rangle$, apart from the time-persistent
terms, as
\begin{equation} \la{e13}
  \langle 12 3 \h3^{\gamma} \rangle = \langle 1 \h3^{\gamma} \rangle
  \langle 3 2 \rangle_\text{c} + \langle 1 3 \rangle_\text{c} \langle
  2 \h3^{\gamma} \rangle - 2 D_0 \int_4 \langle 1 \h4^\delta \rangle
  \langle 4 3 \h3^{\gamma} \rangle \langle 2 \h4^\delta \rangle.
\end{equation}
The bare first-order expression for $\langle 4 3 \h3^{\gamma} \rangle$
is given by [see Eq.~\eqref{d9}]
\begin{equation} \la{e14}
 \langle 4 3 \h3^{\gamma} \rangle = \l \r_0 D_0^2 \int_6 \int_9 [
   \nabla_6^a \nabla_6^b \Phi(69) ] \langle 4 \h6^a \rangle_0 \langle
 6 \h3^{\gamma} \rangle_0 \langle 3 \h9^b \rangle_0 + O(\l^2),
\end{equation}
and can be spotted in Eq.~\eqref{e9}.  Again, the corresponding term
appears associated with the first-order expansion of $\o{R}$, as in
Eq.~\eqref{d4}.

Collecting Eqs.~\eqref{e4} and \eqref{e13}, we have
\begin{equation}\begin{split} \la{e15}
  \nabla_3^{\gamma} [ \r_0 \langle 12 \h3^{\gamma} \rangle & + \langle
    123 \h3^{\gamma} \rangle ] \\
  & = \nabla_3^{\gamma} [ \langle 1
    \h3^{\gamma} \rangle \langle 3 2 \rangle_\text{c} + \langle 1 3
    \rangle_\text{c} \langle 2 \h3^{\gamma} \rangle ] -2 \int_4
  \langle 1 \h4^\delta \rangle [ \r_0 D_0 \nabla_4^2 \langle 4 \h3
    \rangle + D_0 \nabla_3^{\gamma}\langle 4 3 \h3^{\gamma} \rangle ]
  \langle 2 \h4^\delta \rangle \\
  & = \nabla_3^{\gamma} [ \langle 1 \h3^{\gamma} \rangle \langle 3 2
    \rangle_\text{c} + \langle 1 3 \rangle_\text{c} \langle 2
    \h3^{\gamma} \rangle ] + 2i \int_4 \langle 1 \h4^\delta \rangle
  \o{R}(43) \langle 2 \h4^\delta \rangle,
\end{split}\end{equation}
where we used [see Eq.~\eqref{d5}] $\o{R}(43) = i \r_0 D_0 \nabla_4^2
\langle 4 \h3 \rangle + i D_0 \nabla_3^{\gamma} \langle 4 3
\h3^{\gamma} \rangle$.

We are now ready to write down the first-order renormalized dynamical
equation for the density correlation function. It reads, ignoring $2
\o{R}(21)$ and performing the first integral in the right-hand side of
Eq.~\eqref{e1},
\begin{multline} \la{e16}
  ( \partial_t -D_0 \nabla^2 ) C(12) = - D_0 \nabla^2 C_\text{d}(12)
  - i \l D_0^2 \nabla^\alpha \left( \int_3 [ \nabla^\alpha \Phi(13) ]
  \nabla_3^{\gamma} [ \langle 1 \h3^{\gamma} \rangle \langle 3 2
    \rangle_\text{c} + \langle 1 3 \rangle_\text{c} \langle 2
    \h3^{\gamma} \rangle ] \right) \\
  + 2 \l D_0^2 \nabla^\alpha \left( \int_3 [ \nabla^\alpha \Phi(13) ]
  \int_4 \langle 1 \h4^\gamma \rangle \o{R}(43) \langle 2 \h4^\gamma
  \rangle \right),
\end{multline}
or, integrating out the physical response function in the last term
and introducing the connected density correlation function,
\begin{multline} \la{e17}
  ( \partial_t -D_0 \nabla^2 ) F(12) = \\
  - i \l D_0^2 \nabla^\alpha \left( \int_3 [ \nabla^\alpha \Phi(13) ]
  \nabla_3^{\gamma} [ \langle 1 \h3^{\gamma} \rangle \langle 3 2
    \rangle_\text{c} + \langle 1 3 \rangle_\text{c} \langle 2
    \h3^{\gamma} \rangle ] \right) + 2 \l \rho_0 D_0^2 \nabla^\alpha
  \left( \int_4 [ \nabla^\alpha \Phi(14) ] \langle 1 \h4^\gamma
  \rangle \langle 2 \h4^\gamma \rangle \right).
\end{multline}

We then get the Fourier-transformed equation of motion,
\begin{equation}\begin{split} \la{e18}
  ( \partial_t + \Gam_k ) F_k(t-t') & = - \l D_0^2 \int_{-\infty}^t ds
    \int_{\bf q} {\b q} \cdot {\b p} [{\b k} \cdot {\b q} \Phi_q ]
    G_p(t-s) F_k(s-t') \\
  & \quad + \l D_0^2 \int_{-\infty}^{t'} ds \int_{\bf q} {\b k} \cdot
    {\b q} [{\b k} \cdot {\b q} \Phi_q ] F_p(t-s) G_k(t'-s) \\
  & \quad + 2 \l \rho_0 D_0^2 \int_{-\infty}^{t'} ds \int_{\bf q} {\b
      k} \cdot {\b p} [{\b k} \cdot {\b q} \Phi_q ] G_p(t-s)
    G_k(t'-s),
\end{split}\end{equation}
which we rewrite as
\begin{gather} 
  (\partial_t +\Gam_k) F_k(t-t') = - \int_{t'}^t ds \S_k(t-s)
  F_k(s-t') + N_k(t-t'), \la{e19} \\
  \begin{split}
    N_k(t-t') & \equiv - \l D_0^2 \int_{-\infty}^{t'} ds \int_{\bf q}
    {\b q} \cdot {\b p} [{\b k} \cdot {\b q} \Phi_q ] G_p(t-s)
    F_k(t'-s) \\
    & \quad + \l D_0^2 \int_{-\infty}^{t'} ds \int_{\bf q} {\b k}
    \cdot {\b q} [{\b k} \cdot {\b q} \Phi_q ] F_p(t-s) G_k(t'-s) \\
    & \quad + 2 \l \r_0 D_0^2 \int_{-\infty}^{t'} ds \int_{\bf q} {\b
      k} \cdot {\b p} [{\b k} \cdot {\b q} \Phi_q ] G_p(t-s)
    G_k(t'-s). \la{e20}
  \end{split}
\end{gather}

Note that, in the absence of simple relations between the response and
correlation functions beyond the FDR, it is not guaranteed that the
sum of three integrals in Eq.~\eqref{e20} actually reduces to a local
function of time as posited in Eq.~\eqref{e19}.  For the same reason,
it is not obvious that the time derivative of $N_k(t-t')$, given by
\begin{equation}\begin{split} \la{e21}
  \partial_t N_k(t-t') & = \r_0 \S_k(t-t') - \l D_0^2
  \int_{-\infty}^{t'} ds \int_{\bf q} {\b q} \cdot {\b p} [{\b k}
    \cdot {\b q} \Phi_q ] G_p(t-s) \o{R}_k(t'-s) \\
   & \quad - \l D_0^2 \int_{-\infty}^{t'} ds \int_{\bf q} {\b k}
  \cdot {\b q} [{\b k} \cdot {\b q} \Phi_q ] \o{R}_p(t-s)
  G_k(t'-s) \\
  & \quad + 2\l \r_0 D_0^2 \int_{-\infty}^{t'} ds \int_{\bf q} {\b k}
  \cdot {\b p} [{\b k} \cdot {\b q} \Phi_q ] \big[\partial_t
    G_p(t-s)\big] G_k(t'-s),
\end{split}\end{equation}
where the FDR and integrations by parts have been used, equals
$\r_0\S_k(t-t')-L_k(t-t')$, as required for consistency with the FDR.
In fact, general arguments support the exact opposite
\cite{MiyRei05JPA,AndBirLef06JSMTE}.

However, using the following first-order consistent substitutions [see
  Eqs.~\eqref{eqFORT7}, \eqref{eqFORT8}, and \eqref{eqFORT1a}]
\begin{subequations}
\begin{gather}
  F_k(t) = \r_0 G_k (t) + O(\l), \la{sub1} \\
  \o{R}_k(t) = \r_0 D_0 k^2 G_k (t) + O(\l), \la{sub2} \\
  \partial_t G_k(t) = - D_0 k^2 G_k(t) + O(\l), \la{sub3} 
\end{gather}  
\end{subequations}
in the above integrals, one can easily show, with calculations similar
to those performed in Sec.~\ref{sec:FOBT}, that these properties hold
to first order.  One can use, in particular, the identity $k^2 {\b q}
\cdot {\b p} + p^2 ({\b k} \cdot {\b q} + 2 {\b k} \cdot {\b p}) ={\b
  k} \cdot {\b p} (k^2 +p^2)$.

As an example, we may show that, within the first order,
Eq.~\eqref{e20} is indeed compatible with Eq.~\eqref{eqFORT12}, which
is FDR-consistent.  One first uses Eq.~\eqref{sub1} in Eq.~\eqref{e20}
to get
\begin{equation} 
  N_k(t-t') = \l \rho_0 D_0^2 \int_{-\infty}^{t'} ds \int_{\bf q} [{\b
      k} \cdot {\b q} \Phi_q ] [ p^2 G_p(t-s) G_k(t'-s) + G_p(t-s) k^2
    G_k(t'-s)] + O(\l^2).
\end{equation}
Then, Eq.~\eqref{sub3} gives
\begin{equation} 
  N_k(t-t') = \l \rho_0 D_0 \int_{-\infty}^{t'} ds \int_{\bf q} [{\b
      k} \cdot {\b q} \Phi_q ] \partial_s [ G_p(t-s) G_k(t'-s) ] +
  O(\l^2),
\end{equation}
hence
\begin{equation} 
  N_k(t-t') = \l \rho_0 D_0 \int_{\bf q} [{\b k} \cdot {\b q} \Phi_q ]
  G_p(t-t') + O(\l^2).
\end{equation}
Since $G_p(0)=1$ and $\int_{\bf q} {\b k} \cdot {\b q} \Phi_q = 0$ by
isotropy, this can be rewritten as
\begin{equation} 
  N_k(t-t') = \l \rho_0 D_0 \int_{t'}^{t} ds \int_{\bf q} [{\b k}
    \cdot {\b q} \Phi_q ] \partial_s G_p(s-t') + O(\l^2),
\end{equation}
and, with one last use of Eq.~\eqref{sub3}, one gets
\begin{equation} 
  N_k(t-t') = - \l D_0 \int_{t'}^{t} ds \int_{\bf q} [{\b k} \cdot {\b
      q} \Phi_q ] [\rho_0 D_0 p^2 G_p(s-t')] + O(\l^2).
\end{equation}
Truncated to first order, this is nothing but Eq.~\eqref{eqFORT12}.

Finally, $N_k(t-t')$ can also be written as
\begin{equation} \la{e25}
  N_k(t-t') = - \int_{-\infty}^{t'} ds \Sigma_k(t-s) F_k(t'-s)
  + \int_{-\infty}^{t'} ds D_k(t-s) [ \r_0 \Gamma_k G_k(t'-s) ],
\end{equation}
with
\begin{equation} \la{e26}
  D_k(t) = M_k(t) + \frac{2}{\rho_0 \Gamma_k} L_k(t),
\end{equation}
where $M_k(t)$ is the mode-coupling kernel defined in
Eq.~\eqref{eqFORT27}.  One can thus readily transpose the discussion
around Eq.~\eqref{eqFOBTennalt} of the bare theory to the renormalized
framework.

%

\end{document}